\documentclass{article}
\usepackage{dsfont}

\usepackage{amsmath,amssymb,graphicx}
\usepackage{diagbox}
\usepackage[nosort]{cite}

\usepackage{tikz}
\usetikzlibrary{calc} \usetikzlibrary{patterns} \usetikzlibrary{decorations.pathreplacing} \usetikzlibrary{decorations.markings} \usetikzlibrary{decorations.pathmorphing} \usetikzlibrary{positioning}

\usepackage{color}
\definecolor{darkred}{rgb}{0.65,0.15,0}
\definecolor{newgreen}{rgb}{0.2,0.62,0.14}
\usepackage{hyperref}

\hypersetup{pdfborder={0 0 0},colorlinks=true,urlcolor=blue,citecolor=blue,linkcolor=darkred,linktocpage=true}

\numberwithin{equation}{section}

\usepackage[shadow,textwidth=2.7cm]{todonotes}
\usepackage{ifthen}
\reversemarginpar

\DeclareMathSymbol{\shortminus}{\mathbin}{AMSa}{"39}

\newcommand{\seed}[1]{\varphi(#1)}
\newcommand{\ZZ}{\mathbb{Z}}

\newcommand{\poin}{{\rm B}(\mathbb{Z})\backslash {\rm SL}(2,\mathbb{Z})}
\newcommand{\dd}{\mathrm{d}}

\def\half{{\scriptstyle \frac 12}}

\def\sevenh{{\scriptstyle \frac 72}}
\def\threeh{{\scriptstyle \frac 32}}
\def\fiveh{{\scriptstyle \frac 52}}

\def\nineh{{\scriptstyle \frac 92}}

\newcommand{\seedfn}[5]{\upsilon\left(#1,#2,#3,#4\vert #5\right)}
\newcommand{\modfn}[5]{\Upsilon\left(#1,#2,#3,#4\vert #5\right)}
\newcommand{\modfntwo}[4]{\Upsilon\left(#1,#2,#3,#4\right)}
\newcommand{\SLtwoZ}{{\rm SL}(2,\mathbb{Z})}

\newcommand{\Eisenstein}[2]{\mathcal{E}(#1\vert #2)}
\newcommand{\normEisenstein}[2]{{\rm E}(#1 \vert #2)}

\newcommand{\genEisenstein}[4]{\mathcal{E}\left(#3; #1,#2\big\vert#4\right)}
\newcommand{\genEisensteinSeed}[4]{{\rm e}\!\left(#3; #1,#2\big\vert#4\right)}
\newcommand{\average}[1]{\langle #1\rangle}
\newcommand{\fzero}[5]{\Upsilon_0(#1,#2,#3,#4\vert #5)}
\newcommand{\fm}[6]{\Upsilon_{#1}(#2,#3,#4,#5\vert #6)}
\newcommand{\eigen}{{\rm Y}}

\begin{document}
\begin{center}

{\bf {\LARGE \sc 
Two string theory flavours \\
of generalised Eisenstein series}}

\vspace{6mm}
\normalsize
{\large  Daniele Dorigoni and Rudolfs Treilis}

\vspace{10mm}
${}${\it Centre for Particle Theory \& Department of Mathematical Sciences\\
Durham University, Lower Mountjoy, Stockton Road, Durham DH1 3LE, UK}

\vspace{30mm}

\hrule

\vspace{6mm}

 \begin{tabular}{p{14cm}}
 
Generalised Eisenstein series are non-holomorphic modular invariant functions of a complex variable, $\tau$, subject to a particular inhomogeneous Laplace eigenvalue equation on the hyperbolic upper-half $\tau$-plane. Two infinite classes of such functions arise quite naturally within different string theory contexts.
A first class can be found by studying the coefficients of the effective action for the low-energy expansion of type IIB superstring theory, and relatedly in the analysis of certain integrated four-point functions of stress tensor multiplet operators in $\mathcal{N} = 4$ supersymmetric Yang-Mills theory.
A second class of such objects is known to contain all two-loop modular graph functions, which are fundamental building blocks in the low-energy expansion of closed-string scattering amplitudes at genus one.
In this work, we present a Poincar\'e series approach that unifies both classes of generalised Eisenstein series and manifests certain algebraic and differential relations amongst them.
We then combine this technique with spectral methods for automorphic forms to find general and non-perturbative expansions at the cusp $\tau \to i \infty$.   
Finally, we find intriguing connections between the asymptotic expansion of these modular functions as $\tau \to 0$ and the non-trivial zeros of the Riemann zeta function.

\end{tabular}

\vspace{2mm}
\hrule
\end{center}

\thispagestyle{empty}

\newpage
\setcounter{page}{1}

\setcounter{tocdepth}{2}
\tableofcontents

\bigskip

\section{Introduction}
\label{sec:1}

There is a multitude of places where modular invariance plays a crucial  r\^ole in string theory and quantum field theory.  For example, the modular group,~$\SLtwoZ$, appears as U-duality group~\cite{Hull:1994ys} of ten-dimensional type IIB string theory and, via the~${\rm AdS/CFT}$ correspondence, as Montonen-Olive electro-magnetic duality group~\cite{Montonen:1977sn,Witten:1978mh,Osborn:1979tq} of its holographic dual~$\mathcal{N}=4$ supersymmetric Yang-Mills theory (SYM).
Similarly, in closed-string perturbation theory the modular group arises as mapping class group for genus-one world-sheet,~i.e.~for strings whose world-sheet is a two-dimensional torus, strongly constraining the low-energy expansion of string scattering amplitudes~\cite{Green:1999pv,Green:2008uj,DHoker:2015gmr}.

A consequence of modularity particularly relevant for the present work is that physical observables must be invariant, or more generally covariant, under the modular group, i.e. physical observables are automorphic forms with respect to the modular group. 
Although the world of modular forms is extremely vast and diverse, we focus our attention to an interesting class of automorphic forms relevant for string theory and known as \textit{non-holomorphic Eisenstein series} and \textit{generalised Eisenstein series}. For a recent and more general introduction on the broader subject we refer to the beautifully written set of lectures \cite{DHoker:2022dxx}.

    As we will discuss in more detail shortly, generalised Eisenstein series are non-holomorphic modular invariant functions of a complex variable, $\tau$, which parametrises the usual hyperbolic upper-half complex plane. These functions satisfy an inhomogeneous Laplace eigenvalue equation on the $\tau$-plane with sources bilinear in non-holomorphic Eisenstein series. The physical r\^ole of the parameter $\tau$, as well as the spectrum of eigenvalues and the details of the source terms do depend on the particular string theory calculation under consideration.

A first flavour of generalised Eisenstein series arises in the context of higher-derivative corrections to the low-energy effective action of type IIB superstring theory and in related calculations in the holographic dual $\mathcal{N} = 4$ SYM gauge theory.
Thanks to modular invariance, combined with supersymmetric arguments~\cite{Green:1997tv,Kiritsis:1997em,Pioline:1998mn,Green:1998by,Obers:1999um,Green:2005ba,Green:2014yxa}, the coefficients of certain higher-derivative operators in the low-energy effective action of type IIB superstring theory are computed exactly in terms of Eisenstein and generalised Eisenstein series where their argument, $\tau$, is given by the axio-dilaton vacuum expectation value.

On the dual side we consider $\mathcal{N}=4$ SYM theory with gauge group $SU(N)$,
for which it was argued in \cite{Binder:2019jwn} that certain integrated correlation functions of four stress-tensor superconformal primaries, computable via supersymmetric localisation, can be related to four-graviton scattering amplitudes in IIB superstring theory. 
A lot of progress has recently been made in this direction \cite{Chester:2019jas,Chester:2019pvm,Chester:2020vyz,Chester:2020dja,Dorigoni:2021bvj,Dorigoni:2021guq,Alday:2021vfb,Collier:2022emf,Dorigoni:2022zcr} and, using the holographic dictionary, these calculations were shown to reproduce and extend known results for the low-energy type IIB superstring effective action in ten-dimensional flat-space, as well as having striking implications for analogous considerations in ${\rm AdS}_5\times {\rm S}^5$.

In particular \cite{Chester:2020dja} considered the expansion of one such integrated correlator in the large-$N$ limit (with $N$ the rank of the gauge group) with fixed complexified Yang-Mills coupling, $\tau$. 
Order by order in $1/N$, Montonen-Olive duality constrains the coefficients of this expansion to be modular invariant functions of the complexified coupling $\tau$. While at half-integer orders in $1/N$ only Eisenstein series appear, at integer orders in $1/N$ we encounter an infinite class of generalised Eisenstein series akin to the higher-derivative coefficients just mentioned.

From a seemingly very different albeit closely related point of view, a second flavour of generalised Eisenstein series can be found while studying the low-energy expansion of closed-string perturbation theory at genus one.
The low-energy expansion for string amplitudes with toroidal world-sheet leads to the introduction of an infinite class of non-holomorphic and modular covariant building blocks usually named \textit{modular graph functions}~\cite{DHoker:2015gmr} and \textit{modular graph forms}~(MGFs)~\cite{DHoker:2016mwo,DHoker:2016quv}, where to a Feynman world-sheet diagram we associate a modular invariant or covariant function whose argument is the torus complex structure. Similarly to standard Feynman integrals, the number of loop-momenta dictates the complexity of the objects under consideration.
While one-loop MGFs evaluate to non-holomorphic Eisenstein series, two-loop MGFs are contained in a second infinite class of generalised Eisenstein series \cite{DHoker:2015gmr,Dorigoni:2021jfr,Dorigoni:2021ngn}.
 
Given the physical and mathematical importance of generalised Eisenstein series, a crucial problem is understanding their analytic, algebraic and differential properties. 
For this reason, a first approach is to try and represent these modular invariant functions as Poincar\'e series, an extremely convenient way of
rewriting a modular invariant function as a sum over images under the modular group of a simpler function, usually called \textit{seed} function.

In general, passing from a modular invariant function to its Poincar\'e seed reduces the functional complexity. This viewpoint was exploited in~\cite{DHoker:2015gmr,Ahlen:2018wng,Dorigoni:2019yoq,Basu:2020kka} to obtain Poincar\'e-series representations for various two-loop MGFs, then extended to all two-loop MGFs in \cite{DHoker:2019txf}. 
A streamlined approach was presented in \cite{Dorigoni:2021jfr} where a unified description was presented: Poincar\'e seeds for this second infinite family of generalised Eisenstein series, and hence for all two-loop MGFs, can be constructed from iterated integrals over single holomorphic Eisenstein series and their complex conjugates thus considerably simplifying their studies. However, it had already been noted in \cite{Dorigoni:2019yoq}, that this type of seed functions is ill-suited for describing the first class of generalised Eisenstein series relevant for higher-derivative corrections and integrated correlators.

One of our main results is the derivation of a new Poincar\'e series representation unifying both classes of generalised Eisenstein series. We introduce a new space of modular invariant functions embedding quite naturally both flavours of generalised Eisenstein series and manifesting many of their algebraic and differential properties. In particular, we find that this space of functions is closed under the action of the hyperbolic Laplacian and exploit this fact to clarify the origins of the spectrum of eigenvalues and respective source terms for the two classes of generalised Eisenstein series that have relevance to string theory.

We then combine our Poincar\'e series approach with methods coming from spectral analysis for automorphic forms, see e.g.~\cite{Iwaniec:2002,Terras}, to derive the complete asymptotic expansion at the cusp $\tau \rightarrow i \infty$ for both flavours of generalised Eisenstein series in the Fourier zero-mode sector.
As argued in~\cite{Dorigoni:2019yoq,Dorigoni:2020oon,Dorigoni:2022bcx}, resurgence analysis can be used to reconstruct the entire non-perturbative completion of the MGFs, i.e.~the exponentially suppressed terms, from their perturbative expansion around the cusp. With the help of spectral analysis we clarify these resurgence results and extend them to the case of higher-derivative corrections and integrated correlators where such non-perturbative terms can be interpreted as D-instanton/anti-D-instanton events~\cite{Green:2014yxa}.

Finally, from our analysis we can easily derive an asymptotic expansion for $\tau \rightarrow 0$, where surprisingly (and quite mysteriously) we find that the non-trivial zeros of the Riemann zeta function play a key r\^{o}le. While for MGFs the limit $\tau \rightarrow 0$ is simply a particular degeneration case of the toroidal world-sheet, for higher-derivative corrections and integrated correlators this limit corresponds to a strong coupling regime, hence extremely difficult to access by other means.
 
The rest of the paper is organised as follows. In section \ref{eq:Background} we properly define the key characters of our story, namely the generalised Eisenstein series, and review in more detail the string theory origins for these two infinite families thereof. In section \ref{sec:Poin} we propose a new Poincar\'e series representation, which provides a unifying framework to derive analytic and differential properties for both classes of generalised Eisenstein series. We also present various examples coming from higher-derivative corrections as well as MGFs. 
After a brief review of the Roelcke-Selberg spectral decomposition for automorphic forms, section \ref{sec:spectral} is devoted to extracting analytic properties of the modular functions discussed. We derive their asymptotic expansion at the cusp, $\tau \to i\infty$, where a connection is made with previous resurgence results, and their expansion at the origin, $\tau \to 0$, where we find intriguing connections with the non-trivial zeros of the Riemann zeta function. We end in section \ref{sec:Fin} with a brief summary and discussion on future directions. Some more technical details are contained in two appendices.

\section{String theory and generalised Eisenstein series}
\label{eq:Background}

In this section we briefly review two instances where string theory calculations give rise to generalised Eisenstein series. In particular, we firstly discuss how these modular functions are defined, how they arise in the study of the low-energy expansion of type IIB superstring theory and how this connects, via the holographic dictionary, with integrated correlators in $\mathcal{N}=4$ super-Yang-Mills theory. Secondly, we present a different class of generalised Eisenstein series arising in the low-energy expansion of genus-one string-scattering amplitudes, i.e. for a string with a toroidal world-sheet. Most of the topics presented in this section are reviewed in \cite{Dorigoni:2022iem}.

\subsection{Low-energy expansion of type IIB superstring theory}

When discussing the low-energy expansion of type IIB superstring theory, the group $\SLtwoZ$ is interpreted as the non-perturbative U-duality group of ten-dimensional type IIB string theory~\cite{Hull:1994ys}. 
In the classical theory the vacuum expectation value of the axio-dilaton scalar field,
$$\tau =\chi+ \frac{i}{g_s}= \Re(\tau)+i\,\Im(\tau)\,,$$
with $g_s$ the string coupling constant, parametrises the coset space  \begin{equation}
\mathcal{H}:={\rm SL}(2,\mathbb{R})/{\rm U}(1)=\{ \tau\in\mathbb{C}\, \vert\, \Im(\tau) >0 \}\,.\label{eq:FundDom}
\end{equation}
However, quantum corrections \cite{Gaberdiel:1998ui} generate an anomaly in the ${\rm U}(1)$ R-symmetry thus breaking ${\rm SL}(2,\mathbb{R})$ to $\SLtwoZ$.
 This U-duality symmetry group, $\SLtwoZ$, acts on the axio-dilaton in the standard way
\begin{equation}
    \label{eq_mod_trans}
    \gamma =
    \begin{pmatrix}
    a & b\\
    c & d
    \end{pmatrix}
    \in \SLtwoZ\,,
    \qquad
    \gamma\cdot\tau := \frac{a\tau + b}{c\tau + d}\,.
\end{equation}
 Since the axio-dilaton includes the string coupling, $g_s$, U-duality is an extremely powerful and non-perturbative symmetry, relating perturbative and non-perturbative effects in $g_s$. 

 The low-energy expansion of type IIB superstring theory is therefore expected to be invariant (or covariant) under $\SLtwoZ$, where $\tau$ parameterises a fundamental domain that can be chosen to be 
 \begin{align}
 \mathcal{F} &:=\label{eq:FundSL2} \SLtwoZ \backslash \mathcal{H}   \\
 &\notag \phantom{:}=  \Big\lbrace \tau\in\mathcal{H}\,\Big\vert\,  |\tau|> 1\,,\, -\frac{1}{2} < \Re(\tau)\leq \frac{1}{2} \Big\rbrace \cup \Big\lbrace  \tau\in\mathcal{H}\,\Big\vert\, |\tau| = 1\,, \,0\leq \Re(\tau)\leq \frac{1}{2}\Big\rbrace \,. 
 \end{align}

At low energy type IIB supergravity receives corrections coming from excited string states which can be neatly assembled in an effective Lagrangian.
Focusing for simplicity only on four-graviton interactions, we expect to find an effective Lagrangian containing the standard Einstein-Hilbert term, as well as an infinite tower of higher-derivative corrections schematically of the form $d^{2n}R^4$, where $R^4$ is a certain contraction of Riemann tensors and $d$ is the covariant derivative. 
For $n\leq 3$ these terms are fixed by supersymmetry to be of the form (in the string frame)
\begin{equation}\label{SUGRA_Lagran}
    \mathcal{L}_{{\rm eff}} = (\alpha')^{-4}g_s^{-2}R + (\alpha')^{-1}g_s^{-\frac{1}{2}}\pi^{\frac{3}{2}}\Eisenstein{\threeh}{\tau}R^4+\alpha'g_s^{\frac{1}{2}}\pi^{\frac{5}{2}}\Eisenstein{\fiveh}{\tau}d^4R^4-(\alpha')^2g_s \pi^3\genEisenstein{\threeh}{\threeh}{4}{\tau} d^6R^4+... \,,
\end{equation}
where $\alpha' = \ell_s^2$ is the square of the string length scale.

As expected, the leading term when $\alpha'\to0$ is simply given by the Einstein-Hilbert term (where $R$ is the Ricci scalar). Although we only wrote the bosonic part, this term comes with its supersymmetric completion, involving other bosonic as well as fermionic fields, reproducing the type IIB supergravity lagrangian in ten dimensions. 

For the higher-derivative corrections here reviewed, we have that maximal supersymmetry uniquely fixes the Lorentz contractions of the tensor indices and forbids the presence of $R^2$ and  $R^3$ interactions. The first correction is proportional to $R^4$ \cite{Gross:1986iv, Grisaru:1986dk} which is a $1/2$-BPS operator, i.e. it preserves only $16$ of the $32$ supersymmetries associated with ten-dimensional maximal supersymmetry.  
Similarly, the higher-derivative term $d^4R^4$ is $1/4$-BPS while $d^6R^4$ is $1/8$-BPS and it is the last term to be protected by supersymmetry.  

The ellipsis in \eqref{SUGRA_Lagran} represents various supersymmetric completions, as well as higher-order terms and terms contributing to higher-point amplitudes. A particular class of interesting higher-point BPS amplitudes involve the scattering of four gravitons with certain massless fields of type IIB supergravity carrying specific ${\rm U}(1)$ charges~\cite{Schwarz:1983qr,Howe:1983sra} and transforming covariantly under U-duality. The modular properties of these amplitudes have been analysed in~\cite{Green:2019rhz}, while their connection with the holographic dual picture of integrated correlators in $\mathcal{N}=4$ SYM is presented in~\cite{Green:2020eyj,Dorigoni:2021rdo}. We will not be discussing these corrections here.

 Interestingly, the coefficients of the higher-derivative and BPS protected corrections displayed in \eqref{SUGRA_Lagran} can be computed exactly and are expressible in terms of special modular invariant functions. 
In particular, we see that the coefficient of the $R^4$  \cite{Green:1997tv, Green:1997as,Green:1998by} and the $d^4R^4$ \cite{Green:1999pu} interactions involve \textit{non-holomorphic Eisenstein series}:
\begin{align}\
    \Eisenstein{s}{\tau} &\label{Eisenstein_def}\!:= \sum_{(m,n)\neq(0,0)}\frac{(y/\pi)^s}{|m+n\tau|^{2s}} \\
   &\notag= \frac{2\zeta(2s)}{\pi^s}y^s + \frac{2\xi(2s-1)}{\Gamma(s)}y^{1-s}+\frac{4}{\Gamma(s)}\sum_{k\neq 0}|k|^{s-\frac{1}{2}}\sigma_{1-2s}(k)y^{\frac{1}{2}}K_{s-\frac{1}{2}}(2\pi |k|y)e^{2\pi ikx}
\end{align}
with $\tau=x+iy\in\mathcal{H}$ and  $\Re(s)>1$ for now. 
We denote by $\xi(s):=\pi^{-s/2}\Gamma(s/2)\zeta(s)$ the completed zeta function invariant under reflection $\xi(s)=\xi(1-s)$, while $\sigma_a(k):=\sum_{d|k}d^a$ is a divisor sum and $K_\nu(y)$ is a modified Bessel function of the second kind. 
 
The coefficient of $d^6R^4$ \cite{Green:2005ba} is a new kind of object known in the literature as a \textit{generalised non-holomorphic Eisenstein series}, and which is defined as the unique modular-invariant solution to the differential equation
\begin{equation}\label{gen_Eisenstein_defIntro}
    [\Delta -\lambda(\lambda-1)]\,\genEisenstein{s_1}{s_2}{\lambda}{\tau} = \Eisenstein{s_1}{\tau}\Eisenstein{s_2}{\tau},
\end{equation}
where $\Delta := y^2(\partial_x^2+\partial_y^2)$ is the hyperbolic Laplace operator in $\tau$. The solution is taken subject to the boundary condition that the term of order $y^\lambda$ in the Laurent polynomial around the cusp $y\gg 1$ has vanishing coefficient. 
This boundary condition uniquely fixes the modular-invariant solution, since the Eisenstein series is the unique modular-invariant solution with polynomial growth at the cusp to the differential equation
\begin{equation}\label{Eisenstein_eigval}
    [\Delta - \lambda(\lambda-1)]\,\Eisenstein{\lambda}{\tau} = 0\,.
\end{equation}

Beyond $d^6R^4$, higher derivative corrections are not supersymmetrically protected any longer, hence the same methods leading to the exact results presented above cannot be applied.
However, novel results have been obtained by considering the holographic dual of type IIB superstring theory on ${\rm AdS}_5\times{ \rm S}^5$, notoriously given by $\mathcal{N} = 4$ SYM theory with gauge group $SU(N)$. 
Thanks to supersymmetric localisation \cite{Binder:2019jwn}, we can obtain various specific $\mathcal{N}=4$ four-point integrated correlators of superconformal primaries of the stress-energy tensor multiplet\footnote{Very recently exciting results have been obtained in \cite{Paul:2022piq,Brown:2023cpz,Paul:2023rka,Brown:2023why} for a different class of integrated four-point functions of local operators, as well as for integrated two-point functions of superconformal primaries of the stress-energy tensor multiplet in the presence of a half-BPS line defect \cite{Pufu:2023vwo}.} by taking different combinations of four derivatives of the partition function for the $\mathcal{N} = 2^*$ theory (a massive deformation of $\mathcal{N} = 4$) on a squashed ${\rm S}^4$ with respect to different parameters (squashing, mass and complexified coupling).

In \cite{Chester:2019jas,Chester:2020vyz} the authors exploited these supersymmetric localisation results to compute the large-$N$ expansion of such integrated correlators while keeping fixed the modular parameter, $\tau$, now denoting the Yang-Mills complexified coupling $\tau = \theta/ 2\pi +  4\pi i/ g_{_{YM}}^2$. 
Using the ${\rm AdS/CFT}$ dictionary, we identify  $g_{_{Y\!M}}^2  = 4 \pi g_s$ and $( g_{_{Y\!M}}^2 N)^\half= L^2/\alpha'$, where $g_s$ is the string coupling constant and $L$ is the scale of ${\rm AdS}_5\times {\rm S}^5$. Hence the large-$N$ limit of such integrated correlators can help us in understanding higher derivative corrections in type IIB superstring theory on ${\rm AdS}_5\times{ \rm S}^5$ beyond $d^6R^4$ \cite{Chester:2019jas,Chester:2020vyz} as well as non-perturbative effects in $\alpha'$ \cite{Hatsuda:2022enx,Dorigoni:2022cua}.

 As a consequence of $\mathcal{N}=4$ Montonen–Olive duality (also known as ${\rm S}$-duality), order by order at large-$N$ we must have an expansion with coefficients that are non-holomorphic modular invariant functions of $\tau$. 
From \cite{Chester:2019jas,Chester:2020vyz} we know that half-integer orders in $1/N$ produce only non-holomorphic Eisenstein series. However, for integer orders in $1/N$ this expansion is conjectured to involve an infinite class of generalised Eisenstein series, $\genEisenstein{s_1}{s_2}{\lambda}{\tau}$, 
 with half-integer indices $s_1,s_2$ and spectrum of eigenvalues $\lambda\in {\rm Spec}_1(s_1,s_2)$ constrained by
\begin{equation}
\lambda \in {\rm Spec}_1(s_1,s_2) := \{s_1{+}s_2{+}1,\, s_1{+}s_2{+}3,\,s_1{+}s_2{+}5,\,...\}\,,\qquad \qquad s_1,s_2 \in \mathbb{N}+\frac{1}{2}.\label{eq:spec1}
\end{equation} 

The coefficient of the $d^6R^4$ higher-derivative correction, $\genEisenstein{\threeh}{\threeh}{4}{\tau}$, in \eqref{SUGRA_Lagran} is simply the first instance of generalised Eisenstein series belonging to this first class \eqref{eq:spec1}.
For future reference, we notice that within this first flavour of generalised Eisenstein series, $\genEisenstein{s_1}{s_2}{\lambda}{\tau}$, relevant for higher derivative corrections and the large-$N$ expansion of integrated correlators, the eigenvalue $\lambda$ has always opposite even/odd parity when compared to the ``weight'' $w=s_1+s_2$.

As a final comment, we stress again that from the gauge theory side we obtain exact expressions for four-point correlators which are \textit{integrated} against different measures over the four insertion points. A difficult open problem is how to reconstruct from the dual IIB superstring side which higher-derivative corrections are responsible for a given generalised Eisenstein series in the large-$N$ expansion. However, in \cite{Chester:2019jas,Chester:2020vyz} the authors used the gauge theory results to reproduce exactly the known BPS corrections to the low-energy expansion of the four-graviton amplitude  \eqref{SUGRA_Lagran}  in type IIB superstring theory in ten-dimensional flat-space. We expect the generalised Eisenstein series \eqref{eq:spec1} to have important implications in our understanding of flat-space higher derivative corrections as well as for the structure of a similar expansion in ${\rm AdS}_5\times{ \rm S}^5$.

\subsection{Modular graph functions and superstring perturbation theory}

We now turn our attention towards string perturbation theory where a different manifestation of modularity arises and for which a second and distinct flavour of generalised Eisenstein series plays an important r\^ole.  

The study of the low-energy expansion of superstring perturbation theory has broader connections with different areas of algebraic geometry and number theory.  Many recent developments  have appeared both in the theoretical physics literature 
\cite{DHoker:2015gmr,DHoker:2016mwo,DHoker:2016quv, Ahlen:2018wng,Dorigoni:2019yoq,Basu:2020kka,DHoker:2019txf, Green:1999pv, Green:2008uj, Green:2013bza,  DHoker:2015sve, Basu:2015ayg, DHoker:2015wxz,  Basu:2016xrt, Basu:2016kli, Basu:2016mmk, Kleinschmidt:2017ege, Broedel:2018izr, Gerken:2018zcy, Gerken:2018jrq,  DHoker:2019xef, DHoker:2019mib, DHoker:2019blr, Basu:2019idd,  Gerken:2019cxz, Hohenegger:2019tii, Gerken:2020yii , Dorigoni:2022npe}
and the mathematics literature \cite{Brown:mmv, Zerbini:2015rss, Brown:I, Brown:II, DHoker:2017zhq,Zerbini:2018sox, Zerbini:2018hgs, Zagier:2019eus}.   

It is well known that string amplitudes can be computed as perturbative power series expansions in $g_s^2$, in which a term of order $g_s^{2g-2}$ is associated with a functional integral over a genus-$g$ world-sheet.  
For the present work we focus our attention to the well-studied case of the ten-dimensional four-graviton scattering amplitude in type IIB superstring theory at genus one. 

As already seen in \eqref{SUGRA_Lagran}, a consequence of supersymmetry is that the four-graviton amplitude has a prefactor of $R^4$, for a particular scalar contraction of four  linearised Riemann curvature tensors.  This means that the genus-$g$ contribution to the four-graviton amplitude takes the form 
\begin{equation}
 \mathcal{A}_g^{(4)}(\epsilon_i,k_i) =\kappa_{10}^2\, R^4 \, T_g(s,t,u)\,,
\label{genfour}
\end{equation}
where $(\epsilon_i,k_i)$  denotes  the polarisations and momenta of the scattered massless particles and $\kappa_{10}^2$ is  the ten-dimensional Newton constant.
 The function $T_g(s,t,u)$ contains all the non-trivial dynamical structure of the amplitude and is a scalar function of the Mandelstam invariants, conventionally defined by $s_{ij} := -\alpha'  (k_i + k_j )^2 /4$ with  $s:=s_{12}=s_{34}\,,\,t:=s_{13}=s_{24}$ and $u:=s_{14}=s_{23}$ satisfying $s+t+u=0$.

Let us focus our attention to the genus-one contribution in string perturbation theory. A genus-one world-sheet, $\Sigma_\tau$, has the topology of a torus, which is diffeomorphic to $\mathbb{C}/\Lambda$, where the lattice $\Lambda = \mathbb{Z}+\tau \mathbb{Z} $ defines the shape of the torus for $\tau$ in $\mathcal{H}$. Inequivalent tori are parametrised by different complex structures modulo identifications under large diffeomorphisms associated with the modular group, $\SLtwoZ$, i.e. inequivalent tori are parametrised by $\tau$ in $\mathcal{F} =\SLtwoZ \backslash \mathcal{H}$.

The genus-one amplitude $\mathcal{A}_{g=1}^{(4)}(\epsilon_i\,, k_i)$ can then be expressed as an integral over the insertion points $z_i\in \Sigma_\tau$ of the four-graviton punctures and  an integral over $\tau$ in $\mathcal{F}$,
\begin{equation}
\mathcal{A}_{g=1}^{(4)}(\epsilon_i,k_i)  = 2\pi \kappa_{10}^2  R ^4\int_{\mathcal{F}} \frac{d^2\tau}{y^2} \mathcal{M}_4(s_{ij};\tau)\,
,\label{eq:genus1}
\end{equation}
where $\mathcal{M}_4(s_{ij};\tau)$ is a modular function that results from the integral
\begin{equation}\label{string_amplitude}
\mathcal{M}_4(s_{ij};\tau):= \int_{\Sigma_\tau} \Big( \prod_{i=2}^4 \frac{d^2 z_i}{y}\Big) \exp\Big( \sum_{1\leq i < j\leq 4} s_{ij} G(z_i -z_j | \tau) \Big)\,,
\end{equation}
having used translational invariance to fix the insertion point $z_1$. The function $G(z|\tau)$ is the Green function on a torus and it is given by 
\begin{equation}\label{torus_greens_fn} 
    G(z|\tau) := -\log{\Big|\frac{\theta_1(z|\tau)}{\theta_1'(0|\tau)}\Big|^2} - \frac{\pi}{2y}(z-\bar{z})^2 = \frac{ y}{\pi} \sum_{(m,n) \neq (0,0)} \frac{ e^{2\pi i (nu-mv)}}{|m+n\tau|^2},
\end{equation}
where $\tau = x+i y$, $z = u+v\tau $ with $u,v\in [0,1)$, and $\theta_1(z|\tau)$ is a Jacobi theta function. 

Needless to say the string amplitude \eqref{string_amplitude} cannot be computed in closed form, however it can be expanded as an infinite series of low-energy contributions by considering the limit in which the Mandelstam invariants $s_{ij} \to 0$. In this way, we obtain a perturbative expansion in both $\alpha'\to 0$ and $g_s\to0$, directly connecting with the perturbative part of the previously discussed effective action \eqref{SUGRA_Lagran} in type IIB superstring theory. 

There is a nice graphical formalism to compute the low-energy expansion of string amplitudes such as \eqref{string_amplitude}, where different terms in this expansion are represented in terms of Feynman diagrams for a conformal scalar field theory on the torus. 
Each diagram corresponds to a specific way of contracting different Green functions joining pairs of points at positions $z_i$ and $z_j$, which are then integrated over $\Sigma_\tau$, thus from Feynman graphs we obtain associated modular invariant functions called \textit{Modular Graph Functions} (MGFs) \cite{Green:1999pv,Green:2008uj,DHoker:2015gmr,DHoker:2015wxz}. 

It is convenient to represent the propagator as the momentum-space lattice-sum \eqref{torus_greens_fn} and divide diagrams according to the number of independent loop-momenta we sum over.
The simplest class of MGFs is associated with one-loop graphs containing $s\in \mathbb{N}$ propagators, which can be evaluated to non-holomorphic Eisenstein series $\Eisenstein{s}{\tau}$ with integer index $s\in \mathbb{N}$, see~e.g.~\cite{DHoker:2015gmr}.

Two-loop modular graph functions are less familiar and much more interesting.
Surprisingly \cite{DHoker:2015gmr}, the action of the Laplacian $\Delta$ closes on the vector space of two-loop modular graph functions and produces source terms which are either linear or bilinear in integer index Eisenstein series.
In \cite{Dorigoni:2021jfr} it was shown that all two-loop modular graph functions can be expressed in terms of a second flavour of generalised Eisenstein series, $\genEisenstein{s_1}{s_2}{\lambda}{\tau}$,
with integer indices $s_1,s_2\in \mathbb{N}$ and $s_1,s_2\geq 2$, and with a spectrum of eigenvalues $\lambda\in {\rm Spec_2}(s_1,s_2)$ now given by:
\begin{equation}\label{eq:spec2}
\lambda \in  {\rm Spec_2}(s_1,s_2) := \{|s_1{-}s_2|{+}2,\,|s_1{-}s_2|{+}4,\,...\,,s_1{+}s_2{-}2\}\,,\qquad  s_1,s_2\in \mathbb{N}^{\geq 2}\,.
\end{equation}
Oppositely to \eqref{eq:spec1}, all generalised Eisenstein series relevant for two-loop MGFs have eigenvalues $\lambda$ of the same even/odd parity as their ``weight'' $w=s_1+s_2$.

As shown in \cite{Dorigoni:2021ngn}, the space generated by this second flavour of generalised Eisenstein with spectrum \eqref{eq:spec2} actually goes beyond the world of two-loop MGFs considered in \cite{DHoker:2015gmr}. As argued from the generating series of modular graph forms \cite{Gerken:2019cxz,Gerken:2020yii}, all MGFs are conjecturally given by single-valued iterated integrals of holomorphic Eisenstein series, while the space spanned by the generalised Eisenstein series has to be extended to also include iterated integrals of holomorphic cusp forms. The presence of holomorphic cusp forms has deep consequences; in particular we find that the Fourier expansion of these modular functions presents novel coefficients given by $L$-values of these holomorphic cusp forms inside and outside the critical strip\footnote{In \cite{Dorigoni:2021ngn} it was argued that for linear combinations of generalised Eisenstein series corresponding to two-loop MGFs, the holomorphic cusp forms always drop out. More recently in \cite{FKRinprogress}, a similar phenomenon (albeit completely different in nature) has been observed for the generalised Eisenstein with spectrum \eqref{eq:spec1} and their special linear combinations appearing in the large-$N$ expansion of the $\mathcal{N}=4$ integrated correlators. We thank Ksenia Fedosova and Kim Klinger-Logan for related discussions and for sharing their results with us.}.  We will come back to these issues in section \ref{sec:Instanton}.

To conclude this introductory section, we re-emphasise the importance of understanding the spaces of generalised Eisenstein series \eqref{eq:spec1} and \eqref{eq:spec2}. We presented two fundamental instances in non-perturbative and perturbative string theory, where the study of generalised Eisenstein series can provide insight into the possible gravitational interactions of type IIB superstring theory. 
In this work we firstly introduce a unifying framework which incorporates in a natural way both flavours \eqref{eq:spec1} and \eqref{eq:spec2} of generalised Eisenstein series, and subsequently combine Poincar\'e series, resurgence theory and spectral analysis techniques to extract novel results. 

\section{A unifying Poincar\'e series approach}
\label{sec:Poin}

As explained in the previous section, the key player for the rest of the paper is the \textit{generalised Eisenstein series}, non-holomorphic modular invariant solution to the inhomogeneous Laplace equation 
\begin{equation}\label{gen_Eisenstein_def}
    [\Delta -\lambda(\lambda-1)]\,\genEisenstein{s_1}{s_2}{\lambda}{z} = \Eisenstein{s_1}{z}\Eisenstein{s_2}{z}.
\end{equation}
In the rest of the paper we denote the modular parameter by $z=x+iy$ and by $\Delta = y^2(\partial_x^2+\partial_y^2)$ its associated hyperbolic Laplace operator.
Although our studies will be completely general, we will always refer back to the special (i.e.~of string theory origin) cases \eqref{eq:spec1} and \eqref{eq:spec2} for which the modular parameter $z\to \tau$ is respectively the axio-dilaton (or complexified Yang-Mills coupling in the dual gauge theory side) or the genus-one world-sheet complex structure.

The first representation we are going to discuss for these modular invariant functions is in terms of Poincar\'e series, i.e.\ we will express a generalised Eisenstein series, $\genEisenstein{s_1}{s_2}{\lambda}{z}$, as a sum over ${\rm SL}(2,\mathbb{Z})$-images of a special class of seed functions.
The idea behind Poincar\'e series is very natural~\cite{Iwaniec:2002,Fleig:2015vky}: if we are interested in constructing functions which are invariant under a symmetry group, we can start with an arbitrary seed function and then consider the sum over its orbits under said symmetry group. This sum, if it exists, is guaranteed to be invariant under the required symmetry group. 

When the symmetry group is ${\rm SL}(2,\mathbb{Z})$, i.e. for modular invariant functions, we can proceed as follows. Denoting by $\Phi(z)$ a modular invariant function, and by $\seed{z}$ its seed function, the Poincar\'e series representation for $\Phi(z)$ is given by:
\begin{equation}\label{eq:Psum}
    \Phi(z) = \sum_{\gamma\in\poin}\seed{\gamma\cdot z}\,,
\end{equation}
where, as usual, we have defined the ${\rm SL}(2,\mathbb{Z})$ action:
\begin{align}
\label{eq:SLact}
\gamma = \begin{pmatrix} a& b\\ c& d\end{pmatrix} \in {\rm SL}(2,\mathbb{Z})\,,
\quad\quad \gamma\cdot z := \frac{az+b}{cz+d}\,,
\end{align}
and we assumed that the seed function $\varphi(z)$ is periodic in the real direction, i.e.~$\seed{z{+}n}=\seed{z}$ for all $n\in\mathbb{Z}$, thus explainining the presence of the (Borel) stabiliser
\begin{align}
{\rm B}(\mathbb{Z}) := \left\{ \begin{pmatrix} \pm 1 & n\\0 & \pm1\end{pmatrix} \,\middle|\, n\in\mathbb{Z}\right\} \subset {\rm SL}(2,\mathbb{Z})
\end{align}
in~\eqref{eq:Psum}. Note that in general, the Poincar\'e sum~\eqref{eq:Psum} is only absolutely convergent for appropriate seeds, however, will shortly clarify that the representation~\eqref{eq:Psum} can often be understood as a suitable analytic continuation in some complex parameters which $\seed{z}$ depends on.

The simplest example of this construction is the non-holomorphic Eisenstein series:
\begin{align}
    \normEisenstein{s}{z} & \label{Eisen_poinc}\!:= \frac{\pi^s}{2\zeta(2s)}\Eisenstein{s}{z} = \sum_{\gamma\in\poin} \Im(\gamma\cdot z)^s \\
   &\notag =y^s + \frac{\xi(2s-1)}{\xi(2s)}y^{1-s}+\frac{2 \pi^s }{\Gamma(s)\zeta(2s)}\sum_{k\neq 0}|k|^{s-\frac{1}{2}}\sigma_{1-2s}(k) y^{\frac{1}{2}}K_{s-\frac{1}{2}}(2\pi |k|y)e^{2\pi ikx}\,,
\end{align}
where we introduced a different normalization compared to \eqref{Eisenstein_def} for later convenience. 

It is important to note that for a given modular invariant function, its Poincar\'e series representation is far from being unique. Since a Poincar\'e series is just a sum over ${\rm SL}(2,\mathbb{Z})$ images of a particular seed function, we can simply consider as a new seed function any of these images (or even an infinite sum thereof), and the Poincar\'e series associated with this new seed function will produce exactly the same modular invariant function. We stress that this change in seed function does in general change the stabiliser of the cusp from ${\rm B}(\mathbb{Z})$ to some other conjugate Borel subgroup. However, if we allow for formally divergent Poincar\'e series to be interpreted via analytic continuation, we can construct different seeds with the same Borel stabiliser as in \eqref{eq:Psum}, and all giving rise to the same modular invariant function.

The easiest example of such phenomenon can be seen immediately from \eqref{Eisen_poinc}. Firstly, we notice that~\eqref{Eisen_poinc} converges only for $\Re(s)>1$ and then we observe that there exists an analytic continuation for the Eisenstein series, which satisfies the reflection formulae 
\begin{align}
\Gamma(s)\Eisenstein{s}{z} &= \Gamma(1-s)\Eisenstein{1-s}{z}\,,\\
\xi(2s) \normEisenstein{s}{z} &= \xi(2-2s) \normEisenstein{1-s}{z}\,.\label{eq:reflect}
\end{align}
Hence, at least formally, the Poincar\'e series of the two seeds $y^s$ and $y^{1-s}$ give multiples of the very same $\Eisenstein{s}{z}$, even though $y^{1-s}$ cannot be written as a single ${\rm SL}(2,\mathbb{Z})$ image of $y^s$ but only as an infinite sum of images.

Note that yet another Poincar\'e seed for $\Eisenstein{s}{z}$ was given in~\cite[Eq.~(3.10)]{Bossard:2017kfv}:
\begin{align}
\label{Besselsum}
\sum_{\gamma\in\poin}\left[ \sqrt{|k| y} K_{s-\frac{1}{2}}(2\pi|k| y) e^{2\pi i k x} \right]_{\gamma} 
= \frac{\pi^{2s+1/2}  \sigma_{2s-1}(k) \Eisenstein{s}{z}  }{4 |k|^{s-1} \cos(\pi s) \Gamma(s{+}1/2)\zeta(2s-1) \zeta(2s)}  \,,
\end{align}
where the notation $[\cdots]_{\gamma}$ means that $\gamma$ acts on all occurrences of $z$ (and $\bar{z}$) inside the bracket using the fractional linear action~\eqref{eq:SLact}. 
As we can easily see from  \eqref{Eisenstein_def}, the seed appearing in the Poincar\'e sum is given by the generic Fourier non-zero mode of $\Eisenstein{s}{z}$ (or alternatively $\normEisenstein{s}{z}$) and is therefore expected again to be proportional to $\Eisenstein{s}{z}$ (or $\normEisenstein{s}{z}$). 

We stress that the sum \eqref{Besselsum} is divergent, but the result on the right-hand side, as argued for in~\cite{Bossard:2017kfv}, can be obtained via analytic continuation by rewriting the Poincar\'e series as the difference of two Niebur–Poincar\'e series, introduced in \cite{Angelantonj:2012gw}, that are absolutely convergent on the two non-intersecting domains $\Re( s)>1$ and $\Re(1-s) >1$. In the next section we show that a generalisation of such a Niebur–Poincar\'e series \eqref{Besselsum} provides a privileged class of seed functions whose Poincar\'e sums produce all generalised Eisenstein series, thus constructing a unifying framework to discuss both higher derivative corrections and MGFs. 

As already mentioned in the introduction, there are many reasons for seeking Poincar\'e series representations of modular functions.
Perhaps most importantly: Poincar\'e series representations are manifestly modular-invariant expressions which in general reduce the complexity of the objects under consideration,~e.g.~for the Eisenstein $\normEisenstein{s}{z}$ the seed is simply $y^s$.~Similarly, for generalised Eisenstein series relevant for two-loop MGFs, convenient Poincar\'e seeds were proposed in \cite{Dorigoni:2021jfr}, and were given in terms of certain iterated Eisenstein integrals of depth one, while their corresponding Poincar\'e series, i.e. two-loop modular graph forms, have to be built in general from iterated Eisenstein integrals of depth two.

While the Poincar\'e seed is in general of reduced complexity when compared to its modular invariant companion, a drawback is that dealing with generic Poincar\'e series usually makes it rather cumbersome to extract the analytic properties of the objects under study. 
For example, it is not straightforward to obtain from the Poincar\'e series representation \eqref{eq:Psum} the asymptotic expansion at the cusp, $z\to i\infty$, of a modular function $\Phi(z)$. 

For Eisenstein series it is a standard result~\cite{Iwaniec:2002,Fleig:2015vky} to obtain the Fourier mode decomposition from its Poincar\'e series, as presented in \eqref{Eisen_poinc}. However, for general Poincar\'e series, a similar analysis is far more intricate. The general asymptotic expansion for Poincar\'e series of two-loop MGFs was presented in \cite{DHoker:2019txf}, while for alternative seeds analogous results~\cite{Ahlen:2018wng,Dorigoni:2019yoq} involved certain Kloosterman sums. For the generalised Eisenstein series similar to the $d^6 R^4$ higher-derivative coefficient, the asymptotic expansion at the cusp was derived from yet another different ``double''-Poincar\'e series in \cite{Green:2014yxa,Klinger-Logan:2022whi}. In this work we present a unified Poincar\'e series approach thanks to which both classes of generalised Eisenstein series can be treated in parallel.

As it will be useful shortly, we briefly review how to obtain the Fourier mode expansion of a modular function from that of its Poincar\'e seed.
Given the Fourier expansions\begin{align}
\Phi(z)&=\sum_{k \in\mathbb{Z}}a_k(y)e^{2\pi ik x} = \sum_{\gamma\in\poin}\seed{\gamma\cdot z}\,,\\
\seed{z}& = \sum_{k\in\mathbb{Z}}c_k(y)e^{2\pi ikx}\,,
\end{align}
with $x=\Re (z)$ and $y = \Im (z) $, the Fourier modes $a_k(y)$ can be reconstructed from the seed function using the well-known result~\cite{Iwaniec:2002,Fleig:2015vky}:
\begin{align}
a_k(y) &=\label{eq:nonzeromode} c_k(y) + \sum_{d=1}^\infty\sum_{m\in\mathbb{Z}} S(k,m;d) \int_{\mathbb{R}} e^{-2\pi i k \omega -2\pi i m \frac{\omega}{d^2 (y^2+\omega^2)}} c_m\Big(\frac{y}{d^2(y^2+\omega^2)}\Big)\dd \omega\,.
\end{align}
Here $S(k,m;d)$ denotes in general a Kloosterman sum
\begin{equation}
S(k,m;d) := \sum_{r\in (\mathbb{Z}/d\mathbb{Z})^\times} e^{\frac{2\pi i}{d} (k r + m r^{-1}) }\,,\label{eq:Kloos}
\end{equation}
which is a finite sum over all $0\leq r <d$ that are coprime to $d$, such that $r$ has a multiplicative inverse, denoted by $r^{-1}$, in $ (\mathbb{Z}/d\mathbb{Z})^\times$.

In particular, we have that the Fourier zero-mode $a_0(y)$ can be expressed as
\begin{align}
a_0(y) &\label{Fzero_poinc_formula}= c_0(y) + \sum_{d=1}^\infty \sum_{m\in\mathbb{Z}} \sum_{r\in (\mathbb{Z}/d\mathbb{Z})^\times} e^{\frac{2\pi i m r}{d}} \int_{\mathbb{R}} e^{-2\pi i m \frac{\omega}{d^2 (y^2+\omega^2)}} c_m\Big(\frac{y}{d^2(y^2+\omega^2)}\Big)\dd \omega\,,
\end{align}
where we rewrote the Kloosterman sum $S(0,m;d)$ as a Ramanujan sum,
\begin{equation} S(0,m;d) = S(m,0;d) = \sum_{r\in (\mathbb{Z}/d\mathbb{Z})^\times} e^{\frac{2\pi i m r}{d}} \,.\label{eq:KloostEasy}
\end{equation}
Note that although \eqref{Fzero_poinc_formula} is an expression for the whole Fourier zero-mode sector, $a_0(y)$, it is in general quite hard to separate the perturbative terms in the asymptotic expansion at the cusp $y\gg1$, i.e. the power-behaved terms, from the non-perturbative, exponentially suppressed terms $(q\bar{q})^n = e^{-4\pi n y}$.
In \cite{Dorigoni:2019yoq,Dorigoni:2022bcx} it was proven that for two-loop modular graph functions it is actually possible to reconstruct these non-perturbative corrections from the perturbative terms using methods from resurgent analysis~\cite{Dorigoni:2014hea}.
\subsection{A new Niebur-Poincar\'e series}

One way of constructing a Poincar\'e series representation for the generalised Eisenstein series,
\begin{equation}
\genEisenstein{s_1}{s_2}{\lambda}{z} = \sum_{\gamma\in\poin} \genEisensteinSeed{s_1}{s_2}{\lambda}{\gamma\cdot z}\,,
\end{equation}
 relies on rewriting the Laplace equation \eqref{gen_Eisenstein_def} after having replaced one of the Eisenstein series in the source term, say $\Eisenstein{s_1}{z}$, by its Poincar\'e series \eqref{Eisenstein_def}, usually dubbed as \textit{folding} $\Eisenstein{s_1}{z}$. This leads us to consider an auxiliary Laplace equation for the candidate seed function $\genEisensteinSeed{s_1}{s_2}{\lambda}{ z}$:
\begin{equation}
[\Delta-\lambda(\lambda-1)]\, \genEisensteinSeed{s_1}{s_2}{\lambda}{ z} =  \frac{2\zeta(2s_1)}{\pi^{s_1}} y^{s_1} \Eisenstein{s_2}{z}\,.\label{eq:LaplaceSeed}
\end{equation}
We can first rewrite the source term as a Fourier series \eqref{Eisenstein_def} with respect to $x =\Re(z)$, and then find a particular solution for this Laplace equation mode by mode.
For the Fourier zero-mode sector there is no issue in finding such a particular solution. However, for a Fourier non-zero mode it is rather difficult to find a particular solution to \eqref{eq:LaplaceSeed} which is expressible in terms of simple building-block seed functions for generic values of $s_1,s_2$ and $\lambda$ . 

In \cite{Dorigoni:2021jfr} it was shown that all two-loop MGFs, or more broadly all generalised Eisenstein series with spectrum given by \eqref{eq:spec2}, can be written as Poincar\'e series of finite linear combinations of the building-block seed functions introduced in \cite{Dorigoni:2019yoq}
\begin{equation}
\varphi(a,b,r\vert z)=  \sum_{m\neq 0} \sigma_{a}( m ) (4\pi \vert m \vert )^b y ^r e^{-2\pi |m|   y}  e^{2\pi i m x}\,,\label{eq:DDAKseed}
\end{equation} 
for different values of the parameters $(a,b,r)$. It was nonetheless noticed in \cite{Dorigoni:2019yoq} that such seeds are rather ill-suited to describe generalised Eisenstein series relevant for higher-derivative corrections and integrated correlators, where the spectrum is \eqref{eq:spec1}. For these generalised Eisenstein series it is still possible to write a seed function in terms of building-blocks \eqref{eq:DDAKseed}, but one requires an infinite sum over such simple seeds, thus making it quite hard to extract the asymptotic expansion at the cusp or other analytic properties from the corresponding Poincar\'e series.
Other types of Poincar\'e series have been proposed in the literature \cite{Green:2014yxa,Klinger-Logan:2022whi} for the diagonal elements, i.e.~$s_1=s_2$, in the first family \eqref{eq:spec1}, while in \cite{fedosova2022whittaker} other examples in this class are analysed directly from the differential equation point of view.

To construct a class of Poincar\'e seeds suited for discussing both \eqref{eq:spec1}-\eqref{eq:spec2} in a uniform manner, we have to re-examine the Laplace equation \eqref{eq:LaplaceSeed}. From the Fourier decomposition of the Eisenstein series \eqref{Eisenstein_def}, we notice that the $m^{th}$ Fourier mode, with $m\neq0$, of the source term is schematically of the form
$$
\sigma_a(m) |m|^{b-\half} y^{r+\half} K_{s-\half}(2\pi |m| y)e^{2\pi i mx}\,,
$$
for some specific values of the parameters $(a,b,r,s)$.
Thanks to the recurrence relations satisfied by the modified Bessel function $K_s(y)$, for both spectra \eqref{eq:spec1}-\eqref{eq:spec2} it is always possible to find a finite linear combination over different values of the parameters\footnote{In this context the parameter $a$ is rather special, since it is the index of the divisor sum function $\sigma_a(m)$. From the Laplace equation \eqref{eq:LaplaceSeed} and the Fourier mode decomposition \eqref{Eisenstein_def} it is easy to see that $a=1-2s_2$ for the present discussion.} $(a,b,r,s)$ of terms as above which is a solution to \eqref{eq:LaplaceSeed} in the $m^{th}$ Fourier mode sector. 

With this fact at hand, we can now introduce a novel space of Poincar\'e seeds and associated Poincar\'e series which is both general enough, in that every string theory generalised Eisenstein series \eqref{eq:spec1}-\eqref{eq:spec2} can be written as a Poincar\'e series of finite linear combinations of these novel seeds, and simple enough so that we can easily extract asymptotic data both at the cusp $y\gg 1$ and at the origin $y\to0$. 

We define the seed function
\begin{align}
    \seedfn{a}{b}{r}{s}{z} &\notag=\sum_{m\neq0} \upsilon_m(a,b,r,s ; y) e^{2\pi i mx}\\
    &\label{seed_def}\!:= \sum_{m\neq 0}\sigma_a(m)|m|^{b-\half} y^{r+\half} K_{s-\half}(2\pi |m| y)e^{2\pi i mx},
\end{align}
which depends on four complex parameters $(a,b,r,s)$. Given that the Bessel function $K_s(y)$ is exponentially suppressed for large values of its argument, we immediately have that the sum over the Fourier mode, $m$, is absolutely convergent for any values of the parameters $(a,b,r,s)$. This property of the Bessel function implies as well that the seed function is exponentially suppressed for $y\gg 1$, however, the limit $y\to 0$ is more delicate to analyse.

Under the assumption that the Poincar\'e series of such a class of seed functions is well-defined, we can introduce a novel class of modular invariant functions which we denote by
\begin{align}
    \modfn{a}{b}{r}{s}{z} :\label{f_definition} = \sum_{\gamma\in\poin} \seedfn{a}{b}{r}{s}{\gamma\cdot z}\,.
\end{align}
The convergence of this Poincar\'e series is studied in appendix \ref{app:AsyOrigin}, where we prove that absolute convergence is guaranteed when 
\begin{equation}
\min{\{\Re(r+1-s),\,\Re(r+s),\,\Re(r-b),\,\Re(r-a-b)\}}>1\,.\label{eq:AbsConv}
\end{equation}
 In what follows we can relax the requirement of absolute convergence and consider if necessary the Poincar\'e series \eqref{f_definition} in terms of its analytic continuation in some of its complex parameters $(a,b,r,s)$, in direct analogy with the discussion below \eqref{Besselsum}.
 
The keen-eyed reader will notice that the new seeds \eqref{seed_def} are very reminiscent of the rather unconventional Poincar\'e series representation \eqref{Besselsum} for $\Eisenstein{s}{z} $. The reason is that, very much like \eqref{Besselsum}, our expression \eqref{seed_def} can be obtained as an infinite sum over all Fourier non-zero modes, $m\neq 0$, of the difference between two Niebur–Poincar\'e series \cite{Niebur_1973,Angelantonj:2012gw}. We will shortly prove that both string theory generalised Eisenstein series \eqref{eq:spec1}-\eqref{eq:spec2} can be obtained from finite linear combinations of these new Niebur–Poincar\'e series \eqref{f_definition}.

As already stressed, one of the perks of a Poincar\'e series representation is that in general it simplifies the complexity of the objects under consideration. 
In particular, from the seed function definition \eqref{seed_def} we can already deduce various algebraic and differential identities satisfied by the modular objects $ \modfn{a}{b}{r}{s}{z}$.
 Firstly, we note that the seed functions \eqref{seed_def} are invariant under the reflection $s\to 1-s$,
 \begin{equation}
     \seedfn{a}{b}{r}{1-s}{z} = \sum_{m\neq 0}\sigma_a(m)|m|^{b-\half} y^{r+\half}K_{\half-s} (2\pi |m| y)e^{2\pi i mx} =      \seedfn{a}{b}{r}{s}{z}\,,
 \end{equation} 
 due to the Bessel function identity $K_{s}(y)=K_{-s}(y)$.
 Similarly, we have invariance under the transformation $(a,b)\to (-a,b+a)$,
  \begin{equation}
     \seedfn{-a}{a+b}{r}{s}{z} = \sum_{m\neq 0}\sigma_{-a}(m)|m|^{a+b-\half}y^{r+\half}K_{s-\half} (2\pi |m| y)e^{2\pi i mx} =   \seedfn{a}{b}{r}{s}{z}\,,
 \end{equation} 
a straightforward consequence of the identity $\sigma_{-a}(m) = |m|^{-a}\sigma_{a}(m)$. 
From these two observations we deduce that the modular functions must also inherit these symmetries,
\begin{align}
    \modfn{a}{b}{r}{s}{z}&=\label{eq:Yident1}\modfn{a}{b}{r}{1-s}{z}\,,\\
    \modfn{a}{b}{r}{s}{z}&=\label{eq:Yident2}\modfn{-a}{b+a}{r}{s}{z}\,.
\end{align}

More interestingly, given the well-known Bessel function recurrence relation
\begin{equation}
    K_{s+1}(y)-K_{s-1}(y) = \frac{2s}{y}K_s (y)\,,
\end{equation}
we can immediately derive the three-term recursion
\begin{equation}\label{f_recursion_formula}
    \modfn{a}{b}{r}{s+1}{z}-\modfn{a}{b}{r}{s-1}{z} = \frac{2s-1}{2\pi}\modfn{a}{b-1}{r-1}{s}{z}\,.
\end{equation}
Note that even if we consider a seed function whose parameters $(a,b,r,s)$ satisfy the conditions \eqref{eq:AbsConv} for absolute convergence of the Poincar\'e series, repeated applications of this recursion formula \eqref{f_recursion_formula} will inevitably bring us outside of the domain \eqref{eq:AbsConv} where the analytic continuation of \eqref{f_definition} has to be discussed carefully.

Finally, given that our discussion started from the inhomogeneous Laplace equation \eqref{gen_Eisenstein_def}, it is natural to consider the action of the Laplace operator on \eqref{f_definition}. By simply applying the Laplacian to \eqref{seed_def} and using the known identity for the derivative of the Bessel function,
\begin{equation}
K'_s(y)  = - \frac{s}{y} K_{s}(y)  - K_{s-1}(y) \,,
\end{equation}
 we arrive at
 \begin{equation}\label{f_diff_eq_2}
   \big[ \Delta -(r+1-s)(r-s)\big] \modfn{a}{b}{r}{s}{z} = -4\pi r \modfn{a}{b+1}{r+1}{s-1}{z} \,,
\end{equation}
or equivalently making use of \eqref{f_recursion_formula}:
\begin{equation}\label{f_diff_eq}
  \big[  \Delta- (r+s)(r+s-1)\big] \modfn{a}{b}{r}{s}{z} = -4\pi r\modfn{a}{b+1}{r+1}{s+1}{z} \,.
\end{equation}
We have thus obtained that the functions $\modfn{a}{b}{r}{s}{z} $ satisfy a closed system of inhomogeneous Laplace eigenvalue equations where the source term is given by yet another function of the same type, but different parameters $(a,b,r,s)$. 

Both Laplace equations \eqref{f_diff_eq_2}-\eqref{f_diff_eq} simplify dramatically for $r=0$, where they reduce to
\begin{equation}
\big[\Delta - s(s-1) \big]\, \modfn{a}{b}{0}{s}{z}= 0 \,,\label{eq:YEisenLap}
\end{equation} 
and since the function $\modfn{a}{b}{0}{s}{z}$ is manifestly a modular invariant eigenfunction of $\Delta$ with eigenvalue $s(s-1)$ it must be proportional to $\Eisenstein{s}{z}$.

We will shortly prove that $\modfn{a}{b}{r}{s}{z}$ has polynomial growth at the cusp and compute explicitly its asymptotic expansion using the integral representation \eqref{Fzero_poinc_formula}, thus easily fixing the coefficient of proportionality between $\modfn{a}{b}{0}{s}{z}$ and $\Eisenstein{s}{z}$.
Alternatively, we can see from \eqref{seed_def} that each summand with Fourier mode $m=k$ in the seed function $\seedfn{a}{b}{0}{s}{z}$ is proportional to the Poincar\'e seed \eqref{Besselsum} for $\Eisenstein{s}{z}$. The only difference with  \eqref{Besselsum}, is that the sum over $m$ in \eqref{seed_def} will simply produce a particular Dirichlet series which will contribute to the proportionality factor between $\modfn{a}{b}{0}{s}{z}$ and $\Eisenstein{s}{z}$.

 As already mentioned, the novel seeds \eqref{seed_def} are constructed precisely to provide for a broad enough basis of solutions to \eqref{eq:LaplaceSeed}.
 Correspondingly, we will show that it is possible to produce finite linear combinations of $\modfn{a}{b}{r}{s}{z}$ which are solutions to the generalised Eisenstein series differential equation \eqref{gen_Eisenstein_def} relevant for string theory. 
 A central part of this analysis is the observation that the space of functions $\modfn{a}{b}{r}{s}{z}$ contains all products of two Eisenstein series,~i.e.~all possible source terms of~\eqref{gen_Eisenstein_def}.~The proof of this statement is very simple. If we consider the bilinear $ \Eisenstein{s_1}{z}\Eisenstein{s_2}{z}$, we first fold $\Eisenstein{s_1}{z}$ and then re-express $\Eisenstein{s_2}{z}$ in Fourier modes arriving at
 \begin{align}
  &\label{prod_Eisenstein} \Eisenstein{s_1}{z}\Eisenstein{s_2}{z}=\frac{8\xi(2s_1)}{\Gamma(s_1)\Gamma(s_2)}\modfn{1-2s_2}{s_2}{s_1}{s_2}{z}\\
 &\notag  + \frac{2\Gamma(s_1+s_2)\xi(2s_1)\xi(2s_2)}{\Gamma(s_1)\Gamma(s_2)\xi(2(s_1+s_2))}\Eisenstein{s_1+s_2}{z}+\frac{2\Gamma(s_1+1-s_2)\xi(2s_1)\xi(2s_2-1)}{\Gamma(s_1)\Gamma(s_2)\xi(2(s_1+1-s_2))}\Eisenstein{s_1+1-s_2}{z}\,.
\end{align}
Alternatively we can use the reflection formula \eqref{eq:reflect}, combined with \eqref{eq:Yident1}-\eqref{eq:Yident2}, to derive
\begin{align}
  &\label{prod_Eisenstein2}  \Eisenstein{s_1}{z}\Eisenstein{s_2}{z}=\frac{8\xi(2s_2-1)}{\Gamma(s_1)\Gamma(s_2)}\modfn{1-2s_1}{s_1}{1-s_2}{1-s_1}{z} \\
   &\notag + \frac{2\Gamma(s_1{+}s_2{-}1)\xi(2s_1{-}1)\xi(2s_2{-}1)}{\Gamma(s_1)\Gamma(s_2)\xi(2(s_1+s_2){-}3)}\Eisenstein{s_1{+}s_2{-}1}{z}+\frac{2\Gamma(s_1{+}1{-}s_2)\xi(2s_1)\xi(2s_2{-}1)}{\Gamma(s_1)\Gamma(s_2)\xi(2(s_1{-}s_2){+}2)}\Eisenstein{s_1{+}1{-}s_2}{z}\,.
\end{align}
Note that by folding $\Eisenstein{s_1}{z}$ we break the symmetry between $s_1\leftrightarrow s_2$. This comes at a notable price in the diagonal case $s_1=s_2$ where \eqref{prod_Eisenstein}-\eqref{prod_Eisenstein2} have to be regulated. For $s_1=s_2$, the right-hand side of both equations contains the divergent Eisenstein series $\Eisenstein{1}{z}$. However, since the bilinear $\Eisenstein{s_1}{z}^2$ is perfectly regular for $s_1\neq1$, this implies that the modular functions $\modfntwo{1-2s_1}{s_1}{s_1 }{s_1}$ and $\modfn{1-2s_1}{s_1}{1-s_1}{1-s_1}{z}$ must diverge as well. A regularised versions of \eqref{prod_Eisenstein}-\eqref{prod_Eisenstein2} for the case $s_1=s_2$ is easily obtained by  considering the continuous limit away from the diagonal $s_1=s_2$ case:
\begin{equation}\label{diagonal_regul}
\Eisenstein{s_1}{z}^2=\lim_{\epsilon\to 0}\big[ \Eisenstein{s_1+\epsilon}{z}\,\Eisenstein{s_1}{z}\big]\,.   
\end{equation}
When $\epsilon\neq0$ we can safely write the right-hand side using~\eqref{prod_Eisenstein}-\eqref{prod_Eisenstein2}. As $\epsilon \to 0$ our formulae~\eqref{prod_Eisenstein}-\eqref{prod_Eisenstein2} produce a divergent contribution coming from $\Eisenstein{1+\epsilon}{z}$ which cancels against the similarly singular $\Upsilon$ thus leaving us with a regular expression.
 The need for a regularisation of the diagonal case $s_1=s_2$ is an ubiquitous phenomenon  \cite{Ahlen:2018wng,Dorigoni:2019yoq} and it is independent from the particular seeds considered in the present work.

\subsection{Asymptotic expansion at the cusp}

Let us now derive the asymptotic expansion near the cusp $z\to i \infty$ for the modular invariant functions $\modfn{a}{b}{r}{s}{z}$. 
Firstly we perform a Fourier mode decomposition, 
\begin{equation}
\modfn{a}{b}{r}{s}{z} = \sum_{k\in \mathbb{Z}} \fm{k}{a}{b}{r}{s}{y}e^{2\pi ikx}\,,
\end{equation}
and focus on deriving the asymptotic expansion for large $y$ of the Fourier zero-mode $ \fzero{a}{b}{r}{s}{y}$.

In the previous section we have already reviewed how to retrieve the Fourier modes of a Poincar\'e series from an integral transform \eqref{eq:nonzeromode} of the Fourier modes for the corresponding seed function. In particular, if we focus on the Fourier zero-mode sector \eqref{Fzero_poinc_formula} for the specific seeds \eqref{seed_def}  under consideration, we have to compute:
\begin{align}
 &   \fzero{a}{b}{r}{s}{y} \label{eq:fzerostart} \\
 &\notag= \sum_{d=1}^\infty \sum_{m\neq 0} S(m,0;d) \int_{\mathbb{R}} e^{-2\pi im \frac{\omega}{d^2(\omega^2 + y^2)}} \sigma_a(m)|m|^{b-\half} \Big(\frac{y}{d^2(\omega^2 + y^2)}\Big)^{r+\half} K_{s-\half}\Big(\frac{2\pi|m|y}{d^2(\omega^2 + y^2)}\Big)\dd\omega.
\end{align}
In appendix \ref{app:AsyCusp} we show how the above integral transform can be rewritten as a nicer Mellin-Barnes type of contour integral, thus making the task of extracting the asymptotic expansion at the cusp much more manageable. 

Relegating the more technical details to the appendix, we present here the key result of our analysis: the integral representation \eqref{eq:fzerostart} can be rewritten as the Mellin-Barnes integral
\begin{equation}\label{modfn_Fourierzero_mode}
    \fzero{a}{b}{r}{s}{y}=  \int_{\frac{1}{2}-i\infty}^{\frac{1}{2}+i\infty} U(a,b,r,s\vert t)\,y^t \frac{\dd t}{2\pi i}\,,
\end{equation}
where we define
\begin{multline}\label{U_definition}
   U(a,b,r,s\vert t) :=     \frac{\Gamma\big(\frac{r+1-s-t}{2}\big)\Gamma\big(\frac{r+s-t}{2}\big)\Gamma\big(\frac{t+r-s}{2}\big)\Gamma\big(\frac{t+r+s-1}{2}\big)}{2\pi^r\, \Gamma(r)\xi(2-2t)}\\ \times
   \frac{\zeta(r+1-b-t)\zeta(r+1-a-b-t)
    \zeta(t+r-b)\zeta(t+r-a-b)}{\zeta(2r+1-a-2b)}\,.
\end{multline}
For this section, unless otherwise specified, we restrict ourselves to the range of parameters~\eqref{eq:AbsConv} for which the Poincar\'e series is absolutely convergent. However, at the end of appendix \ref{app:AsyCusp} we explain that for parameters,~$(a,b,r,s)$, which do not produce convergent Poincar\'e series, the Mellin-Barnes representation~\eqref{modfn_Fourierzero_mode} is still perfectly valid provided the contour of integration is modified from the straight line $\Re(t)=\half$ to a contour separating the two families of poles we are about to discuss.

\begin{figure}
    \centering
    \includegraphics[height=5cm, width=5cm]{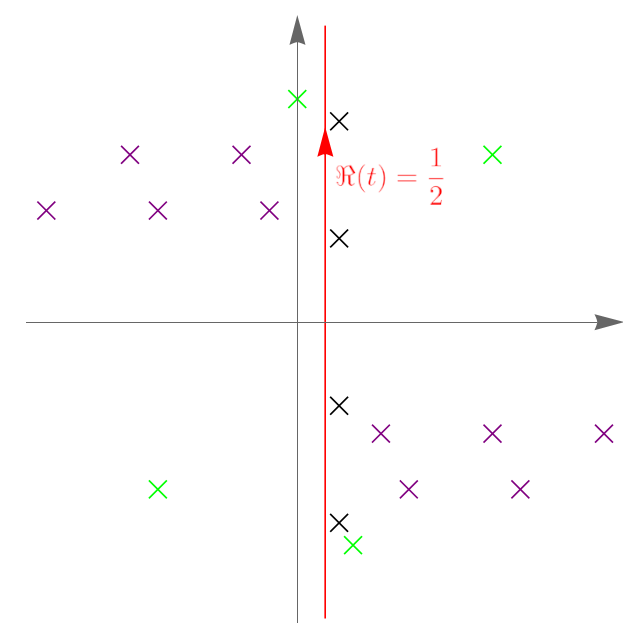}
    \caption{Schematic pole structure of $U(a,b,r,s\vert t)$. The infinite family of poles from the gamma functions is given in purple while the four poles from the zeta functions are given in green. In black, we have an infinite family of poles with $\Re(t) = \frac{3}{4}$ (if Riemann hypothesis is correct) coming from the non-trivial zeroes of the Riemann zeta. The contour of integration, $\Re(t) = \frac{1}{2}$, is indicated in red.}
    \label{fig:polesU}
\end{figure}

It is now fairly straightforward to extract from the Mellin-Barnes integral \eqref{U_definition} the asymptotic expansion of $\fzero{a}{b}{r}{s}{y}$ as $y \gg 1$ by closing the contour of integration at negative infinity on the left semi-half plane $\Re(t)<\frac{1}{2}$ and collecting residues from the different singular terms in \eqref{U_definition}.
As we can see in Figure \ref{fig:polesU}, when the parameters $(a,b,r,s)$ defines an absolutely convergent Poincar\'e series, i.e. when they satisfy \eqref{eq:AbsConv}, closing the contour at negative infinity in the half-plane $\Re(t)<\frac{1}{2}$ selects two different types of poles:
\begin{itemize}
\item From the zeta functions $\zeta(t+r-b)$ and $\zeta(t+r-a-b)$ we have two poles located respectively at $t = b+1-r$ and $ t=a+b+1-r$;
\item From the gamma functions $\Gamma\big(\frac{t+r-s}{2}\big)$ and $\Gamma\big(\frac{t+r+s-1}{2}\big)$ we have two infinite families of poles located respectively at $t= s - r -2n$ and $t=1-s-r-2n$ with $n\in \mathbb{N}$.
\end{itemize}
It is easy to see that under the assumption \eqref{eq:AbsConv} of an absolutely convergent Poincar\'e series, the above poles are all located in the half-plane $\Re(t)<\frac{1}{2}$, while all remaining poles in \eqref{U_definition} are located in the half-plane $\Re(t)>\frac{1}{2}$.

Computing the residues at said poles and summing over all of them produces the wanted asymptotic expansion for large-$y$ of the Fourier zero-mode,
\allowdisplaybreaks{\begin{align}\label{full_asymptotics}
    \fzero{a}{b}{r}{s}{y}\sim&\notag  \frac{\zeta(2r-a-2b)}{2\Gamma(r)\zeta(2r-a-2b+1)}  \Big[ c^{(1)}(a,b,r,s)  y^{b+1 \shortminus r} \!+\! c^{(1)}(\shortminus a,a\!+\!b,r,s)y^{a+b+1 \shortminus r} \Big]\\
   & +\sum_{n=0}^\infty y^{-r- 2n} \big[ y^{s} c^{(2)}_n(a,b,r,s)\!+\! y^{1- s}c^{(2)}_n(a,b,r,1{\shortminus}s)\big] \,,
\end{align}}
where for convenience of presentation we defined the coefficients
\allowdisplaybreaks\begin{align}
&c^{(1)}(a,b,r,s) =  \frac{\Gamma\big(\frac{b+1-s}{2}\big)\Gamma\big(\frac{2r-b-s}{2}\big)\Gamma\big(\frac{b+s}{2}\big)\Gamma\big(\frac{2r+s-b-1}{2}\big)\zeta(1-a)}{\pi^{b}\,\Gamma(r-b)}\,,\\
&\notag c^{(2)}_n(a,b,r,s) \label{eq:c2pert}= \frac{(-1)^n \pi ^{2 n+1-s}  \Gamma \left(n+r\right) \Gamma \left(s-n-\frac{1}{2}\right) \Gamma \left(n+r-s+\frac{1}{2}\right)  }{n! \Gamma \left( r \right) \Gamma (2 n+r+1-s)}\\*
& \times \frac{ \zeta (s-b-2 n)\zeta (s-a-b-2 n) \zeta (2 n+2 r+1-b-s) \zeta (2 n+2 r+1-a-b-s)}{ \zeta (2r-a-2 b+1) \zeta (4 n+2 r+2-2 s)}\,.
\end{align}

Besides the first two terms $y^{b+1 \shortminus r}$ and $y^{a+b+1 \shortminus r}$, coming from the isolated poles of the two Riemann zeta functions, the remaining perturbative series is, for general parameters, $(a,b,r,s)$, an asymptotic factorially divergent power series.
From~\eqref{eq:c2pert}, the growth of the perturbative coefficients is $c^{(2)}_n(a,b,r,s)  = O( (2n)!)$ which combined with the power-like growth $(4\pi y )^{-2n}$ immediately suggests the presence of exponentially suppressed corrections $(q\bar{q}) = e^{-4\pi y}$. 

While the modular functions $\modfn{a}{b}{r}{s}{z}$ provide for a natural extension of the generalised Eisenstein series, unlike the generalised Eisenstein series they  have, for non-specific values of the parameters, non-terminating and factorially divergent formal power series expansions at the cusp $y \gg 1$.
Crucially, at non-generic and physically relevant points in parameter space, i.e. for special values of $a,b,r,s$ corresponding to generalised Eisenstein series, the asymptotic tail of $\modfn{a}{b}{r}{s}{z}$ vanishes and \eqref{full_asymptotics} reduces to a sum of finitely many terms. In the next section, we show that this happens for $a\in\mathbb{Z}$ and $(b,r,s)$ either all integers or all half-integers.

This dramatic change of the asymptotic series \eqref{full_asymptotics} from a factorially divergent formal power series to a finite sum can be understood quite easily from the contour integral representation given in \eqref{modfn_Fourierzero_mode}. From the definition \eqref{U_definition} of the function $U(a,b,r,s\vert t)$ we notice that the gamma  functions generate two infinite families of poles in $t$ on both side of the integration contour $\Re(t) = \half$. At the same time, the four Riemann zeta functions present two pairs of identically spaced, infinite families of (trivial-)zeros in $t$ again on both side of the integration contour $\Re(t) = \half$. The truncation of the asymptotic series \eqref{full_asymptotics} to a finite sum happens precisely for special values of $a,b,r,s$ for which these families of poles and zeroes start overlapping at some point. As we will show in the next section, the case of interest - the generalised Eisenstein series - neatly falls into this category.

The analytic continuation in $(a,b,r,s)$ is crucial for fixing the exponentially suppressed $(q\bar{q})$-terms from the formal and factorially divergent perturbative expansion at the cusp, since for generic $(a,b,r,s)$ the requirement of a well-defined Borel-Ecalle resummation of \eqref{full_asymptotics} allows for calculation of all $(q\bar{q})$-terms, similar to \cite{Dorigoni:2019yoq,Dorigoni:2022bcx}.

Surprisingly, even when at special values of the parameters $(a,b,r,s)$ the series \eqref{full_asymptotics} becomes a finite sum, such non-perturbative resurgent corrections do survive. In the literature, this is usually dubbed Cheshire Cat resurgence \cite{Dunne:2016jsr,Kozcaz:2016wvy,Dorigoni:2017smz} from the eponymous feline of Alice in Wonderland with a disappearing body but a lingering enigmatic grin. 
Since such a resurgence analysis is akin to the one carried out in \cite{Dorigoni:2022bcx} for a different general class of seed functions \eqref{eq:DDAKseed}, we will not repeat this calculation here. Later in the paper we will however revisit the calculation of exponentially suppressed terms from the spectral analysis point of view.

We conclude this section with a simpler ``special'' example, namely the case of the standard Eisenstein series. 
As previously remarked, since $\modfn{a}{b}{r=0}{s}{z}$ is a modular solution to the Laplace equation \eqref{eq:YEisenLap} it must proportional to  $\normEisenstein{s}{z}$. 
Given the generic asymptotic expansion \eqref{full_asymptotics}, we can now fix the constant of proportionality.

Firstly, it is a well-known result \eqref{Eisen_poinc} that the asymptotic expansion at the cusp for $\normEisenstein{s}{z}$ has only two power-behaved terms: $y^{s}$ and $y^{1-s}$.
These two terms are easily recognisable in \eqref{full_asymptotics} as regulated versions of the $n=0$ terms $y^{s-r} c_0(a,b,r,s)\!+\! y^{1- s-r}c_0(a,b,r,{\shortminus}s)$, while all other terms vanish.
More precisely, from the definition \eqref{U_definition} we notice in the denominator the factor $\Gamma(r)$ is singular for $r=0$, but easily regulated by considering $r=\epsilon$ and taking the limit $\epsilon \to 0$ at the very end. Only the poles of \eqref{U_definition} located at $t=s+\epsilon$ and $t=1-s-\epsilon$ have a non-vanishing residue in the limit $\epsilon \to 0$ and produce precisely a multiple of the expected Eisenstein series Laurent polynomial  \eqref{Eisen_poinc}.
This allows us to fix the proportionality factor between $\modfn{a}{b}{r=0}{s}{z}$ and  $ \normEisenstein{s}{z}$ as such
\begin{equation}\label{eq:YEisen}
    \modfn{a}{b}{0}{s}{z} =  \frac{2 \tan(\pi s) \Gamma(s) \zeta(1-b-s)\zeta(1-a-b-s)\zeta(s-b)\zeta(s-a-b)}{(2s-1)\pi^{s-1}\,\zeta(1-a-2b)\zeta(2-2s)}\, \normEisenstein{s}{z}\,.
\end{equation}
We stress again that we could have reached the same result from a direct comparison between each Fourier mode of the Poincar\'e seed \eqref{seed_def} and the unusual Poincar\'e series \eqref{Besselsum} for $\normEisenstein{s}{z}$. Applying \eqref{Besselsum} to each Fourier mode in \eqref{seed_def}, leaves us with a particular Dirichlet sum over the Fourier non-zero modes $m \in \mathbb{Z}\setminus\{0\}$ which, once evaluated, brings us back \eqref{eq:YEisen}.

Lastly, an easy application of the recursion formula \eqref{f_recursion_formula} shows that all of $\modfn{a}{b}{-n}{s}{z}$, with $n\in \mathbb{N}$, are also finite sums of Eisenstein series,
\begin{equation}
\modfn{a}{b}{{-} n}{s}{z} =\pi^n n! \sum_{k=0}^n\frac{(\shortminus1)^{ k+1} (s+n\!-\!2 k-\half) \Gamma(s\!-\!k-\half)}{k! \Gamma (n\!-\!k+1) \Gamma(n+s\!-\!k+\half)} \gamma(a,b+n,s+n\!-\!2k)\normEisenstein{s+n\shortminus2k}{z}\,,
\end{equation}
where the coefficient $\gamma(a,b,s)$ is the proportionality constant appearing in \eqref{eq:YEisen}, i.e.
\begin{equation}
\gamma(a,b,s) =  \frac{2 \tan(\pi s) \Gamma(s) \zeta(1-b-s)\zeta(1-a-b-s)\zeta(s-b)\zeta(s-a-b)}{(2s-1)\pi^{s-1}\,\zeta(1-a-2b)\zeta(2-2s)}\,.
\end{equation}

\subsection{A ladder of inhomogeneous Laplace equations}

We have just seen that this newly defined space \eqref{f_definition} of modular invariant functions does contain both single Eisenstein series \eqref{eq:YEisen} and products of two Eisenstein series \eqref{prod_Eisenstein}-\eqref{prod_Eisenstein2}. We now show that the functions $\modfn{a}{b}{r}{s}{z}$ are also closed under the action of the Laplace operator in $z$.
In particular, we describe a method of generating solutions to an infinite ladder of Laplace equations where the source term is a fixed function $\modfn{a}{b}{r}{s}{z}$ and the eigenvalues lie in the spectrum 
\begin{equation}
{\rm Spec}(r{+}s) = \{r{+}s{-}2,\,r{+}s{-}4,\,r{+}s{-}6,\,...\}\,,
\end{equation}
i.e.~they take the form
\begin{equation}
\lambda_n(r+s):=r+s-2(n+1)\,,\label{eq:eigenLad}
\end{equation}
with and $n\in\mathbb{N}$. 
Once the source term $\modfn{a}{b}{r}{s}{z}$ is properly chosen, this spectrum reduces to the string theory spectra \eqref{eq:spec1}-\eqref{eq:spec2} and the constructed solution produces precisely a given generalised Eisenstein series expressed as a finite linear combination of novel Poincar\'e series \eqref{f_definition}.
Not to clutter the notation, in this section we will suppress the explicit $z$-dependence.

The starting point of our analysis is the differential equation \eqref{f_diff_eq}, rewritten here in a more convenient form
\begin{equation}\label{Diff_eq_maxeigen}
    \big[\Delta -\lambda_0(r+s)(\lambda_0(r+s)-1)\big] \frac{\modfntwo{a}{b-1}{r-1}{s-1}}{4\pi (1-r)}=\modfntwo{a}{b}{r}{s}\,.
\end{equation}
To construct this ladder of Laplace equations, we view this equation as the top element in a tower of similar equations with decreasing eigenvalues.
We now look for linear combinations, $\eigen_n(a,b,r,s)$, of functions $\modfntwo{a'}{b'}{r'}{s'}$ with different parameters $(a',b',r',s')$ and solutions to
\begin{equation}\label{ladder_of_diff}
    \big[ \Delta-\lambda_n(r+s)(\lambda_n(r+s)-1)\big]\eigen_n(a,b,r,s)= \modfntwo{a}{b}{r}{s}\,.
\end{equation}
The starting Laplace equation \eqref{Diff_eq_maxeigen} gives us the initial condition
\begin{equation}
\eigen_0(a,b,r,s) = \frac{\modfntwo{a}{b-1}{r-1}{s-1}}{4\pi (1-r)}\,,\label{eq:Y0}
\end{equation}
while the rest of the ladder is generated from here by exploiting the crucial recursion relation \eqref{f_recursion_formula} as we now show.

To simplify the discussion we introduce a linear operator $\mathcal{D}$ which acts on the space of modular functions \eqref{f_definition} as 
\begin{equation}
    \mathcal{D}\modfntwo{a}{b}{r}{s}:=\modfntwo{a}{b}{r}{s-2}+\frac{2s-3}{2\pi}\modfntwo{a}{b-1}{r-1}{s-1}\,,
\end{equation}
for which the recursion relation \eqref{f_recursion_formula} can then be written in the compact form 
$$\mathcal{D}\modfntwo{a}{b}{r}{s}=\modfntwo{a}{b}{r}{s}\,.$$
One can easily check by induction that an $n$-fold application of this operator produces a sum of $n+1$ modular functions given by
\begin{equation}
    \mathcal{D}^n\modfntwo{a}{b}{r}{s}=\sum_{k=0}^n {n \choose k} \Big(\prod_{i=0}^{k-1}\frac{2(s+i-n)-1}{2\pi}\Big)\modfntwo{a}{b-k}{r-k}{s+k-2n}\,.
\end{equation}

While it is not immediately obvious how to use the Laplace equation \eqref{f_diff_eq} to invert \eqref{ladder_of_diff} and find $\eigen_n(a,b,r,s)$, we can use the recursion relation to rewrite \eqref{ladder_of_diff} as
\begin{align}
   & \big[ \Delta-\lambda_0(r+s-2n)(\lambda_0(r+s-2n)\notag-1)\big]\eigen_n(a,b,r,s)=\modfntwo{a}{b}{r}{s}\\
    &\label{ladder_of_diff2}= \mathcal{D}^n\modfntwo{a}{b}{r}{s}=\sum_{k=0}^n {n \choose k} \Big(\prod_{i=0}^{k-1}\frac{2(s+i-n)-1}{2\pi}\Big)\modfntwo{a}{b-k}{r-k}{s+k-2n}\,.
\end{align}
Although $ \mathcal{D}^n \modfntwo{a}{b}{r}{s}$ is a linear combination of modular functions $\modfntwo{a}{b'}{r'}{s'}$ with different parameters $(a,b',r',s')$, we notice that the action of $ \mathcal{D}^n$ produces a uniform shift on $r+s$, i.e. for every term in this linear combination we have $ r'{+}s' =  r{+}s{-}2n$.
This means that if we consider the left-hand side of \eqref{ladder_of_diff2} term by term, we have reduced the problem to a collection of equations \eqref{Diff_eq_maxeigen} for different values of parameters $(a,b',r',s')$ but all satisfying $r'{+}s' =  r{+}s{-}2n$.
We can then use the inversion of the Laplacian \eqref{eq:Y0} term by term to arrive at
\begin{align}\label{eq:Ynsol}
\eigen_n(a,b,r,s) = \sum_{k=0}^n {n \choose k} \Big(\prod_{i=0}^{k-1}\frac{2(s+i-n)-1}{2\pi}\Big) \frac{\modfntwo{a}{b-k-1}{r-k-1}{s+k-2n-1}}{4\pi(k+1-r)}\,,
\end{align}
which is the sought-after solution to the ladder of Laplace equations \eqref{ladder_of_diff} with eigenvalue $\lambda_n(r+s) = r{+}s{-}2(n+1)$ and source $\modfntwo{a}{b}{r}{s}$.

Note that while in general this ladder does not terminate, whenever the parameter $r$ is a strictly positive integer, which will be the relevant case for the MGFs spectrum \eqref{eq:spec2}, the ladder does in fact terminate after finitely many steps.
This is easy to see from \eqref{eq:Ynsol}, let us assume that $r=n+1$ with $n\in \mathbb{N}$ for which \eqref{eq:Ynsol} becomes ill-defined. In \eqref{eq:Ynsol} the would-be $k=n$ term reduces to $\modfntwo{a}{b-n-1}{0}{s-n-1}$ and according to the differential equation \eqref{ladder_of_diff2}
the action of the Laplace eigenvalue operator on such a factor should produce the corresponding source proportional to $\modfntwo{a}{b-n}{1}{s-n}$. However, this is not possible since $\modfntwo{a}{b-n-1}{0}{s-n-1}$ is precisely proportional \eqref{eq:YEisen} to the Eisenstein series $\normEisenstein{s-n-1}{z}$ which is annihilated by \eqref{ladder_of_diff} in the case $r=n+1$. We will come back to this point when discussing this ladder of equations for the case of MGF generalised Eisenstein series.

In the context of this paper, we are particularly interested in generating solutions to Laplace eigenvalue equations with sources given by products of two Eisenstein series.
One of the perks of our approach is that the ladder of Laplace equations \eqref{ladder_of_diff} just found precisely reduces to the desired inhomogeneous Laplace 
eigevalue equations when the source term $\modfn{a}{b}{r}{s}{z}$ is suitably chosen as to reproduce the wanted bilinear in Eisenstein series as given in 
\eqref{prod_Eisenstein}-\eqref{prod_Eisenstein2}. \vspace{0.0cm}
\vspace{0.2cm}

\textbf{The first flavour of generalised Eisenstein series}
\vspace{0.2cm}

Let us now use the ladder \eqref{ladder_of_diff} just discussed to reconstruct the first string theory flavour of generalised Eisenstein series \eqref{eq:spec1}.
We then consider half-integer indices $s_1,s_2\in\mathbb{N}+\frac{1}{2}$ and we want to reproduce the non-terminating spectrum of eigenvalues 
$${\rm Spec_1}(s_1,s_2)=\{s_1{+}s_2{+}1,\,s_1{+}s_2{+}3,\,s_1{+}s_2{+}5,\,...\}\,.$$
To this end, we specialise the ladder \eqref{ladder_of_diff} to the case for which the source term $\modfn{a}{b}{r}{s}{z}$ produces the second representation we found for the product of two Eisenstein series \eqref{prod_Eisenstein2}, i.e. we specialise our ladder to 
\begin{equation}
(a,b,r,s) = \Big( 1-2 s_1, \, s_1,\, 1 -s_2,\, 1-s_1\Big)\,,\label{eq:Param1}
\end{equation}
and assume that $s_1,s_2$ are fixed half-integers, in which case \eqref{ladder_of_diff} can be reduced to
\allowdisplaybreaks{
\begin{align}
    &\label{diff_eq_genEisen2} \Big[\Delta-\lambda^{(1)}_n(\lambda^{(1)}_n-1)\Big]\frac{8\xi(2s_2{-}1)}{\Gamma(s_1)\Gamma(s_2)}\eigen_n(1-2s_1,\, s_1,\,1-s_2,\,1-s_1\vert z)= \Eisenstein{s_1}{z}\Eisenstein{s_2}{z}\\*
    &\notag {-}\frac{2\Gamma(s_1{+}s_2-1)\xi(2s_1{-}1)\xi(2s_2{-}1)}{\Gamma(s_1)\Gamma(s_2)\xi(2(s_1{+}s_2){-}3)}\Eisenstein{s_1{+}s_2{-}1}{z}{-}\frac{2\Gamma(s_1{+}1{-}s_2)\xi(2s_1)\xi(2s_2{-}1)}{\Gamma(s_1)\Gamma(s_2)\xi(2(s_1{+}1{-}s_2))}\Eisenstein{s_1{+}1{-}s_2}{z}\,.
\end{align}}
If we apply directly the ladder procedure with fixed parameters \eqref{eq:Param1}, we find that the ladder eigenvalues (dropping their explicit dependence from the fixed source indices $s_1,s_2$) are now ${\tilde{\lambda}}^{(1)}_{n}=-s_1-s_2-2n$, however, the exchange ${\tilde{\lambda}}^{(1)}_{n} \to \lambda^{(1)}_n = 1-{\tilde{\lambda}}^{(1)}_{n}$ leaves the equation invariant and produces the expected spectrum of eigenvalues 
$$\lambda^{(1)}_{n}  = s_1{+}s_2{+}2n{+}1\,.$$ 

This change is not without consequences: the constructed modular invariant solution,~$\eigen_n$, does not quite land (modulo single Eisenstein terms) on $\mathcal{E}(\lambda^{(1)}_n;s_1,s_2\vert z)$, the generalised Eisenstein series we are interested in, but rather on the reflected~$\mathcal{E}(1{-}\lambda^{(1)}_n;s_1,s_2\vert z)$. We can use the general expression \eqref{full_asymptotics} to compute the asymptotic expansion of $\eigen_n(1{-}2s_1, s_1,1{-}s_2,1{-}s_1\vert z)$  at large-$y$ and confirm that the homogeneous solution $y^{1-\lambda^{(1)}_n}$ has vanishing coefficient, i.e. we land exactly on the opposite boundary condition compared to the wanted generalised Eisenstein series $\mathcal{E}(\lambda^{(1)}_n;s_1,s_2\vert z)$.
This can be fixed by adding a suitable multiple of the modular invariant homogeneous solution, $\Eisenstein{\lambda^{(1)}_n}{z}$, such that the new solution satisfies the desired boundary condition of a vanishing coefficient for the homogeneous solution $y^{\lambda^{(1)}_n}$.

Lastly,  with the help of the differential equation \eqref{Eisenstein_eigval} we can easily invert the single Eisenstein source terms in \eqref{diff_eq_genEisen2}. 
With all these considerations in mind, we arrive to the final expression
\begin{align}
    \genEisenstein{s_1}{s_2}{\lambda^{(1)}_n}{z}=&\,\notag \frac{8\xi(2s_2-1)}{\Gamma(s_1)\Gamma(s_2)}\eigen_n(1-2s_1,\, s_1,\,1-s_2,\,1-s_1\vert z)\\
    &\notag \!\!\!\!\! \!\!\!\!\!\!\!\! -\frac{2\Gamma(\lambda^{(1)}_n)\xi(2n{+}2)\xi(2s_1{+}2n{+}1)\xi(2s_2{+}2n{+}1)\xi(2(s_1{+}s_2{+}n))}{(2\lambda^{(1)}_n-1)\Gamma(s_1)\Gamma(s_2)\xi(2\lambda^{(1)}_n-1)\xi(2\lambda^{(1)}_n)}\Eisenstein{\lambda^{(1)}_n}{z}\\
    &\label{GenEisen_Spec1_formula}+\frac{2\Gamma(s_1+s_2-1)\xi(2s_1-1)\xi(2s_2-1)}{\Gamma(s_1)\Gamma(s_2)\xi(2(s_1+s_2)-3)\mu(s_1+s_2-1,\lambda^{(1)}_n)}\Eisenstein{s_1+s_2-1}{z}
    \\
&\notag    +\frac{2\Gamma(s_1+1-s_2)\xi(2s_1)\xi(2s_2-1)}{\Gamma(s_1)\Gamma(s_2)\xi(2(s_1+1-s_2))\mu(s_1+1-s_2,\lambda^{(1)}_n)}\Eisenstein{s_1+1-s_2}{z}\,,
\end{align}
where we defined $\mu(s,\lambda):=s(s-1)-\lambda(\lambda-1)$.
\vspace{0.2cm}

\textbf{The second flavour of generalised Eisenstein series}
\vspace{0.2cm}

 We turn our focus to the second flavour of generalised Eisenstein series. The indices $s_1,s_2\geq2$ are now integers and without loss of generality we assume $s_1\geq s_2$. We want to  use the ladder \eqref{ladder_of_diff} to reproduce the finite spectrum of eigenvalues
 $${\rm Spec_2}(s_1,s_2)=\{|s_1{-}s_2|{+}2,\,|s_1{-}s_2|{+}4,\,...\,,\,s_1{+}s_2{-}2\}\,.$$
Consequently, we specialise the ladder \eqref{ladder_of_diff} to the case for which the source term $\modfn{a}{b}{r}{s}{z}$ produces the first representation we found for the product of two Eisenstein series \eqref{prod_Eisenstein}, i.e. we specialise our ladder to 
\begin{equation}
(a,b,r,s) = \Big( 1-2 s_2,\,s_2,\,s_1, \,s_2 \Big)\,.\label{eq:Param2}
\end{equation}
 
With this choice of parameters the ladder equation \eqref{ladder_of_diff} reduces to
\allowdisplaybreaks{
\begin{align}
 &\label{diff_eq_genEisen1}  \Big[\Delta-\lambda^{(2)}_n(\lambda^{(2)}_n-1)\Big]\frac{8\xi(2s_1)}{\Gamma(s_1)\Gamma(s_2)}\eigen_n(1-2s_2,\,s_2 ,\,s_1 ,\,s_2 \vert z)  = \Eisenstein{s_1}{z}\Eisenstein{s_2}{z}\\*
 &\notag  -\frac{2\Gamma(s_1+s_2)\xi(2s_1)\xi(2s_2)}{\Gamma(s_1)\Gamma(s_2)\xi(2(s_1+s_2))}\Eisenstein{s_1+s_2}{z}-\frac{2\Gamma(s_1+1-s_2)\xi(2s_1)\xi(2s_2-1)}{\Gamma(s_1)\Gamma(s_2)\xi(2(s_1+1-s_2))}\Eisenstein{s_1+1-s_2}{z}\,,
\end{align}}
and the ladder eigenvalues, $\lambda^{(2)}_n=s_1+s_2-2(n+1)$, reproduce immediately the desired spectrum.

In this second setup there is no issue with the large-$y$ asymptotic behaviour for the solution $\eigen_n(1-2s_2,s_2 ,s_1 ,s_2 \vert z)$: using the general expression \eqref{full_asymptotics} we can confirm that our ladder solution satisfies the desired boundary condition for which the coefficient of the homogeneous solution $y^{\lambda^{(2)}_n}$ vanishes. This means that for the specific parameters \eqref{eq:Param2} the ladder solution \eqref{eq:Ynsol} must reproduce (modulo single Eisenstein terms) the second flavour of generalised Eisenstein series, $\mathcal{E}(\lambda^{(2)}_n;s_1,s_2\vert z)$.
Proceeding as we did before, we use \eqref{Eisenstein_eigval} to invert the single Eisenstein source terms in \eqref{diff_eq_genEisen1} and arrive at
\begin{align}
    \genEisenstein{s_1}{s_2}{\lambda^{(2)}_n}{z}=&\notag\frac{8\xi(2s_1)}{\Gamma(s_1)\Gamma(s_2)}\eigen_n(1-2s_2,\,s_2 ,\,s_1,\,s_2 \vert z)\\
 &\label{GenEisen_Spec2_formula}   +\frac{2\Gamma(s_1+s_2)\xi(2s_1)\xi(2s_2)}{\Gamma(s_1)\Gamma(s_2)\xi(2(s_1+s_2))\mu(s_1+s_2,\lambda^{(2)}_n)}\Eisenstein{s_1+s_2}{z}\\
 &\notag  +\frac{2\Gamma(s_1+1-s_2)\xi(2s_1)\xi(2s_2-1)}{\Gamma(s_1)\Gamma(s_2)\xi(2(s_1+1-s_2))\mu(s_2-s_1,\lambda^{(2)}_n)}\Eisenstein{s_1+1-s_2}{z},
\end{align}

Unlike what happens in the previous case, when the sources have integer indices, $s_1,s_2$, we notice that the spectrum of eigenvalues is bounded both from above and below. 
There is a maximal eigenvalue in the ladder which is given by $\lambda^{(2)}_0{=}s_1{+}s_2{-}2$ and agrees with the maximal eigenvalue obtained in the study of MGFs in the second spectrum \eqref{eq:spec2}. 
However the minimal eigenvalue in the ladder does not quite reproduce the minimal eigenvalue expected from \eqref{eq:spec2}. 

As discussed below equation \eqref{eq:Ynsol}, in the case when the parameter $r=\tilde{n}+1$, with $\tilde{n}\in \mathbb{N}$, the ladder terminates after $\tilde{n}$ steps. In the present case \eqref{GenEisen_Spec2_formula}, the parameter $r= s_1 $ has precisely this property, hence the ladder terminates after $\tilde{n} = s_1-1$ steps, i.e.~we have constructed generalised Eisenstein solutions \eqref{GenEisen_Spec2_formula} for $n=0,...,s_1-2$ and fixed sources.
The minimal eigenvalue we obtain is then $\lambda^{(2)}_{s_1-2}{=}s_2{-}s_1{+}2$, in general lower than the minimal eigenvalue expected from the spectrum \eqref{eq:spec2}. These ladder solutions \eqref{GenEisen_Spec2_formula} with eigenvalues lower than the MGFs spectrum \eqref{eq:spec2} correspond precisely to the modular objects discussed in section 7.3 of \cite{Dorigoni:2021ngn} and constructed from certain ``overly-integrated seed functions''.

In summary, the ladder of Laplace equations \eqref{ladder_of_diff} includes in a natural and uniform way the two string theory flavours of generalised Eisenstein series \eqref{eq:spec1}-\eqref{eq:spec2}. In both cases \eqref{GenEisen_Spec1_formula}-\eqref{GenEisen_Spec2_formula}, we expressed these generalised Eisenstein series as linear combination of finitely many novel Poincar\'e series \eqref{f_definition}. We now discuss some concrete examples for both flavours.

\subsection{Examples}

In this section we present some concrete, and string theory relevant, examples of our general construction. 
We begin with the generalised Eisenstein series $\genEisenstein{\threeh}{\threeh}{4}{z}$, coefficient of the higher derivative correction $d^6R^4$ in the effective low-energy action of type IIB superstring theory~\eqref{SUGRA_Lagran}. For the given indices, $s_1=s_2=\threeh$, the eigenvalue is $\lambda=\lambda_0^{(1)}=s_1+s_2+1 = 4$, hence~$\genEisenstein{\threeh}{\threeh}{4}{z}$ is the function with smallest eigenvalue in the spectrum~\eqref{eq:spec1} for these sources. 

Since this is a diagonal example where the indices $s_1$ and $s_2 $ coincide, we need to use the regularisation scheme described in \eqref{diagonal_regul}. 
Substituting the regularised parameters $s_1 = \threeh+\epsilon$, $s_2=\threeh$ and $\lambda_0^{(1)} =4+\epsilon$ in the general expression \eqref{GenEisen_Spec1_formula} we derive
\begin{align}
    \genEisenstein{\threeh}{\threeh}{4}{z} =
    &\notag\lim_{\epsilon\to 0}\Big[\,\frac{4}{9\sqrt{\pi}\,\Gamma(\frac{3}{2}+\epsilon)}\modfn{\shortminus 2(1+\epsilon)}{\half+ \epsilon}{\shortminus\threeh }{\shortminus\threeh\shortminus\epsilon}{z}-\frac{\zeta(3+2\epsilon)}{9(2+\epsilon)\zeta(2+2\epsilon)}\Eisenstein{1+\epsilon}{z}\Big]\\
&   -\frac{32\pi^6}{127\,575\zeta(7)}\Eisenstein{4}{z}  -\frac{2\pi^2}{45\zeta(3)}\Eisenstein{2}{z}\,.
\end{align}
As previously stated each term inside the limit is separately singular at $\epsilon=0$, however, this combination is such that the divergences in $1/\epsilon$ cancel out and produce a finite expression for $\epsilon=0$. 
We can substitute this regulated expression into the general formula \eqref{full_asymptotics} to recover the well-known asymptotic expansion \cite{Green:2014yxa} of the $d^6R^4$ correction
\begin{equation}
    \genEisenstein{\threeh}{\threeh}{4}{z}\sim -\frac{2\zeta(3)^2y^3}{3\pi^3}-\frac{2\zeta(3)y}{9\pi}-\frac{2\pi}{45y}-\frac{4\pi^3}{25\,515y^3} \qquad \text{as} \quad y\gg 1\,.
\end{equation}

A second related example is the modular invariant function $\genEisenstein{\fiveh}{\threeh}{7}{z}$ which arises at order $O(N^{\shortminus2})$ in the large-$N$ expansion of the particular $\mathcal{N}=4$ SYM integrated correlator discussed in \cite{Chester:2020vyz}. This case falls again into the spectrum \eqref{eq:spec1}, the indices are $s_1= \fiveh\,,\,s_2 = \threeh$ while the eigenvalue is $\lambda =\lambda_1^{(1)}= s_1+s_2+3 =7$ hence one step above the lowest eigenvalue in our Laplace tower \eqref{GenEisen_Spec1_formula}.  
If we substitute these specific values for $s_1,s_2$ and $\lambda_1^{(1)}$ in \eqref{GenEisen_Spec1_formula}  we obtain the Poincar\'e series representation
\begin{align}
    \genEisenstein{\fiveh}{\threeh}{7}{z}=&-\frac{16}{15\pi^2}\modfn{-4}{\half}{-\fiveh }{-\sevenh}{z}+\frac{16}{27\pi}\modfn{-4}{\threeh}{-\threeh}{-\nineh}{z}\\
    &\notag -\frac{4096\pi^{12}}{46\,414\,974\,375\zeta(13)}\Eisenstein{7}{z}-\frac{8\pi^4}{10\,935\zeta(5)}\Eisenstein{3}{z}-\frac{3\zeta(5)}{2\pi^4}\Eisenstein{2}{z}\,.
\end{align}
Substituting this expression in the general formula \eqref{full_asymptotics} we obtain the asymptotic expansion
\begin{equation}
    \genEisenstein{\threeh}{\fiveh}{7}{z}\sim -\frac{2\zeta(3)\zeta(5)y^4}{15\pi^4}-\frac{\zeta(5)y^2}{30\pi^2}-\frac{4\zeta(3)}{2835}-\frac{2\pi^2}{3645y^2}-\frac{8\pi^6}{200\,930\,625y^6}\qquad \text{as}\quad y\gg1 .
\end{equation}

Finally, we discuss an example of generalised Eisenstein series belonging to the second spectrum \eqref{eq:spec2}.
We consider the function $\genEisenstein{3}{2}{3}{z}$ which captures the genuine dept-two part of the two-loop MGF usually denoted by $C_{3,1,1}(z)$,
\begin{equation}
C_{3,1,1}(z) = - 4 \, \genEisenstein{3}{2}{3}{z} + \frac{43}{35} \Eisenstein{5}{z} - \frac{ \zeta_5}{60}\,.
\end{equation} 
The indices are $s_1=3,s_2=2$ while the eigenvalue is $\lambda=\lambda_0^{(2)}=s_1+s_2-2=3$ hence $\genEisenstein{3}{2}{3}{z}$ is the function with largest eigenvalue in the second spectrum \eqref{eq:spec2} for these sources.
Substituting the specific values for $s_1,\,s_2$ and $\lambda_0^{(2)}$ in the general solution \eqref{GenEisen_Spec2_formula} we obtain
\begin{equation}\label{eq:ex1}
    \genEisenstein{3}{2}{3}{z} =-\frac{\pi^2}{945}\modfn{-3}{1}{2}{1}{z}+ \frac{11}{70}\Eisenstein{5}{z}-\frac{\zeta(3)}{42}\Eisenstein{2}{z}\,.
\end{equation}
It is interesting to compare the present Poincar\'e series representation \eqref{eq:ex1} with a different one (finely tuned to represent all two-loop MGFs) considered in \cite{Dorigoni:2019yoq} for which we have
 \begin{equation}
    \genEisenstein{3}{2}{3}{z} = \sum_{\gamma \in B(\ZZ)\backslash {\rm SL}(2,\ZZ)}  \left[ \frac{(\pi y)^5}{297\,675}-\frac{(\pi y)^2 \zeta(3)}{1890}-  \frac{(\pi y)^2}{1890} \sum_{m=1}^{\infty} \sigma_{-3}(m) ( q^m + \bar{q}^m)\right]_{\gamma}\,.
 \end{equation}
Again thanks to the general expression \eqref{full_asymptotics}, starting from \eqref{eq:ex1} we can retrieve the known asymptotic expansion
\begin{equation}
    \genEisenstein{3}{2}{3}{z} \sim \frac{\pi^5y^5}{297\,675}-\frac{\zeta(3)\pi^2y^2}{1890}-\frac{\zeta(5)}{360}-\frac{7\zeta(7)}{64\pi^2y^2}+\frac{\zeta(3)\zeta(5)}{8\pi^3y^3}\qquad \text{as}\quad y\gg 1.
\end{equation}

Compared to previous results in the literature, one novelty of our Poincar\'e series \eqref{f_definition} is that all the examples here considered, and more broadly all generalised Eisenstein with spectra \eqref{eq:spec1}-\eqref{eq:spec2} can be expressed as linear combinations of finitely many $\modfn{a}{b}{r}{s}{z}$.

\section{Spectral analysis point of view}
\label{sec:spectral}

The second representation we wish to discuss for the modular objects under consideration relies on $\SLtwoZ$ spectral theory.
The key idea behind spectral theory is to decompose any modular invariant function as a linear combination of ``good'' basis elements, i.e. normalisable eigenfunctions of the hyperbolic Laplace operator.

This has been extremely fruitful in the study of two-dimensional conformal field theories~\cite{Benjamin:2021ygh,Benjamin:2022pnx,DiUbaldo:2023qli} and integrated correlators in $\mathcal{N}=4$ SYM~\cite{Collier:2022emf,Paul:2022piq,Paul:2023rka}. In particular, appendix B of~\cite{Collier:2022emf} presents a self-contained spectral analysis discussion for some of the generalised Eisenstein series appearing in our work, while in~\cite{Klinger:2018} a more general study is presented.
A complete treatment of spectral analysis is beyond the scope of the present work and we refer to~\cite{Iwaniec:2002,Terras} for a thorough introduction to the subject while presenting here only some of the key details.

 We remind the reader that the standard fundamental domain of $\SLtwoZ$ is defined by 
 \allowdisplaybreaks{
 \begin{align*}
 \mathcal{F} &:=\SLtwoZ \backslash \mathcal{H}  \\*
 &\phantom{:}=\Big\lbrace z\in\mathcal{H}\,\Big\vert\,  |z|> 1\,,\, -\frac{1}{2} < \Re(z) \leq \frac{1}{2} \Big\rbrace \cup \Big\lbrace  z\in\mathcal{H}\,\Big\vert\, |z| = 1\,, \,0\leq \Re(z)\leq \frac{1}{2}\Big\rbrace \,,
 \end{align*}}
endowed with the natural hyperbolic metric
 \begin{equation}
 \dd s^2 = \frac{\dd x^2 +\dd y^2}{y^2}\,,
 \end{equation}  
 where $z = x+i y$.
 
Given that any point $z$ in the upper half-plane $\mathcal{H}$ is conjugate to a point $\gamma\cdot z \in \mathcal{F}$ by a suitable $\gamma \in \SLtwoZ$, we have that modular invariant functions $f(z)= f(\gamma\cdot z)$ with $f:\mathcal{H}\to \mathbb{C}$ can be considered simply as functions defined on $\mathcal{F}$.
 We can then define the Hilbert space $L^2(\mathcal{F})$ of square-integrable functions with respect to the Petersson inner product
\begin{equation}\label{inner_product}
    (f,g) = \int_\mathcal{F} f(z)\,\overline{g(z)} \,\dd \mu ,
\end{equation}
where the invariant Haar measure is $\dd\mu = y^{-2}\dd x\, \dd y$. 

Note that for a function $f$ to be an element of $L^2(\mathcal{F})$, its growth at the cusp $y\gg1$ must be at most  $|f(z)| = O(y^{\frac{1}{2}})$. In what follows, we will often encounter modular invariant functions $f$ violating such bound, i.e. non-$L^2(\mathcal{F})$ normalisable functions. 
Although this growth condition seems quite restrictive, and in sharp conflict with the asymptotic expansion \eqref{full_asymptotics} previously found, spectral analysis methods can be extended from square-integrable functions to a broader class of functions that have moderate growth at the cusp.

If a function $f$ has cuspidal growth $|f(z)| = O(y^{\alpha})$ with $\Re(\alpha)>\frac{1}{2}$, we can find a coefficient $\beta$ such that the new modular invariant combination $f(z) - \beta \,\normEisenstein{\alpha}{z}$ has a tamer growth at the cusp. More generally, we will be discussing modular invariant functions whose asymptotic expansion at the cusp is controlled by finitely many non-integrable power-like terms $y^{\alpha_i}$ with $\Re(\alpha_i)>\frac{1}{2}$. Although such functions $f(z)$ are not elements  of $L^2(\mathcal{F})$, we can find coefficients $\beta_i$ for which the linear combination 
\begin{equation}
f_{{\rm new}}(z) = f(z) - \sum_i \beta_i \,\normEisenstein{\alpha_i}{z} \in L^2(\mathcal{F})\,,
\end{equation}
is $L^2$-normalisable.

Modulo the caveat just mentioned, we now consider in more detail the Hilbert space $ L^2(\mathcal{F})$ with inner product \eqref{inner_product}. One of the main benefits of working with a vector space is that we can always express a generic element in terms of a basis. Furthermore, since we are interested in solving differential equations with respect to the hyperbolic Laplacian\footnote{Note that in the mathematics literature, the hyperbolic Laplacian considered is usually $\tilde{\Delta} = -y^2(\partial_x^2+\partial_y^2)$ so that the spectrum of its $L^2(\mathcal{F})$ eigenfunctions is non-negative.}$\Delta = y^2(\partial_x^2+\partial_y^2)$, and since this operator is self-adjoint with respect to the inner product \eqref{inner_product}, it is natural to use the Laplace eigenfunctions as a basis for $ L^2(\mathcal{F})$.

The spectrum of the hyperbolic Laplacian decomposes into three distinct eigenspaces (again we refer to \cite{Iwaniec:2002,Terras} for details): 
\begin{itemize}
\item The constant function $f(z)=1$ is clearly an eigenfunction of $\Delta$ with eigenvalue $0$, and it is an element of $ L^2(\mathcal{F})$, since $(1,1) = {\rm Vol}(\mathcal{F}) = \frac{\pi}{3}$ is the volume of the fundamental domain.
\item The continuous part of the spectrum is spanned by $\normEisenstein{s}{z}$ with $\Re( s) = \frac{1}{2}$ and eigenvalue $s(s-1)$ given \eqref{Eisenstein_eigval}.
\item The discrete part of the spectrum is spanned by the Maass cusp forms, $\phi_n(z)$ with $n\in \mathbb{N}^{>0}$\,. 
\end{itemize}

While the non-holomorphic Eisenstein series $\normEisenstein{s}{z}$ with $\Re( s) = \frac{1}{2}$ are simply meromorphic continuations in $s$ of \eqref{Eisen_poinc}, the Maas cusp forms $\phi_n(z)$ are different beasts altogether.
These are modular invariant eigenfunctions of the Laplacian 
\begin{equation}
    \Delta \phi_n(z) = \mu_n\phi_n(z),\qquad \text{where}\qquad \mu_n=-\Big( \frac{1}{4}+t_n^2\Big)\,,\quad 0<t_1<t_2<...\,,
\end{equation}
with the spectral parameters $t_n$, specifying the eigenvalue $\mu_n$, forming an infinite and unbounded set of sporadic positive numbers.
Similarly to \eqref{Eisen_poinc}, they admit a Fourier mode decomposition 
\begin{equation}
\phi_n(z) =  \sum_{k\neq 0} a_k^{(n)}  y^{\frac{1}{2}}K_{it_n}(2\pi |k|y)e^{2\pi ikx}\,,\label{eq:MaassFourier}
\end{equation}
and the Fourier coefficients $a_k^{(n)}$ are once more a set of sporadic real numbers.
Given the outer automorphism of order two $z\to -\bar{z}$, we can divide the Maass forms into even forms, i.e. $\phi_n(z)  = \phi_n(-\bar{z})$, and odd forms, i.e. $\phi_n(z)  = -\phi_n(-\bar{z})$. 

Presently, we are working with the convention that $\phi_n(z)$ is normalised in the sense of the Petersson inner product \eqref{inner_product}, i.e.~we have $(\phi_n,\phi_n)=1$. However, another common choice for $\phi_n(z)$ is to be Hecke normalised, i.e.~to have $a_1^{(n)}=1$. Clearly the two normalisation are just a scalar multiple of one another.

Note that from the Fourier decomposition \eqref{eq:MaassFourier}, and as suggested by their name, the Maass cusp forms are indeed cuspidal objects, i.e.~they have vanishing Fourier zero-mode and decay exponentially fast as $y\gg 1$:
$$
\phi_n(z)  \sim e^{-2\pi y}\qquad\qquad {\rm for }\qquad y\gg 1\,.
$$
Since the Maass cusp forms only contribute to the Fourier non-zero mode sector, we will not be discussing their effects in what follows as our focus will be primarily on the Fourier zero-mode sector.~The interested reader can find both spectral parameters and Fourier coefficients for various even/odd Maass cusp forms on the L-functions and modular forms database (LMFDB)~\cite{lmfdb}.

Once the basis of eigenfunctions for the Laplacian is understood, we are naturally led to consider the Roelcke-Selberg spectral decomposition: 
\begin{equation}\label{eq:spectral}
    f(z) = \average{f} +\int_{\Re(t)=\frac{1}{2}}  (f,{\rm E}_{t})\,\normEisenstein{t}{z} \frac{ \dd t}{4\pi i}+\sum_{n=1}^\infty (f,\phi_n)\phi_n(z)\,,
\end{equation}
for a generic $f\in L^2(\mathcal{F})$ (inside the inner product we use the short-hand notation ${\rm E}_{t}:=\normEisenstein{t}{z}$).

The first term is simply $\average{f} = \int_\mathcal{F} f(z)\, {\rm d}\mu$, which can be understood as the average of the function over the fundamental domain, or equivalently as the spectral overlap with the constant function $\average{f} = (f,1)$. The remaining part of the decomposition can be understood as a ``linear'' combination of orthonormal basis elements whose coefficients are simply given by the inner product of the function $f(z)$ under consideration and the respective basis element.

We will shortly focus on analysing the Fourier zero-mode of generalised Eisenstein series $\genEisenstein{s_1}{s_2}{\lambda}{z}$ and the functions $\modfn{a}{b}{r}{s}{z}$, or rather suitable $L^2(\mathcal{F})$ versions thereof, by using spectral analysis. To this end, we notice that if we Fourier decompose $f\in L^2(\mathcal{F})$ as 
$$f(z) = \sum_{k\in\mathbb{Z}}f_k(y)e^{2\pi ikx}\,,$$
the spectral decomposition \eqref{eq:spectral} immediately provides for a nice contour integral representation for the Fourier zero-mode $f_0(y)$. 
Since from \eqref{eq:MaassFourier} we know that the Maass cusp forms have vanishing Fourier zero-mode, we conclude that only the Eisenstein series can contribute. 
Furthermore, from the Fourier decomposition \eqref{Eisen_poinc} for $\normEisenstein{t}{z}$, we know that the Fourier zero-mode of the Eisenstein series contains only two power-behaved terms, $y^t$ and $y^{1-t}$. We can however combine the reflection property \eqref{eq:reflect}, relating $\normEisenstein{t}{z}$ to  $\normEisenstein{1-t}{z}$, with a change of variables $t\to 1-t$ to show that both terms $y^t$ and $y^{1-t}$ give an equal contribution, arriving at
\begin{equation}\label{spectral_Fzero_mode}
    f_0(y) = \average{f} + \int_{\Re(t)=\frac{1}{2}}(f,{\rm E}_{t})\,y^t \frac{\dd t}{2\pi i }\,.
\end{equation}

This formula may appear rather useless since to extract the Fourier zero-mode $f_0(y)$ it would seem necessary to know already the full modular function $f(z)$ to be able to compute its spectral overlap $(f,{\rm E}_{t})$. However, in the next section we will show that for both generalised Eisenstein $\genEisenstein{s_1}{s_2}{\lambda}{z}$ and $\modfn{a}{b}{r}{s}{z}$ equation \eqref{spectral_Fzero_mode} becomes extremely useful and the overlap $(f,{\rm E}_{t})$ can be neatly computed using an ``unfolding-trick'' involving the Poincar\'e series representations.

Finally, we notice that once the spectral overlap $(f,{\rm E}_{t})$ is known, the integral representation \eqref{spectral_Fzero_mode} enables us to explore both the ``weak-coupling'' asymptotic regime $y\gg 1$ as well as the ``strong-coupling'' regime $y\to 0$ by a suitable choice on how we close the $t$-contour of integration at infinity.

\subsection{Back to the Fourier zero-mode}

Let us briefly review how one can exploit the differential equation \eqref{gen_Eisenstein_def} to compute the  spectral decomposition of the generalised Eisenstein series  and in particular obtain a useful integral representation \eqref{spectral_Fzero_mode} for its Fourier zero-mode. Since the generalised Eisenstein series, $\genEisenstein{s_1}{s_2}{\lambda}{z}$, is not an element of $L^2(\mathcal{F})$, one has to be a little careful in defining a proper regularised version for the spectral overlaps when dealing with functions not of rapid decay.
This problem was addressed in a beautiful and classic paper by Don Zagier \cite{ZagierRS} from which we present here a few key details; we also refer to \cite{Klinger:2018} and appendix B of \cite{Collier:2022emf} for more details on the generalised Eisenstein series.
 
Firstly we want to understand the behaviour at the cusp $y\gg1$ of the generalised Eisenstein series by exploiting its differential equation \eqref{gen_Eisenstein_def}, repeated here for convenience
\begin{equation*}
    [\Delta -\lambda(\lambda-1)]\genEisenstein{s_1}{s_2}{\lambda}{z} = \Eisenstein{s_1}{z}\Eisenstein{s_2}{z}.
\end{equation*}
As usual we perform the Fourier decomposition in $x = \Re(z)$,
\begin{equation}
\genEisenstein{s_1}{s_2}{\lambda}{z}=\sum_{k\in\mathbb{Z}}e_k(\lambda; s_1,s_2\vert y)e^{2\pi ikx}\,,\label{eq:FourierEgen}
\end{equation} 
and thanks to linearity, we can solve the inhomogeneous Laplace equation mode by mode. 

From \eqref{Eisenstein_def} we easily extract the Fourier zero-mode contribution to the bilinear source term $ \Eisenstein{s_1}{z}\Eisenstein{s_2}{z}$, comprised of power-behaved terms and exponentially suppressed terms $e^{-4\pi y}$.
Thus we find a solution to the differential equation for the Fourier zero-mode $e_0(s_1,s_2;\lambda\vert y)$:
\begin{align}\label{FZero_genEisen_asymp}
  &  e_0(\lambda;s_1,s_2 \vert y)  =  \frac{4\pi^{-s_1-s_2}\zeta(2s_1)\zeta(2s_2)}{(s_1+s_2-\lambda)(s_1+s_2+\lambda-1)}y^{s_1+s_2}\\
  &\notag+\frac{4\pi^{-s_1}\xi(2s_2-1)\zeta(2s_1)}{(s_1+1-s_2-\lambda)(s_1-s_2+\lambda)\Gamma(s_2)}y^{s_1+1-s_2} +\frac{4\pi^{-s_2}\xi(2s_1-1)\zeta(2s_2)}{(s_2+1-s_1-\lambda)(s_2-s_1+\lambda)\Gamma(s_1)}y^{s_2+1-s_1}\\
&\notag+\frac{4\xi(2s_1-1)\xi(2s_2-1)}{(s_1+s_2-\lambda-1)(s_1+s_2+\lambda-2)\Gamma(s_1)\Gamma(s_2)}y^{2-s_1-s_2} + \alpha(\lambda;s_1,s_2)y^{1-\lambda} +O(e^{-4\pi y})\,.
\end{align}

The constant $\alpha(\lambda; s_1,s_2)$ parametrises the homogeneous solution, $y^{1-\lambda}$, and can not be determined by solely analysing the differential equation.
However, the coefficient $\alpha(\lambda;s_1,s_2)$ will be promptly fixed by requiring modular invariance for the solution. Furthermore, since we are dealing with a second-order differential equation, we must have two linearly independent homogeneous solutions, which in the Fourier zero-mode sector are $y^{1-\lambda}$ and $y^\lambda$. It is conventional to choose a vanishing coefficient for the second homogeneous solution, $y^{\lambda}$. Once the modular invariant solution, $\genEisenstein{s_1}{s_2}{\lambda}{z}$, subject to this boundary condition has been found, we can always consider $\genEisenstein{s_1}{s_2}{\lambda}{z} + a \,\Eisenstein{\lambda}{z}$, with $a\neq0 $, which is a different modular invariant solution to the same Laplace system, but this time with a non-vanishing coefficient for $y^\lambda$.  

As anticipated, from the Fourier zero-mode analysis \eqref{FZero_genEisen_asymp} we immediately deduce that the generalised Eisenstein series is not an element of $L^2(\mathcal{F})$. To simplify the discussion, we can assume that the eigenvalue $\lambda$ is such that $\Re(\lambda)>\frac{1}{2}$, a condition that is satisfied by both spectra  \eqref{eq:spec1} and \eqref{eq:spec2}.
With this assumption, from \eqref{FZero_genEisen_asymp} we have full control over all power-behaved terms that might grow faster than $y^{\frac{1}{2}}$ at the cusp, and subsequently we can subtract suitable Eisenstein series in order to cancel all non-integrable terms thus obtaining a modular invariant and square-integrable function. 

We are then led to consider the ``regularised'' linear combination
\begin{equation}\label{reg_gen_Eisenstein}
    \tilde{\mathcal{E}}(\lambda;s_1,s_2\vert z) = \genEisenstein{s_1}{s_2}{\lambda}{z}-\sum_{I}\beta_I \Eisenstein{I}{z},
\end{equation}
where $I\in\{s_1+s_2,s_1+1-s_2,s_2+1-s_1,2-s_1-s_2\}$ and $\beta_I$ are chosen such that the term of order $y^I$ in $\tilde{\mathcal{E}}(\lambda;s_1,s_2\vert z)$ has a vanishing coefficient if $\Re(I)>\frac{1}{2}$ and $\beta_I=0$ otherwise. By construction, we clearly have $\tilde{\mathcal{E}}(\lambda;s_1,s_2)\in L^2(\mathcal{F})$, hence its Fourier zero-mode can be given in terms of the contour integral representation \eqref{spectral_Fzero_mode}. 

Now that we have modified the generalised Eisenstein series to obtain a nice and square-integrable function, $\tilde{\mathcal{E}}(\lambda;s_1,s_2\vert z)$, we can combine the spectral methods described in the previous section with the Laplace equation \eqref{gen_Eisenstein_def}.
It is fairly easy to see from our definition \eqref{reg_gen_Eisenstein} that the inhomogeneous Laplacian equation is modified to
\begin{equation}\label{Laplacian_gen_reg_eis}
    [\Delta-\lambda(\lambda-1)]\tilde{\mathcal{E}}(\lambda;s_1,s_2\vert z) = \Eisenstein{s_1}{z}\Eisenstein{s_2}{z}+\sum_I \big[\lambda(\lambda-1)-I(I-1)\big]\beta_I\Eisenstein{I}{z}\,.
\end{equation}

Since both sides of this equation are in $L^2(\mathcal{F})$, we can now take the Petersson inner product against the constant function, the continuous part and the discrete part of the spectrum on both sides of \eqref{Laplacian_gen_reg_eis} to obtain the spectral overlaps previously discussed.
A slight complication arises from the fact that, although both sides of \eqref{Laplacian_gen_reg_eis} are square-integrable, the source term is made of non-square integrable objects, hence a suitable regularisation is required to discuss the Petersson inner product for functions not of rapid decay. 

To this end, we follow \cite{ZagierRS} and introduce a specific regularisation for the divergent integral
\begin{equation}
    \mathcal{I}(s) = \int_0^\infty y^s \dd y = \int_0^1 y^s \dd y+\int_1^\infty y^s \dd y = \mathcal{I}_1(s)+\mathcal{I}_2(s)\,.
\end{equation}
Clearly the starting integral does not converge for any $s\in \mathbb{C}$, but the two parts it splits into do converge on disjoint regions. Namely for $\Re(s)>-1$ the integral $\mathcal{I}_1(s)$ is well-defined and we have $\mathcal{I}_1(s)=\frac{1}{s+1}$, while similarly for $\Re(s)<-1$ the second integral is well-defined and we have $\mathcal{I}_2(s)=-\frac{1}{s+1}$. Since both integrals admit an analytic continuation in $s \in \mathbb{C}\setminus\{-1\}$, we may define $\mathcal{I}(s)=\mathcal{I}_1(s)+\mathcal{I}_2(s) = 0$. 

As a direct application of this formula, we compute the average $\average{{\rm E}_{r}} = ({\rm E}_{r},1)$, i.e.~the spectral overlap of an Eisenstein series with the constant function, as well as the spectral overlap $({\rm E}_{r},{\rm E}_{t})$ for $r\neq t$:
\begin{align}
    \average{{\rm E}_{r}}&=\int_{\mathcal{F}} \Big[\sum_{\gamma\in\poin}\Im(\gamma\cdot z)^r\Big] \dd\mu=  \int_{{\rm B}(\mathbb{Z}) \backslash\mathcal{H}} y^r \frac{\dd x\, \dd y}{y^2}=\int_0^\infty y^{r-2}\dd y=0\,, \label{Eisenstein_constant}\\
   \notag({\rm E}_{r},{\rm E}_{t})   &= \int_{\mathcal{F}}  \normEisenstein{r}{z}\Big[\sum_{\gamma\in\poin}\Im(\gamma\cdot z)^{\bar{t}}\,\Big]\dd\mu =\int_{{\rm B}(\mathbb{Z}) \backslash\mathcal{H}}  \normEisenstein{r}{z} y^{\bar{t}}\, \frac{\dd x\, \dd y}{y^2} \\
&= \label{EE_overlap}  \int_0^\infty  \Big(y^{\bar{t}+r-2}+\frac{\xi(2r-1)\pi^r}{\Gamma(r)\zeta(2r)}y^{\bar{t}-r-1}\Big) \dd y=0
\end{align}
In both calculations we make crucial use of what is usually called the ``\textit{unfolding trick}'', namely we write part of the integrand as a Poincar\'e series and then use this sum over images under ${\rm B}(\mathbb{Z}) \backslash \SLtwoZ$ to unfold the starting domain of integration $\mathcal{F}=\SLtwoZ \backslash \mathcal{H}$ onto the strip
\begin{equation}
{\rm B}(\mathbb{Z}) \backslash\mathcal{H} = \left\lbrace z\in \mathcal{H}: |x| \leq \frac{1}{2}\,,\,y>0\right\rbrace\,,
\end{equation}
after which we can easily integrate over $x$ and subsequently over $y$.
Note that all of the above integrals are ill-defined and need to be regularised in the same way as the original integral $\mathcal{I}(s)$. We will shortly see more interesting examples where the unfolding procedure produces convergent integrals, which can nevertheless be treated via the same type of analytic continuation. 

In particular, we can use the differential equation \eqref{Laplacian_gen_reg_eis} to show the vanishing of the spectral overlap of $ \tilde{\mathcal{E}}(s_1,s_2;\lambda\vert z)$ with the constant function,
\begin{multline}
    \average{\tilde{\mathcal{E}}(\lambda;s_1,s_2)}=\int_{\mathcal{F}} \tilde{\mathcal{E}}(\lambda;s_1,s_2\vert z)\, \dd \mu\\
    =\frac{1}{\lambda(\lambda-1)}\int_\mathcal{F}\left\lbrace\Delta\tilde{\mathcal{E}}(\lambda;s_1,s_2\vert z)-\Eisenstein{s_1}{z}\Eisenstein{s_2}{z}+\sum_I\big[I(I-1)-\lambda(\lambda-1)\big]\beta_I\Eisenstein{I}{z}\right\rbrace \dd \mu=0\,.
\end{multline}
The first term vanishes since it is an integral of a total derivative over a closed surface, while the second and third term vanish due to the previously derived identities \eqref{Eisenstein_constant}-\eqref{EE_overlap}. 

As a result, to derive a useful expression for the Fourier zero-mode integral representation \eqref{spectral_Fzero_mode} of $\tilde{\mathcal{E}}(\lambda;s_1,s_2\vert z)$, we only need considering the spectral overlap with the Eisenstein series $\normEisenstein{t}{z}$ with $\Re(t)=\frac{1}{2}$:
\begin{align}
    &\label{eq:Eeoverlap}(\tilde{\mathcal{E}}(\lambda;s_1,s_2),{\rm E}_t) = \int_{\mathcal{F}} \tilde{\mathcal{E}}(\lambda;s_1,s_2\vert z)\, \normEisenstein{1-t}{z} \dd \mu= \int_{\mathcal{F}}  \tilde{\mathcal{E}}(\lambda;s_1,s_2\vert z) \frac{\Delta\normEisenstein{1-t}{z}}{t(t-1)}\dd \mu \\
 &\notag   = \int_{\mathcal{F}} \Big\lbrace \Eisenstein{s_1}{z}\Eisenstein{s_2}{z} +\lambda(\lambda-1)\tilde{\mathcal{E}}(s_1,s_2;\lambda\vert z) + \sum_I [\lambda(\lambda-1)-I(I-1)]\beta_I \Eisenstein{I}{z}\Big\rbrace \frac{\normEisenstein{1-t}{z}}{t(t-1)}\dd \mu\,,
\end{align}
where in the Petersson inner product we used the fact that $\overline{\normEisenstein{t}{z}} = \normEisenstein{\overline{t}}{z}=\normEisenstein{1-t}{z}$ on the critical line $\Re(t)=\frac{1}{2}$ for which $\overline{t} = 1-t$.

In the first line of \eqref{eq:Eeoverlap} we used the differential equation satisfied by the Eisenstein series \eqref{Eisenstein_eigval}, while in the second line we integrated by parts and then used the inhomogeneous Laplace equation \eqref{Laplacian_gen_reg_eis}. Since we have already shown that the integral over the fundamental domain $\mathcal{F}$ of a product of two Eisenstein series vanishes \eqref{EE_overlap}, the overlap $(\tilde{\mathcal{E}}(\lambda;s_1,s_2),{\rm E}_t)$ can be expressed simply as an integral of a triple product of Eisenstein series.

 Once again this integral can be evaluated  \cite{ZagierRS} via the unfolding trick by rewriting one of the Eisenstein series as a Poincar\'e series and then using the sum over images to unfold the fundamental domain $\mathcal{F}$ onto the strip ${\rm B}(\mathbb{Z})\backslash \mathcal{H}$:
\begin{equation}
\begin{split} \label{eq:EeOverlap}
    (\tilde{\mathcal{E}}(\lambda;s_1,s_2),{\rm E}_t) &= \frac{1}{(t-\lambda)(t+\lambda-1)}\int_\mathcal{F}  \Eisenstein{s_1}{z}\Eisenstein{s_2}{z}\normEisenstein{1-t}{z} \dd \mu\\
    &=\frac{4\xi(t+s_1+s_2-1)\xi(t+s_1-s_2)\xi(t+s_2-s_1)\xi(t+1-s_1-s_2)}{(t-\lambda)(t+\lambda-1)\Gamma(s_1)\Gamma(s_2)\xi(2t-1)}.   
\end{split}
\end{equation}

We can then write the spectral decomposition \eqref{eq:spectral} for the generalised Eisenstein series
\begin{align}
  \notag \genEisenstein{s_1}{s_2}{\lambda}{z}=& \int_{\Re(t)=\frac{1}{2}} \!\!\!\!\! \frac{4\xi(t+s_1+s_2-1)\xi(t+s_1-s_2)\xi(t+s_2-s_1)\xi(t+1-s_1-s_2)}{(t-\lambda)(t+\lambda-1)\Gamma(s_1)\Gamma(s_2)\xi(2t-1)}\,\normEisenstein{t}{z} \frac{ \dd t}{4\pi i} \\
&\label{eq:SpecEe}+\sum_{I}\beta_I \Eisenstein{I}{z}  + \sum_{n=1}^\infty (\tilde{\mathcal{E}}(\lambda;s_1,s_2),\phi_n)\phi_n(z),
\end{align}
where the spectral overlap with the Maass cusp forms can be made more explicit, but it is of little concrete use given the poor analytic control over these objects.

We are now in the position of specialising the integral representation \eqref{spectral_Fzero_mode} to the case of $\mathcal{E}(s_1,s_2;\lambda)$ thus arriving at the useful expression for its Fourier zero-mode
\begin{align}
\label{gen_eisen_zero_mode}
    e_0(\lambda;s_1,s_2&\vert y) = \sum_I \beta_I\Big[\frac{2\zeta(2I)}{\pi^I}y^I+\frac{2\xi(2I-1)}{\Gamma(I)}y^{1-I}\Big]\\
    &\notag+\int_{\Re(t)=\frac{1}{2}} \frac{4\xi(t+s_1+s_2-1)\xi(t+s_1-s_2)\xi(t+s_2-s_1)\xi(t+1-s_1-s_2)}{(t-\lambda)(t+\lambda-1)\Gamma(s_1)\Gamma(s_2)\xi(2t-1)} y^t  \frac{\dd t}{2\pi i }\,,
\end{align}
where again $I\in\{s_1+s_2,s_1+1-s_2,s_2+1-s_1,2-s_1-s_2\}$ and $\beta_I$ was defined in \eqref{reg_gen_Eisenstein}.

The integrand of \eqref{gen_eisen_zero_mode} is a meromorphic function of $t$ for which it is rather easy to understand the structure of singularities. Firstly, we note that the completed Riemann function $\xi(s) = \pi^{-\frac{s}{2}} \Gamma(\frac{s}{2})\zeta(s)$ is meromorphic with simple poles at $s=0$ and $s=1$, while it vanishes only at the non-trivial zeros of the Riemann zeta function, which, from the conjectural Riemann hypothesis, are of the form $s=\frac{1}{2}+i\rho_n$ with $\rho_n$ real. We then deduce that the integrand of \eqref{gen_eisen_zero_mode} has poles located at:
\begin{itemize}
\item $t= 1-I,I$ with $I\in\{s_1+s_2,s_1+1-s_2,s_2+1-s_1,2-s_1-s_2\}$, for which one the completed Riemann zeta functions in the numerator has argument equal to $0$ or $1$ respectively;
\item $t=\lambda, 1-\lambda$, coming from the two rational terms $[(t-\lambda)(t+\lambda-1)]^{-1}$;
\item $t=\frac{3}{4}+i\frac{\rho_n}{2}$, coming from the non-trivial zeroes of $\xi(2t-1)$ present in the denominator. 
\end{itemize}

We can now use \eqref{gen_eisen_zero_mode} to distinguish between the different contributions arising in the asymptotic expansions of the Fourier zero-mode $e_0(\lambda;s_1,s_2\vert y)$ as $y\gg 1$ or as $y\to0$.
Focusing for the present time on the asymptotic expansion at the cusp $y\gg 1$, we see that the integral contour in \eqref{gen_eisen_zero_mode} can be closed in the left half-plane $\Re(t)<0$. In doing so, we pick up the residues for the poles located at $\Re(t)<\frac{1}{2}$, which are: 
\vspace{0.1cm}

(i) $t=I$ for $I\in\{s_1+s_2,s_1+1-s_2,s_2+1-s_1,2-s_1-s_2\}$ with $\Re(I)<\frac{1}{2}$;
\vspace{0.1cm}

(ii) $t=1-I$ for $I\in\{s_1+s_2,s_1+1-s_2,s_2+1-s_1,2-s_1-s_2\}$ with $\Re(I)>\frac{1}{2}$;
\vspace{0.1cm}

(iii) $t = 1-\lambda$ under the original assumption $\Re(\lambda)>\frac{1}{2}$.
\vspace{0.1cm}

The end result can be made more concrete by considering the case relevant for our spectra \eqref{eq:spec1}-\eqref{eq:spec2}, where $s_1,s_2\geq\frac{3}{2}$ and without loss of generality $s_1\geq s_2$. Under these conditions and considering the non-diagonal case where $s_1-s_2\geq 1$, we simply collect the residues from the poles at $t\in \{s_2+1-s_1,s_2-s_1,2-s_1-s_2,1-s_1-s_2\}$ and $t=1-\lambda$. 

Note that, for this range of parameters, the square-integrable function $\tilde{\mathcal{E}}(\lambda;s_1,s_1)$ in \eqref{reg_gen_Eisenstein} is obtained by removing suitable multiples of the Eisenstein series $\Eisenstein{I}{z}$ with $I\in \{s_1+s_2,s_1+1-s_2\}$. From the Fourier zero-mode \eqref{FZero_genEisen_asymp}, we see that this subtraction indeed removes the non-square integrable powers $y^{s_1+s_2}$ and $y^{s_1+1-s_2}$. However, since at the cusp $\Eisenstein{I}{z}\sim \# y^I +\# y^{1-I}$, we also introduce ``unwanted'' reflected powers $y^{1-s_1-s_2}$ and $y^{1-(s_1+1-s_2)}=y^{s_2-s_1}$. These unwanted terms are exactly cancelled by the residues coming from the above-mentioned poles located at $t\in \{s_2-s_1,1-s_1-s_2\}$. The remaining poles at $t\in \{s_2+1-s_1,2-s_1-s_2\}$ produce the remaining powers for the particular solution \eqref{FZero_genEisen_asymp}, while the pole at $t=1-\lambda$ produces the homogeneous solution term.

The diagonal case, $s_1=s_2$, requires some extra care since to define the square-integrable function $\tilde{\mathcal{E}}(\lambda;s_1,s_1)$ in \eqref{reg_gen_Eisenstein}  we need to subtract a regularised version for the divergent Eisenstein series $\Eisenstein{1}{z}$, see e.g.~appendix B of \cite{Collier:2022emf}. At the same time, we see that the spectral overlap \eqref{eq:EeOverlap} develops a double pole at $t=0$ and $t=1$ precisely for $s_1=s_2$. To avoid these complications, we can obtain the diagonal case as the off-diagonal limit $s_2=s_1-\epsilon$ with $\epsilon\to0$.

We can directly use \eqref{gen_eisen_zero_mode} to determine the previously unknown coefficient $\alpha(\lambda;s_1,s_2)$ multiplying the homogeneous solution $y^{1-\lambda}$.
This coefficient was first computed in \cite{Green:2008bf} with a similar method, and can now be calculated by simply picking up the pole of \eqref{gen_eisen_zero_mode} at $t=1-\lambda$, giving us
\begin{equation}\label{alpha_def}
    \alpha(\lambda;s_1,s_2)=-\frac{4\xi(s_1+s_2-\lambda)\xi(s_1-s_2+\lambda)\xi(s_2-s_1+\lambda)\xi(s_1+s_2+\lambda-1)}{(2\lambda-1)\Gamma(s_1)\Gamma(s_2)\xi(2\lambda)}\,.
\end{equation}

In the next section, we discuss the asymptotic expansion of \eqref{gen_eisen_zero_mode} as $y\to0$ where the contour of integration has to be closed instead in the right half-plane $\Re(t)>0$. This will select the ``complementary'' poles to the ones just discussed, and a new infinite family of poles coming from the non-trivial zeros of the Riemann zeta function will also play an essential r\^ole.
 
We conclude this section by analysing the spectral decomposition for the novel functions $\modfn{a}{b}{r}{s}{z}$.
Firstly, from the previously determined asymptotic expansion at the cusp \eqref{full_asymptotics}, we see that all $\modfn{a}{b}{r}{s}{z}$ are directly square integrable functions in the region of parameters $a,b,r,s$ where the Poincar\'e series converges \eqref{eq:AbsConv}, i.e. we have immediately $\modfntwo{a}{b}{r}{s} \in L^2(\mathcal{F})$ when  \eqref{eq:AbsConv} is satisfied. 

As a consequence, we can directly compute the spectral overlaps without any need for subtracting Eisenstein series. 
We start by observing that the spectral overlap with the constant function vanishes
$$\average{\modfntwo{a}{b}{r}{s}}= \int_\mathcal{F} \modfn{a}{b}{r}{s}{z} \dd \mu= 0\,,$$
 since we can use the Poincar\'e series representation \eqref{f_definition} for $\modfn{a}{b}{r}{s}{z}$ to unfold the integral from the fundamental domain $\mathcal{F}$ to the strip ${\rm B}(\mathbb{Z})\backslash \mathcal{H}$, and we conclude that the integral over $x$ vanishes since the seed function $\seedfn{a}{b}{r}{s}{z}$ does not have a Fourier zero-mode. 
 
 We proceed by computing the spectral overlap with the Eisenstein series. A calculation very similar to \eqref{eq:EeOverlap} yields
\begin{align}
 &\notag   (\modfntwo{a}{b}{r}{s},{\rm E}_t) = \int_\mathcal{F} \modfn{a}{b}{r}{s}{z}\normEisenstein{1-t}{z} \dd\mu 
   = \int_{{\rm B}(\mathbb{Z}) \backslash \mathcal{H}} \seedfn{a}{b}{r}{s}{ z}\normEisenstein{1-t}{z} \frac{\dd x\,\dd y}{y^2}\\ 
   &\label{Eisenstein_overlap}= \frac{\Gamma\big(\frac{r+1-s-t}{2}\big)\Gamma\big(\frac{r+s-t}{2}\big)\Gamma\big(\frac{t+r-s}{2}\big)\Gamma\big(\frac{t+r+s-1}{2}\big)}{2\pi^r\, \Gamma(r)\xi(2-2t)}\\ 
   &\notag\quad 
   \times
   \frac{\zeta(r+1-b-t)\zeta(r+1-a-b-t)
    \zeta(t+r-b)\zeta(t+r-a-b)}{\zeta(2r+1-a-2b)}\,.
\end{align}
The keen-eyed reader will notice that if we now plug the spectral overlap just derived into the integral representation formula for the Fourier zero-mode \eqref{spectral_Fzero_mode}, we obtain exactly the same expression \eqref{modfn_Fourierzero_mode} previously derived from the Poincar\'e series representation. This is a significantly simpler derivation of \eqref{modfn_Fourierzero_mode} when compared to the Mellin-Barnes discussion presented in appendix \ref{app:AsyCusp}. However, we need to stress that without having already obtained the result \eqref{full_asymptotics}, we could have not inferred immediately that the functions $\modfn{a}{b}{r}{s}{z}$ are in $L^2(\mathcal{F})$.

\subsection{Non-perturbative terms and small-$y$ behaviour}

So far our analysis of the Fourier zero-mode \eqref{gen_eisen_zero_mode} only concerned with the power-behaved terms at the cusp $y\gg 1$.
In this section we show how the exponentially suppressed corrections $e^{-4\pi y}$ are encoded in \eqref{gen_eisen_zero_mode} and clarify how the resurgent analysis carried out in \cite{Dorigoni:2022bcx} nicely connects with the present discussion. 
In the limit $y\to 0$, the non-perturbative terms stop being exponentially suppressed and produce instead an infinite sum of perturbative corrections related to the non-trivial zeros of the Riemann zeta function.

As discussed in the previous section, we can easily evaluate the perturbative expansion for the Fourier zero-mode integral representation \eqref{gen_eisen_zero_mode} as $y\gg 1$ by closing the contour of integration in the left half-plane $\Re(t)<0$. Picking up various residues allows us to reproduce all power-behaved terms present in \eqref{FZero_genEisen_asymp}, however, the integral does not vanish when we push the contour of integration to infinity, but rather it produces the remaining exponentially suppressed corrections in the Fourier zero-mode sector.
   
We follow this procedure and push the contour of integration to the left half-plane $\Re(t)<0$, while collecting the residues to arrive at
\allowdisplaybreaks{
\begin{align}
\label{gen_eisen_zero_mode_asymp}
   & e_0(\lambda;s_1,s_2\vert y) =  \frac{4\pi^{-s_1-s_2}\zeta(2s_1)\zeta(2s_2)}{(s_1+s_2-\lambda)(s_1+s_2+\lambda-1)}y^{s_1+s_2}\\*
  &\notag+\frac{4\pi^{-s_1}\xi(2s_2-1)\zeta(2s_1)}{(s_1+1-s_2-\lambda)(s_1-s_2+\lambda)\Gamma(s_2)}y^{s_1+1-s_2} +\frac{4\pi^{-s_2}\xi(2s_1-1)\zeta(2s_2)}{(s_2+1-s_1-\lambda)(s_2-s_1+\lambda)\Gamma(s_1)}y^{s_2+1-s_1}\\*
&\notag+\frac{4\xi(2s_1-1)\xi(2s_2-1)}{(s_1+s_2-\lambda-1)(s_1+s_2+\lambda-2)\Gamma(s_1)\Gamma(s_2)}y^{2-s_1-s_2} + \alpha(\lambda;s_1,s_2)y^{1-\lambda} \\
    &\notag+\int_{\Re(t)=\tilde{\gamma}} \frac{4\xi(t+s_1+s_2-1)\xi(t+s_1-s_2)\xi(t+s_2-s_1)\xi(t+1-s_1-s_2)}{(t-\lambda)(t+\lambda-1)\Gamma(s_1)\Gamma(s_2)\xi(2t-1)} y^t  \frac{\dd t}{2\pi i }\,,
\end{align}}
where $\tilde{\gamma}<\min\{\Re(I),\Re(1-I),\Re(\lambda),\Re(1-\lambda)\}$. The integrand in \eqref{gen_eisen_zero_mode_asymp} is manifestly analytic for $t$ in the half-plane $\Re(t)\leq \tilde{\gamma}$ and we claim that the corresponding integral is exponentially suppressed at the cusp $y\gg 1$ thus containing all of the non-perturbative, $e^{-4 \pi y}$, terms.

For aesthetic reasons we perform the change of variables $t\to 1-t$ and use the reflection identity $\xi(s)=\xi(1-s)$ to rewrite the above integral as
\begin{equation}\label{eq:NP}
  \!\!  {\rm NP}^{(\lambda)}_{s_1,s_2}(y)\!:=\! \!\int_{\Re(t)=\gamma} \!\!\!\!\! \frac{4\xi(t+s_1+s_2-1)\xi(t+s_1-s_2)\xi(t+s_2-s_1)\xi(t+1-s_1-s_2)}{(t-\lambda)(t+\lambda-1)\Gamma(s_1)\Gamma(s_2)\xi(2t)} y^{1-t}\frac{\dd t}{2\pi i}\,,
\end{equation}
where $\gamma>\max\{ \Re(I),\Re(1-I),\Re(\lambda),\Re(1-\lambda)\}$ is arbitrary as long as it lies to the right of all the poles location. We can expand all the completed Riemann functions as $\xi(s) = \pi^{-\frac{s}{2}} \Gamma(\frac{s}{2}) \zeta(s)$ and use Ramanujan identity \eqref{Ramanujan_sigma} in reverse to rewrite the particular combination of Riemann zeta functions appearing in \eqref{eq:NP} as a Dirichlet series for the product of two divisor functions, arriving at
\begin{equation}\label{NP_con_expr}
   \begin{split}
   {\rm NP}^{(\lambda)}_{s_1,s_2}(y)&= \sum_{n=1}^\infty \frac{4\sigma_{1-2s_1}(n)\sigma_{1-2s_2}(n)n^{s_1+s_2-1}y}{\Gamma(s_1)\Gamma(s_2)}\\
   &\phantom{=}\times \int_{\Re(t)=\gamma} \frac{\Gamma(\frac{t+s_1+s_2-1}{2})\Gamma(\frac{t+s_1-s_2}{2})\Gamma(\frac{t+s_2-s_1}{2})\Gamma(\frac{t+1-s_1-s_2}{2})}{(t-\lambda)(t+\lambda-1)\Gamma(t)} (\pi ny)^{-t}\frac{\dd t}{2\pi i }\,.
   \end{split}
\end{equation}

We have not managed to evaluate \eqref{NP_con_expr} in closed form, however, its asymptotic expansion as $y\gg 1$ can be obtained via saddle point approximation. 
Firstly we can use the Stirling approximation for the gamma functions to confirm that the integrand has a stationary point at $t=4\pi ny$.
Hence a simple steepest descent calculation produces the required asymptotic expansion,
\begin{align} \label{eq:NPphi}
    {\rm NP}^{(\lambda)}_{s_1,s_2}(y) &=  \sum_{n=1}^\infty \frac{ \sigma_{1-2s_1}(n)\sigma_{1-2s_2}(n)n^{s_1+s_2-2}}{\Gamma \left(s_1\right) \Gamma \left(s_2\right) } e^{-4\pi n y} \phi^{(\lambda)}_{s_1,s_2}(4\pi n y) \,,
\end{align}
where the first few perturbative corrections are given by
\allowdisplaybreaks{
\begin{align}\label{eq:NPpert}
   & \phi^{(\lambda)}_{s_1,s_2}(y)= \\*
   &\notag \frac{8}{y^2}+ \frac{8[s_1(s_1-1){+}s_2(s_2-1){-}4]}{y^3}+ 4 \frac{\big\{ [s_1(s_1-1){+}s_2(s_2-1){-}7]^2{+}2\lambda(\lambda-1){-}13\big\}}{y^4}+O(y^{-5})\,.
\end{align}}

A few comments are in order.
Firstly, although for general values of the parameters $s_1,s_2$ and $\lambda$ the function $ \phi^{(\lambda)}_{s_1,s_2}(y)$ contains infinitely many perturbative terms, we find that for some special cases, this series has only a finite number of terms. For example, if we fix $s_1=3,s_2=2$ and $\lambda=2$, corresponding to the modular invariant function \eqref{eq:ex1} belonging to the spectrum \eqref{eq:spec2}, the non-perturbative sector \eqref{NP_con_expr} simplifies to
\begin{align}
    {\rm NP}^{(3)}_{3,2}(y) \notag&= \sum_{n=1}^\infty  16\pi y\,\sigma_{-3}(n)\sigma_{-5}(n)n^{4}\int_{\Re(t)=5}t\,\Gamma(t-4)\,(4\pi ny)^{-t}\frac{\dd t}{2\pi i}\\
    &=\sum_{n=1}^\infty \frac{\sigma_{-3}(n)\sigma_{-5}(n)n^{3}}{2}e^{-4\pi ny}\Big[ \frac{8}{(4\pi ny)^2}+\frac{32}{(4\pi ny)^3}\Big]\,.
\end{align}

Although we have not proven that for generic values of $s_1,s_2$ and $\lambda$ the perturbative series $\phi^{(\lambda)}_{s_1,s_2}(y)$ contains infinitely many terms, it is easy to see that for the case associated with the spectrum \eqref{eq:spec2} (corresponding to depth-two modular graph functions) the series $\phi^{(\lambda)}_{s_1,s_2}(y)$ is always a polynomial. This is expected from the Laplace equation \eqref{gen_Eisenstein_def} given that for the spectrum \eqref{eq:spec2} the Eisenstein series appearing in the source term have integer index, hence the corresponding Bessel functions, which appear in the Fourier decomposition \eqref{Eisenstein_def} and which are responsible for the non-perturbative terms, have half-integer index thus producing only finitely many perturbative terms in the non-perturbative sector.

A second comment we want to stress is that our expression \eqref{NP_con_expr} can be shown to be the exact solution to the Laplace equation \eqref{gen_Eisenstein_def} for the non-perturbative part of the Fourier zero-mode sector. Given the Laplace equation \eqref{gen_Eisenstein_def} and the Fourier decomposition \eqref{Eisenstein_def} we must have
\begin{equation}
[y^2 \partial_y^2 -\lambda(\lambda-1)]    {\rm NP}^{(\lambda)}_{s_1,s_2}(y)    =\sum_{n=1}^\infty  \frac{32  n^{s_1+s_2-1} \sigma_{1-2s_1}(n)\sigma_{1-2s_2}(n) }{\Gamma(s_1)\Gamma(s_2)} y\, K_{s_1-\frac{1}{2}}(2\pi n y)K_{s_2-\frac{1}{2}}(2\pi n y)\,.
\end{equation}
If we rewrite the source term using the Mellin-Barnes type integral representation for the product of two Bessel function
\begin{equation}
 y \,K_{s_1-\frac{1}{2}}( 2y)K_{s_2-\frac{1}{2}}( 2y) =\int_{\Re(t) = \gamma} \!\!\! \frac{\Gamma(\frac{t+s_1+s_2-1}{2})\Gamma(\frac{t+s_1-s_2}{2})\Gamma(\frac{t+s_2-s_1}{2})\Gamma(\frac{t+1-s_1-s_2}{2})}{\Gamma(t)}  y^{1-t} \frac{\dd t  }{16 \pi i}\,,
\end{equation}
where $\gamma>\max\{\Re(I),\Re(1-I)\}$, and then simply solve the differential equation for $ {\rm NP}^{(\lambda)}_{s_1,s_2}(y)$ by inverting the differential operator as 
$$\frac{1}{[y^2 \partial_y^2 -\lambda(\lambda-1)]} y^{1-t} = \frac{y^{1-t}}{(t-\lambda)(t+\lambda-1)}\,,$$
we find the exact integral representation \eqref{NP_con_expr}.

Furthermore, we note that the formula for the perturbative series expansion \eqref{eq:NPpert} in the non-perturbative sector reproduces exactly the results obtained in \cite{Dorigoni:2022bcx} for the modular invariant functions associated with the spectrum \eqref{eq:spec2}.
We stress that in \cite{Dorigoni:2022bcx}, the authors started from the seed functions \eqref{eq:DDAKseed} and obtained the non-perturbative sector for the generalised Eisenstein series with spectrum \eqref{eq:spec2} from a careful resummation of an evanescent, yet factorially divergent formal perturbative expansion in an example of Cheshire cat resurgence, very similar to our discussion below \eqref{eq:c2pert}.
We now see that the results obtained in \cite{Dorigoni:2022bcx} are actually more general than what originally stated in said reference, and in particular \eqref{eq:NPpert} appears to be valid for all values of $s_1,s_2$ and $\lambda$ and not just for the spectrum \eqref{eq:spec2}. 

Finally, as discussed in \cite{Dorigoni:2022bcx}, it is easy to see that while for $y\gg1$ the Fourier zero-mode contribution \eqref{eq:NP} is non-perturbative and exponentially suppressed, its nature changes dramatically when $y\to 0$. 
Rather than splitting the complete Fourier zero-mode $e_0(s_1,s_2;\lambda\vert y)$ in perturbative plus non-perturbative terms as in \eqref{gen_eisen_zero_mode}, we can analyse directly the integral representation \eqref{gen_eisen_zero_mode} in the limit $y\to 0$.

As previously anticipated just below \eqref{gen_eisen_zero_mode}, in the limit $y\to 0$ we can close the contour of integration to the right half-plane $\Re(t)>0$ and collect the residues from the various ``complementary'' poles plus the infinite set of completely novel poles located at $t=\frac{3}{4}+i\frac{\rho_n}{2}$ and coming from the non-trivial zeroes of $\xi(2t-1)$ present in the denominator of the integrand in \eqref{gen_eisen_zero_mode}. Once again, after we have pushed the contour of integration past all the poles, the remaining integral captures all exponentially suppressed contributions\footnote{We are extremely grateful to Nathan Benjamin and Cyuan-Han Chang for pointing out that such non-perturbative corrections have to be present and for bringing~\cite{Benjamin:2022pnx} to our attention where very similar results were obtained in the spectral decomposition for the partition function of certain $2$-d conformal field theories.} now of the form $e^{-\frac{4\pi }{y}}$. The asymptotic expansion of \eqref{gen_eisen_zero_mode} as $y\to 0$ is simply given by the sum over the residues of all the poles located at $\Re(t)>\frac{1}{2}$ plus a remaining contour integral,
\allowdisplaybreaks\begin{align}
    e_0(\lambda;s_1,s_2\vert y) &\notag =  \frac{4\xi(2s_1-1)\xi(2s_2-1)\xi(2s_1+2s_2-2)}{(s_1+s_2-\lambda-1)(s_1+s_2+\lambda-2)\Gamma(s_1)\Gamma(s_2)\xi(2s_1+2s_2-3)}y^{s_1+s_2-1}\\*
    &\notag+\frac{4\xi(1-2s_1)\xi(2s_2-1)\xi(2s_2-2s_1)}{(s_1+1-s_2-\lambda)(s_1-s_2+\lambda)\Gamma(s_1)\Gamma(s_2)\xi(2s_2-2s_1-1)}y^{s_2-s_1}\\*
    &\label{FZero_genEisen_zero}+\frac{4\xi(1-2s_2)\xi(2s_1-1)\xi(2s_1-2s_2)}{(s_2+1-s_1-\lambda)(s_2-s_1+\lambda)\Gamma(s_1)\Gamma(s_2)\xi(2s_1-2s_2-1)}y^{s_1-s_2}\\*
    &\notag+\frac{4\xi(1-2s_1)\xi(1-2s_2)\xi(2-2s_1-2s_2)}{(s_1+s_2-\lambda)(s_1+s_2+\lambda-1)\Gamma(s_1)\Gamma(s_2)\xi(1-2s_1-2s_2)}y^{1-s_1-s_2} \\*
    &\notag-\alpha(1-\lambda;s_1,s_2)y^{\lambda} \\*
    &\notag+\sum_{\rho_n} \frac{2\xi(t+s_1+s_2-1)\xi(t+s_1-s_2)\xi(t+s_2-s_1)\xi(t+1-s_1-s_2)}{(1-t-\lambda)(t-\lambda)\Gamma(s_1)\Gamma(s_2)\pi^{\frac{1}{2}-t}\Gamma(t-\frac{1}{2})\zeta'(2t-1)} y^t\bigg\vert_{t=\frac{3}{4}+i\frac{\rho_n}{2}}\\*
    &\notag+ \widetilde{{\rm NP}}^{(\lambda)}_{s_1,s_2}(y)\,.     
\end{align}
where the coefficient $\alpha(\lambda;s_1,s_2)$ is given in \eqref{alpha_def}.

Similar to our large-$y$ discussion, at small-$y$ the non-perturbative terms, $\widetilde{{\rm NP}}^{(\lambda)}_{s_1,s_2}(y)$, come from having pushed the contour of integration past all the poles on the right $t$-half-plane,
\begin{equation}\label{eq:NPtilde}
  \!\!  \widetilde{{\rm NP}}^{(\lambda)}_{s_1,s_2}(y)\!:=\! \!\int_{\Re(t)=\gamma} \frac{4\xi(t+s_1+s_2-1)\xi(t+s_1-s_2)\xi(t+s_2-s_1)\xi(t+1-s_1-s_2)}{(t-\lambda)(t+\lambda-1)\Gamma(s_1)\Gamma(s_2)\xi(2t-1)} y^t  \frac{\dd t}{2\pi i }\,,
\end{equation}
where $\gamma>\max\{\Re(I),\Re(1-I),\Re(\lambda),\Re(1-\lambda)\}$. We proceed as before and expand all the completed Riemann functions as $\xi(s) = \pi^{-\frac{s}{2}} \Gamma(\frac{s}{2}) \zeta(s)$, however, we notice that this time the ratio of Riemann zeta functions we obtain,
$$
\frac{\zeta(t+s_1+s_2-1)\zeta(t+s_1-s_2)\zeta(t+s_2-s_1)\zeta(t+1-s_1-s_2) }{\zeta(2t-1)}\,,
$$
cannot be written immediately as a Dirichlet series using Ramanujan identity \eqref{Ramanujan_sigma}.
However, the present discussion is very similar to the spectral decomposition analysis considered in \cite{Benjamin:2022pnx} for the study of certain partition functions in $2$d CFTs. Building on \cite{Benjamin:2022pnx}, we can combine  \eqref{Ramanujan_sigma} with
\begin{equation}
\frac{\zeta(2t)}{\zeta(2t-1)} = \sum_{n=1}^\infty \frac{\varphi^{\shortminus1}(n)}{n^{2t}}\,,
\end{equation}
where $\varphi^{\shortminus1}(n)$ denotes the Dirichlet inverse\footnote{The Dirichlet inverse, $f^{\shortminus 1}$, of an arithmetic function, $f$, is defined such that the Dirichlet convolution of $f$ with its inverse produces the multiplicative identity,~i.e. $\sum_{d|n} f(d) f^{\shortminus1}(n/d) = \delta_{n,1}$. The Dirichlet series $L(f;s) := \sum_{n=1}^\infty \frac{f(n)}{n^s}$ has the property that $L(f^{\shortminus 1};s) = (L(f;s))^{-1}$. The Dirichlet inverse, $\varphi^{\shortminus1}$, of Euler totient function, $\varphi$, is given by $\varphi^{\shortminus 1}(n) = \sum_{d|n} d\, \mu(d)$ where $\mu$ is M\"obius function.} of Euler totient function, $\varphi(n)$, so that \eqref{eq:NPtilde} can be rewritten in terms of a double Dirichlet series and an easier contour integral,
\begin{align}
 \widetilde{{\rm NP}}^{(\lambda)}_{s_1,s_2}(y)= &\label{NPtilde_con_expr} \sum_{m=1}^\infty \sum_{n=1}^\infty \frac{4\sigma_{1-2s_1}(m)\sigma_{1-2s_2}(m) m^{s_1+s_2-1} \varphi^{\shortminus1}(n)}{\sqrt{\pi}\Gamma(s_1)\Gamma(s_2)}\\
  &\notag \times \int_{\Re(t)=\gamma} \frac{\Gamma(\frac{t+s_1+s_2-1}{2})\Gamma(\frac{t+s_1-s_2}{2})\Gamma(\frac{t+s_2-s_1}{2})\Gamma(\frac{t+1-s_1-s_2}{2})}{(t-\lambda)(t+\lambda-1)\Gamma(t -\frac{1}{2})} \Big(\frac{y}{\pi m n^2}\Big)^{t}\frac{\dd t}{2\pi i }\,.
\end{align}

We can now evaluate the asymptotic expansion as $y\to0$ of \eqref{NPtilde_con_expr} via saddle point approximation. 
The integrand has a stationary point at $t= \frac{4\pi m n^2}{y}$ and a simple steepest descent calculation yields the required asymptotic expansion,
\begin{equation}
\widetilde{{\rm NP}}^{(\lambda)}_{s_1,s_2}(y)= \sum_{m=1}^\infty \sum_{n=1}^\infty \frac{\sigma_{1-2s_1}(m)\sigma_{1-2s_2}(m) m^{s_1+s_2-\threeh} \varphi^{\shortminus1}(n)}{\Gamma(s_1)\Gamma(s_2)\,n} \sqrt{4 y}\, e^{-\frac{4 \pi m n^2}{y}} \widetilde{\phi}^{(\lambda)}_{s_1,s_2}\big(\frac{y}{4\pi m n^2}\big) \,,\label{eq:NPtildephi} 
\end{equation}
where the first few perturbative corrections are given by
\allowdisplaybreaks{
\begin{align}\label{eq:NPtildepert}
   & \widetilde{\phi}^{(\lambda)}_{s_1,s_2}(y)= \\*
   &\notag 8 y^2+ 8\Big[s_1(s_1-1){+}s_2(s_2-1){-}\frac{11}{4}\Big]y^3+ 4 \Big\{ \Big[s_1(s_1-1){+}s_2(s_2-1){-}\frac{42}{8}\Big]^2{+}2\lambda(\lambda-1){-}\frac{31}{4}\Big\} y^4+O(y^{5})\,.
\end{align}}
We note the striking similarity between the small-$y$ exponentially suppressed terms \eqref{eq:NPtildephi}-\eqref{eq:NPtildepert} and the parallel large-$y$ expressions \eqref{eq:NPphi}-\eqref{eq:NPpert}. Equation \eqref{eq:NPtildepert} is directly analogous to the crossing equation (3.22) derived in \cite{Benjamin:2022pnx}.

 Going back to the perturbative terms in \eqref{FZero_genEisen_zero}, we see that the infinite series over $\rho_n$ comes precisely from having collected the residues from the poles\footnote{The contribution from these poles was inadvertently missed in \cite{Dorigoni:2022bcx}.} of $1/\xi(2t-1)$ in \eqref{gen_eisen_zero_mode}.
 Under the assumption that Riemann hypothesis is correct, these poles are associated with all non-trivial zeros of the Riemann zeta function $\zeta(s)$ located at $s{=} \frac{1}{2}+i\rho_n$ and $\rho_n\in \mathbb{R}$. Hence in the small-$y$ limit, the last line in equation \eqref{FZero_genEisen_zero} behaves as the power $y^{\frac{3}{4}}$ modulated by oscillatory terms in $y$ with frequencies determined by the $\rho_n$. A similar behaviour was already observed in \cite{ZagierRS} for the modular-invariant function $f(z) = y^{12} |\Delta(z)|^2$ with $\Delta(z)$ Ramanujan discriminant cusp form.
 Similarly, in a different string theory context \cite{Angelantonj:2010ic} and from a two-dimensional CFT context \cite{Benjamin:2022pnx}, the non-trivial zeros of the Riemann zeta function do appear from the asymptotic expansion for the spectral decomposition of different physical quantities.

 \begin{figure}
\centering
\begin{minipage}{.5\textwidth}
  \centering
  \includegraphics[width=0.9\linewidth]{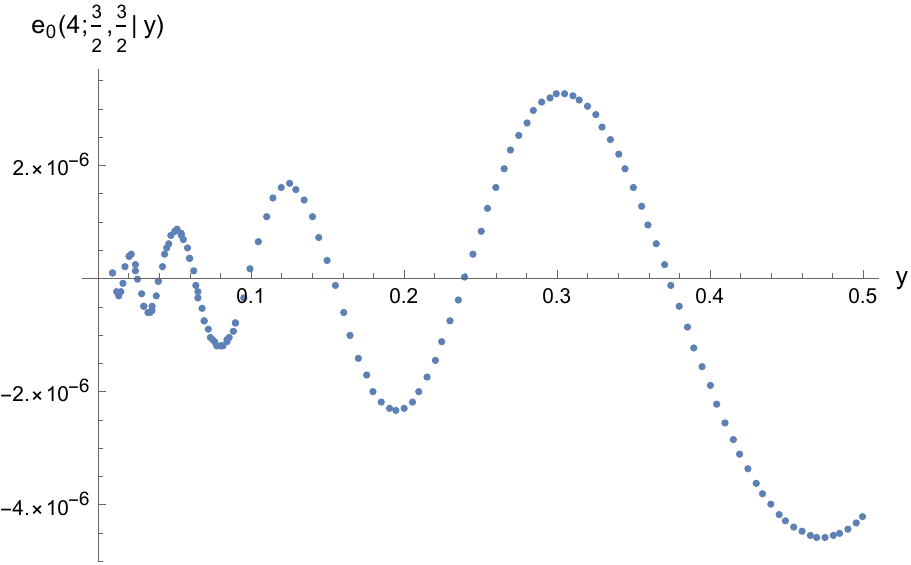}
\end{minipage}%
\begin{minipage}{.5\textwidth}
  \centering
  \includegraphics[width=0.9\linewidth]{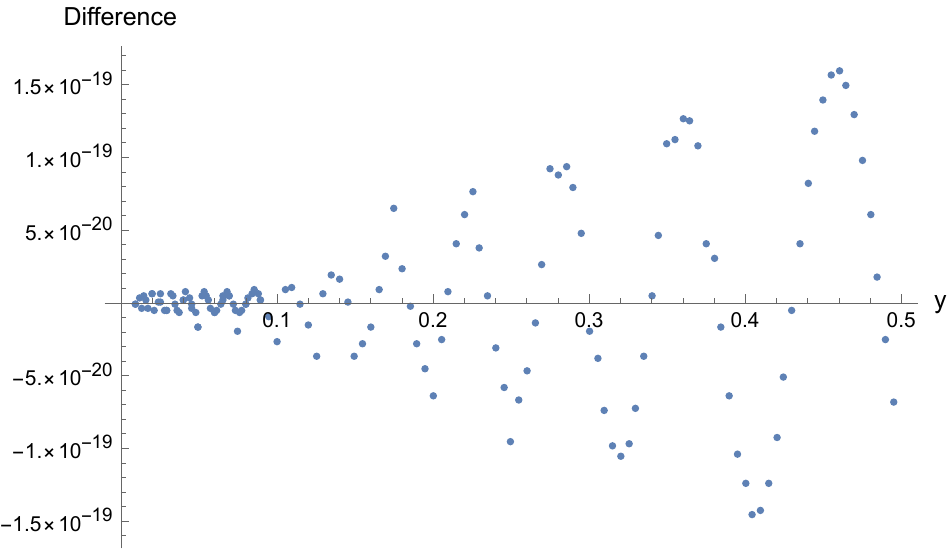}
\end{minipage}
\caption{Comparison between the numerical evaluation of \eqref{gen_eisen_zero_mode} and the small-$y$ expansion \eqref{FZero_genEisen_zero}. On the left, we plot $e_0(4;\frac{3}{2},\frac{3}{2}\vert y)$ after having subtracted all the terms in \eqref{FZero_genEisen_zero} but the Riemann zeta contributions. On the right, we plot the difference between the left data set and the predicted series of contributions in \eqref{FZero_genEisen_zero} from the first $10$ non-trivial zeros of the Riemann zeta.}
    \label{fig:smally}
\end{figure}

\begin{figure}
    \centering
    \includegraphics[scale=0.42]{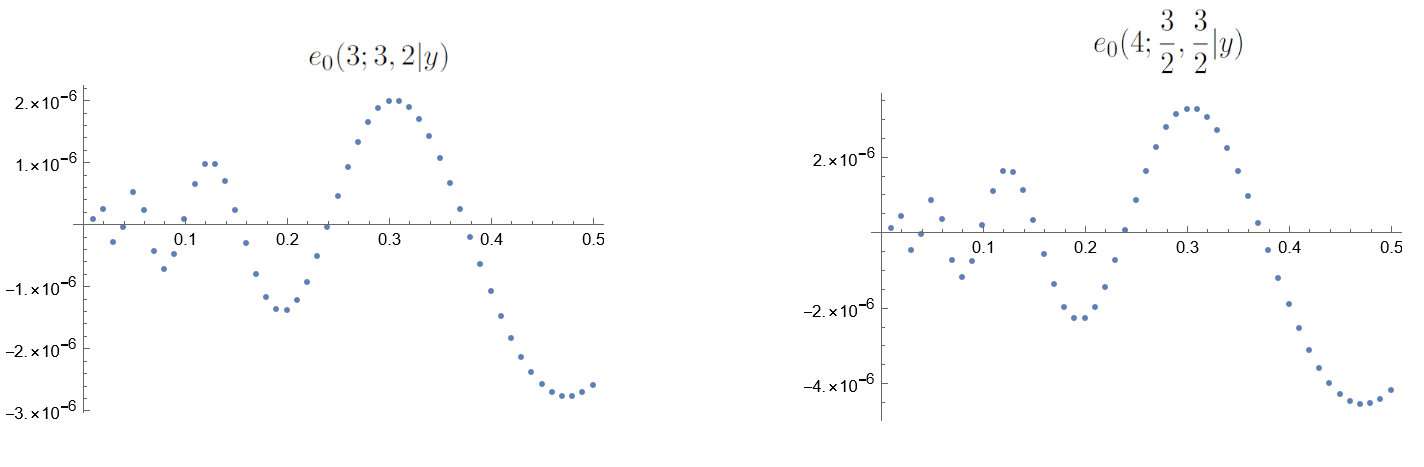}
    \caption{Comparison between the numerical evaluation of \eqref{gen_eisen_zero_mode} and the small-$y$ expansion \eqref{FZero_genEisen_zero}. On the left, we plot $e_0(4;\frac{3}{2},\frac{3}{2}\vert y)$ after having subtracted all the terms in \eqref{FZero_genEisen_zero} but the Riemann zeta contributions. On the right, we plot the difference between the left data set and the predicted series of contributions from the first $50$ non-trivial zeros \eqref{FZero_genEisen_zero} of the Riemann zeta.}
    \label{fig:smally}
\end{figure}
 
In figure~\ref{fig:smally}, we present numerical evidences for the small-$y$ expansion~\eqref{gen_eisen_zero_mode} of the $d^6 R^4$ case, $e_0(4;\frac{3}{2},\frac{3}{2}\vert y)$. We have numerically evaluated to high precision the integral representation \eqref{gen_eisen_zero_mode} for $e_0(4;\frac{3}{2},\frac{3}{2}\vert y)$ at small $y$ and subtracted from it all the terms in \eqref{FZero_genEisen_zero} but the Riemann zeta contributions. In figure \ref{fig:smally} we first plot this quantity and then  subtract from it the predicted series of contributions from the first $10$ non-trivial zeros \eqref{FZero_genEisen_zero} of the Riemann zeta and plot this difference. As the second plot shows, our formula \eqref{FZero_genEisen_zero} is consistent with the numerical data within a $10^{-19}$ error over the whole range of $y$ considered.

Although from a physical point of view, the limit $y\to0$ for the MGFs spectrum \eqref{eq:spec2} corresponds simply to a particular degeneration limit of the worldsheet torus, for the generalised Eisenstein series associated with the spectrum \eqref{eq:spec1}, and in particular for the coefficient of the $d^6 R^4$ in the low-energy expansion of type IIB superstring theory, this limit corresponds to the strong coupling regime $g_s\to \infty$. It would be extremely interesting, and equally difficult, to understand the  string theory origins at strong coupling for the appearance of the non-trivial zeroes of the Riemann zeta.

\subsection{The instanton sectors}
\label{sec:Instanton}

So far we have focused our attention entirely on the Fourier zero-mode sector, while the spectral decomposition \eqref{eq:spectral} in principle allows us  to reconstruct all of the Fourier modes, in particular the Fourier non-zero modes which we will refer to as \textit{instanton sectors}.

Given \eqref{eq:SpecEe}, we can extract the $k$-instanton sector $ e_k(\lambda;s_1,s_2\vert y)$, i.e.~the Fourier mode $e^{2\pi i k x}$, for the generalised Eisenstein series $\genEisenstein{s_1}{s_2}{\lambda}{z}$:
\allowdisplaybreaks\begin{align}
  &\label{eq:SpecEeInst} e_k(\lambda;s_1,s_2\vert y)=  \sum_{I}\beta_I \frac{4}{\Gamma(I)}|k|^{I-\frac{1}{2}}\sigma_{1-2I}(k)y^{\frac{1}{2}}K_{I-\frac{1}{2}}(2\pi |k|y) \\*
&\notag+ \int_{\Re(t)=\frac{1}{2}} \!\Big[ \frac{4\xi(t+s_1+s_2-1)\xi(t+s_1-s_2)\xi(t+s_2-s_1)\xi(t+1-s_1-s_2)}{(t-\lambda)(t+\lambda-1)\Gamma(s_1)\Gamma(s_2)\xi(2t-1)} \frac{4|k|^{t-\frac{1}{2}}\sigma_{1-2t}(k)}{\Gamma(t)} \\*
&\notag \qquad \qquad\,\,\,\, \times y^{\frac{1}{2}}K_{t-\frac{1}{2}}(2\pi |k|y) \frac{ \dd t}{4\pi i}\Big]  +\sum_{n=1}^\infty (\tilde{\mathcal{E}}(\lambda;s_1,s_2),\phi_n)a_k^{(n)}  y^{\frac{1}{2}}K_{it_n}(2\pi |k|y)\,.
\end{align}

Although a complete analysis of the instanton sector is beyond the scope of the present paper, we note that a naive attempt at extracting the large-$y$ behaviour of \eqref{eq:SpecEeInst} would produce an incorrect result. At first glance we may try and expand directly the different Bessel functions for large argument, thus immediately obtaining the expected exponential suppression factor $e^{-2\pi |k| y}$, hallmark of the $k$-instanton sector. However, by doing so the perturbative expansion on top of the instanton factor $q^k$, for $k>0$ with $q= e^{2\pi i z}$, or anti-instanton factor $\bar{q}^k$, for $k<0$, would start at order $y^0$ with sub-leading corrections $O(y^{-1})$, which turns out to be incorrect.

In \cite{Dorigoni:2021jfr,Dorigoni:2021ngn}, a representation for all generalised Eisenstein series with spectrum \eqref{eq:spec2} was provided in terms of iterated integrals of holomorphic Eisenstein series. This representation is extremely convenient for extracting all of the instanton expansions and, by comparing with the examples discussed in \cite{Dorigoni:2021jfr,Dorigoni:2021ngn}, we can clearly see that the above naive argument cannot possibly provide the correct answer for the generalised Eisenstein series with spectrum \eqref{eq:spec2}.  

Furthermore, in the same references, the authors discovered that amongst the coefficients of the perturbative expansion in the instanton sector, $e_k(\lambda;s_1,s_2\vert y)$, besides rational numbers and odd-zeta values, a new class of numbers appears whenever the eigenvalue $\lambda$ is such that the the vector space of \textit{holomorphic} cusp forms of modular weight $w=2\lambda$ has non-zero dimension. For these special eigenvalues the perturbative expansion at large-$y$ of $e_k(\lambda;s_1,s_2\vert y)$ contains non-critical completed L-values of holomorphic cusp forms.
Very recently in \cite{FKRinprogress} a very similar (albeit so far completely different in nature) phenomenon was discovered for the generalised Eisenstein series with spectrum \eqref{eq:spec1} for exactly the same eigenvalues.

It would be extremely interesting to extract the asymptotic expansion as $y\gg 1$ of the $k$-instanton sector \eqref{eq:SpecEeInst} and understand the origin of these completed L-values for holomorphic cusp forms from the spectral decomposition point of view \eqref{eq:SpecEeInst}. In particular, it is tantalising to conjecture some interplay between the non-holomorphic cusp forms and the appearance of holomorphic cusp forms.

\section{Conclusions}
\label{sec:Fin}

In this work, we have presented a family of Poincar\'e series \eqref{f_definition} which contains both string theory flavours of generalised Eisenstein series, namely higher derivative corrections in the low-energy effective action of type IIB superstring theory and integrated correlators coefficients from the gauge theory dual counter-part \eqref{eq:spec1}, as well as all two-loop modular graph functions \eqref{eq:spec2} from the low-energy expansion of perturbative string amplitudes at genus-one.

Besides giving a unifying picture, the newly introduced family of modular invariant functions manifest a variety of algebraic and differential relations. In particular, since this space is closed under the action of the Laplace operator \eqref{f_diff_eq_2}-\eqref{f_diff_eq}, we find a natural explanation \eqref{diff_eq_genEisen2}-\eqref{diff_eq_genEisen1} for the string theory spectra of eigenvalues and possible source terms~\eqref{eq:spec1}-\eqref{eq:spec2}.

From the Poincar\'e series integral representation~\eqref{modfn_Fourierzero_mode}, or equivalently from the spectral decomposition~\eqref{gen_eisen_zero_mode}, we derive in~\eqref{full_asymptotics} the complete asymptotic expansion as~$y\gg1$ for the Fourier zero-mode sector, as well as all non-perturbative, exponentially suppressed terms~\eqref{NP_con_expr}, which in the context of higher derivative corrections and integrated correlators correspond to instanton/anti-instanton events.

It would be interesting to repeat a similar analysis in the instanton sector, i.e. for Fourier non-zero mode, starting from the integral representation~\eqref{eq:SpecEeInst}. As shown in~\cite{Dorigoni:2021jfr,Dorigoni:2021ngn,FKRinprogress}, for particular eigenvalues the large-$y$ perturbative expansion of any Fourier non-zero mode coefficient for both flavours of generalised Eisenstein series \eqref{eq:spec1}-\eqref{eq:spec2} does contain L-values of holomorphic cusp forms. Obtaining these results from a Poincar\'e series or spectral function decomposition is as interesting as challenging. From the Poincar\'e series side this involves tackling infinite sums involving general Kloosterman sums \eqref{eq:nonzeromode}, while from the spectral decomposition side~\eqref{eq:SpecEeInst} we have to sum all contributions from non-holomorphic Maass cusp forms. 

Finally, we have also presented in \eqref{FZero_genEisen_asymp} the general expansion as $y\to0$, which crucially involves the non-trivial zeros of the Riemann zeta function. For the generalised Eisenstein series \eqref{eq:spec2} corresponding to two-loop MGFs this limit corresponds to a particular degeneration of the toroidal world-sheet. However, for the higher derivative corrections and integrated correlators coefficients~\eqref{eq:spec1} the limit $y\to0$ corresponds to the strong coupling regime $g_s\to\infty$, or equivalently $g_{_{YM}}^2\to \infty$ on the gauge theory dual side. We do not know why Riemann hypothesis should play any r\^{o}le in the strong coupling limit of string theory, nonetheless, we find this observation extremely fascinating and in need of further exploration.

\subsection*{Acknowledgements}

We would like to thank Nathan Benjamin, Cyuan-Han Chang, Ksenia Fedosova, Axel Kleinschmidt, Kim Klinger-Logan, Eric Perlmutter, Oliver Schlotterer, and Don Zagier for useful discussions. In particular we would like to thank Nathan Benjamin and Cyuan-Han Chang for helping us correct one of our results and Axel Kleinschmidt for comments on the draft. We are grateful to the organisers of the Pollica Summer Workshop ``New Connections between Physics and Number Theory'' supported by the Regione Campania, Università degli Studi di Salerno, Università degli Studi di Napoli "Federico II", the Physics Department "Ettore Pancini" and "E.R. Caianiello", and Istituto Nazionale di Fisica Nucleare. DD would also like to thank the Albert Einstein Institute, Golm, for the hospitality during the final stages of this project. 

\appendix

\section{Convergence of the Poincar\'e series}\label{app:AsyOrigin}

In this appendix we discuss the region in parameter space, $(a,b,r,s)$, for which the Poincar\'e series \eqref{f_definition} converges absolutely. 
This will be achieved by constructing an auxiliary Poincar\'e series which has the same domain of absolute convergence but it is easier to analyse. 

We start by observing that under a modular transformation $\gamma\in\SLtwoZ$ the magnitude of the seed functions $\seedfn{a}{b}{r}{s}{z}$ is bounded from above by an $x$-independent function  
$$|\seedfn{a}{b}{r}{s}{\gamma\cdot z}| \leq \sum_{m\neq 0}\Big|\big[\sigma_a(m)|m|^{b-\half} y^{r+\half} K_{s-\half}(2\pi |m|y)\big]_{\gamma}\Big|\,,$$
simple consequence of triangle inequality combined with
 $$\Big|\big[e^{2\pi ix}\big]_{\gamma}\Big|=1\,.$$
 Motivated by this observation, we define the auxiliary Poincar\'e series
 \allowdisplaybreaks{
\begin{align}
    \psi(a,b,r,s\vert y)&:= \sum_{m=1}^\infty \sigma_a(m)m^{b-\half}y^{r+\half} K_{s-\half}(2\pi my)\,,\\*
     \label{Phi_definition}\Psi(a,b,r,s\vert z)&:=\sum_{\gamma\in\poin}\big[\psi(a,b,r,s\vert y)\big]_{\gamma}\,,
\end{align}}
and notice that the auxiliary Poincar\'e series \eqref{Phi_definition} converges absolutely if and only if the original Poincar\'e series \eqref{f_definition} does.

We continue by showing that \eqref{Phi_definition} can be written in terms of a contour integral thus manifesting the convergence properties of the Poincar\'e series. 
Given a function $f(t)$ we define its Mellin transform as
\begin{equation}
\mathcal{M} [ f ] (t) := \int_0^\infty f(t)\, y^{t}\, \frac{\dd y}{y}\,,\label{eq:Mellin}
\end{equation}
and proceed to compute the Mellin transform of our new seed function
\begin{align}
\tilde{\psi}(a,b,r,s\vert t)&:=\label{eq:Mellphi}\mathcal{M}[\psi(a,b,r,s)](t) =\int_{0}^\infty \psi(a,b,r,s\vert y)\, y^{t}\, \frac{\dd y}{y}\\
&\notag\phantom{:}=  \frac{1}{4\pi^{t+r+\half}}\Gamma\Big(\frac{t+r+1-s}{2}\Big)\Gamma\Big(\frac{t+r+s}{2}\Big)\zeta(t+r+1-b)\zeta(t+r+1-a-b)\,,
\end{align}
using the identities
\begin{align}
    \int_0^\infty K_s(y)y^b \dd y &= 2^{b-1}\Gamma\big(\frac{b+1-s}{2}\big)\Gamma\big(\frac{b+s+1}{2}\big)\,,\\
    \sum_{m=1}^\infty \sigma_a(m)m^b &= \zeta(-a-b)\zeta(-b)\,.
\end{align}
The Mellin transform \eqref{eq:Mellphi} is well-defined in the strip 
\begin{equation}
\Re(t)>\alpha = \max{\{\Re(s-r-1),\Re(-s-r),\Re(b-r),\Re(a+b-r)\}}\,.\label{eq:strip}
\end{equation}

We can now apply Mellin inversion formula to obtain the integral representation
\begin{equation}\label{Z_int_rep}
    \psi(a,b,r,s\vert y) = \mathcal{M}^{-1}[\tilde{\psi}(a,b,r,s)](y)= \int_{\beta-i\infty}^{\beta+i\infty} \tilde{\psi}(a,b,r,s\vert t)\,y^{-t} \frac{\dd t}{2\pi i}\,,
\end{equation}
where $\beta$ is an arbitrary constant such that $\beta>\alpha$. The reason to derive \eqref{Z_int_rep} is that all of the explicit $y$ dependence has now been reduced to the simple term $y^{-t}$. At this point we can easily perform the Poincar\'e series \eqref{Phi_definition} arriving at
\begin{equation}\label{Phi_int_rep}
    \Psi(a,b,r,s\vert z) = \int_{\beta-i\infty}^{\beta+i\infty} \!\!  \tilde{\psi}(a,b,r,s\vert t)\, \normEisenstein{{-}t}{z} \frac{\dd t}{2\pi i}\,.
\end{equation}
The absolute convergence of the auxiliary Poincar\'e series \eqref{Phi_definition}, and hence of the original Poincar\'e series \eqref{f_definition}, is then equivalent to understanding the conditions for which the Poincar\'e series of the integral representation \eqref{Z_int_rep} is absolutely convergent. This question is much easier to answer: with \eqref{Z_int_rep} the problem has been reduced to the convergence of the Poincar\'e series for Eisenstein series \eqref{Eisen_poinc}. 
We conclude that absolute convergence of \eqref{Phi_definition} and \eqref{f_definition} is guaranteed whenever
$$
\Re(-t) = -\beta > 1 \qquad\Rightarrow  \qquad \alpha <\beta < -1\,,
$$  
which, upon use of the condition \eqref{eq:strip} for a well-defined Mellin transform, reproduces precisely the domain in parameters space \eqref{eq:AbsConv}  stated in the main text  
$$
 \min{\{\Re(r+1-s),\Re(r+s),\Re(r-b),\Re(r-a-b)\}}>1\,.
$$

It is interesting to note that the integral representation \eqref{Phi_int_rep} implies that the spectral overlap $(\Psi(a,b,r,s),\phi_n)$ vanishes for all Maass cusp forms $\phi_n(z)$; such a result is not expected to hold for the more complicated $\modfntwo{a}{b}{r}{s}$.

\section{Mellin-Barnes representation}\label{app:AsyCusp}

In this appendix we derive a Mellin-Barnes representation for the Fourier zero-mode $ \fzero{a}{b}{r}{s}{y}$, starting from the general integral representation  \eqref{Fzero_poinc_formula} specialised to the seed function \eqref{seed_def} under consideration.
Hence we start by considering
\begin{align}
 &   \fzero{a}{b}{r}{s}{y} \label{eq:fzerostartApp} \\
 &\notag= \sum_{d=1}^\infty \sum_{m\neq 0} S(m,0;d) \int_{\mathbb{R}} \! e^{-2\pi im \frac{\omega}{d^2(\omega^2 + y^2)}} \sigma_a(m)|m|^{b-\half} \Big(\frac{y}{d^2(\omega^2 + y^2)}\Big)^{r+\half} K_{s-\half}\Big(\frac{2\pi|m|y}{d^2(\omega^2 + y^2)}\Big)\dd\omega.
\end{align}
The Bessel function can now be substituted by its Mellin-Barnes integral representation
\begin{equation}
    K_s(y) =\left(\frac{y}{2}\right)^s \int_{\alpha - i\infty}^{\alpha+i\infty} \Gamma(t)\Gamma(t-s)\Big(\frac{y}{2}\Big)^{-2t} \frac{\dd t}{4\pi i},
\end{equation}
where $\alpha$ is a real parameter such that $\alpha > \max{\{\Re (s),0\}}$. 
To perform the integral over $\omega$ we furthermore expand the exponential as
\begin{equation}
    e^{-2\pi im \frac{\omega}{d^2(\omega^2 + y^2)}} = \sum_{k=0}^\infty \frac{1}{k!} \Big( \frac{-2\pi i m\omega}{d^2(\omega^2+y^2)}\Big)^k.
\end{equation}

Substituting both the Mellin-Barnes representation for the Bessel function and the above convergent expansion in \eqref{eq:fzerostartApp}, we obtain
\begin{multline}
    \fzero{a}{b}{r}{s}{y} =\sum_{d=1}^\infty  \sum_{m\neq 0}\sum_{k=0}^\infty \int_{\mathbb{R}}\int_{\alpha-i\infty}^{\alpha +i\infty} S(m,0;d) \sigma_a(m) |m|^{b-\half}\Big(\frac{y}{d^2(\omega^2 + y^2)}\Big)^{r+\half}\\
    \Big(\frac{-2\pi im\omega}{d^2(\omega^2+y^2)}\Big)^k\Big(\frac{\pi |m|y}{d^2(\omega^2+y^2)}\Big)^{s-2t-\half}\,\frac{\Gamma(t)\Gamma(t-s+\half)}{k!} \frac{\dd t \,\dd\omega}{4\pi i}\,.
\end{multline}
The integral over $\omega$ can be performed
\begin{equation}\label{omega_integral}
   \int_{\mathbb{R}} \frac{\omega^k}{(\omega^2+y^2)^{k+r+s-2t }} \,\dd\omega = \frac{[1+(-1)^k]y^{4t+1-k-2r-2s}\,\Gamma(\frac{k+1}{2})\Gamma(\frac{k-1}{2}+r+s-2t)}{2\Gamma(k+r+s-2t )},
\end{equation}
provided that the integrand falls-off sufficiently fast as $\omega \to \pm \infty$, which in turns requires the parameter $\alpha$ to be bounded from above by $4\alpha < k+2\Re (r+s)-2$. 

Under the conditions \eqref{eq:AbsConv} for absolute convergence of the Poincar\'e series, we can easily see that for all $k\in \mathbb{N}$ the constraints on the parameter $\alpha$: 
$$ \max{\{\Re (s),0\}} < \alpha< \frac{k+2\Re (r+s)-1}{4}\,,$$
always admit a non-vanishing strip of allowed integration contours in $t$.

At this point, we focus on the series in $m,\,d$ and  $k$. Firstly, given the explicit expression \eqref{eq:KloostEasy} for the Kloosterman sum $S(m,0;d)$ we use that $r\in(\mathbb{Z}/d\mathbb{Z})^\times$ implies $-r\in(\mathbb{Z}/d\mathbb{Z})^\times$ to derive $S(m,0;d)=S(-m,0;d)$. We can then replace the sum over all non-zero integers $m$ by twice the sum over the positive integers $m>0$. Secondly, it is possible to evaluate explicitly the sum over $d$, which takes the form of a well-known Dirichlet series for the Ramanujan sum $S(m,0;d)$,
\begin{equation}
   \sum_{d=1}^\infty \frac{S(m,0;d)}{d^{\tilde{s}}} = \frac{\sigma_{1-\tilde{s}}(m)}{\zeta(\tilde{s})}\,,
\end{equation}
specialised to $\tilde{s}=2r+2k+2s-4t$.
Finally, we note that the term $[1+(-1)^k]$ in the numerator of \eqref{omega_integral} restricts the sum over $k$ to only run over even integers $2k$. 

When the dust settles and after performing the change of variables $t\to \frac{t+ r+s-1}{2}$, we are left with the expression

\begin{align}
   &  \fzero{a}{b}{r}{s}{y} \\
     &\notag =\!\!\sum_{m=1}^\infty \sum_{k=0}^\infty \int_{\frac{1}{2}-i\infty}^{\frac{1}{2}+i\infty} \!\!\frac{\sigma_a(m)\sigma_{2t-4k-1}(m)}{m^{t+r-2k-b}}
     \frac{(\shortminus1)^k \pi^{2k+1-r-t}\Gamma(k\!+\!\frac{1}{2}\!-\!t)\Gamma\big(\frac{t+r-s}{2}\big)\Gamma\big(\frac{t+r+s-1}{2}\big)}{\Gamma(2k+1-t)\zeta(4k+2-2t)k!} y^{t-2k} \frac{\dd t}{4\pi i} \,.
\end{align}

The next sum to evaluate is that over $k$. To this end, we begin by making the change of variable $t\to t'=t-2k$, thus shifting the contour of integration from $\Re(t)=\frac{1}{2}$ to $\Re(t')=\frac{1}{2}-2k$ and, after having changed the integration variable back to $t$, we are left with
\allowdisplaybreaks{
\begin{align}
   &  \fzero{a}{b}{r}{s}{y} \\*
     &\notag =\!\!\sum_{m=1}^\infty \sum_{k=0}^\infty \int_{\frac{1}{2}-2k-i\infty}^{\frac{1}{2}-2k+i\infty} \!\frac{\sigma_a(m)\sigma_{2t-1}(m)}{m^{t+r-b}}
     \frac{(\shortminus 1)^k \pi^{1\shortminus t \shortminus r}\Gamma(\frac{1}{2}{-}k{-}t)\Gamma(\frac{t+2k+r-s}{2})\Gamma(\frac{t+2k+r+s-1}{2})}{\Gamma(1-t)\zeta(2-2t)k!}y^{t}\frac{\dd t}{4\pi i}\,.
\end{align}}

\begin{figure}
\centering
\begin{minipage}{.5\textwidth}
  \centering
  \includegraphics[width=.4\linewidth]{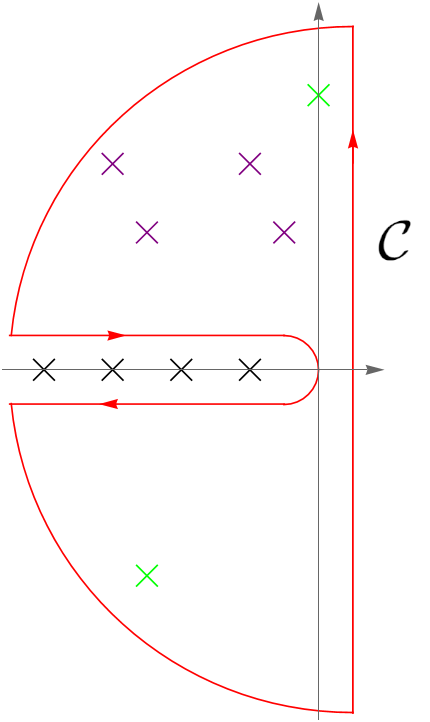}
\end{minipage}%
\begin{minipage}{.5\textwidth}
  \centering
  \includegraphics[width=.37\linewidth]{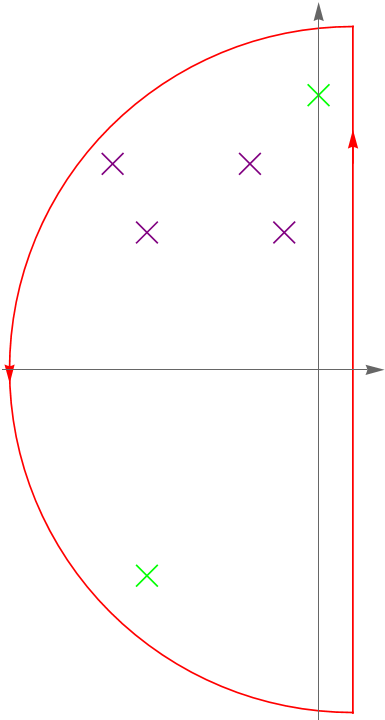}
\end{minipage}
\caption{On the left we show the auxiliary integration contour $\mathcal{C}$ and on the right the deformed contour.  The poles from the gamma functions are depicted in purple, from the zeta functions in green and from the trigonometric functions in black.} \label{eq:contour}
\end{figure}

We would like to translate the shifted integration contour back to its original position at $\Re(t)=\frac{1}{2}$, however, additional poles originating from $\Gamma(\frac{1}{2}-k-t)$ appear at $t=\frac{1}{2}-\ell$, with $\ell \in \mathbb{N}$ and $0<\ell\leq k$. Although the shifted contour cannot be moved back immediately to its initial place, we can nevertheless rewrite it as a sum of two different contours: the original one along $\Re(t)=\frac{1}{2}$ and a new contour encircling these new poles along the negative $t$-axis. As depicted in figure \ref{eq:contour}, these two contours can be connected at infinity to form a single auxiliary contour of integration $\mathcal{C}$ which is independent from the summation variable $k$. We exchange the sum over $k$ with the integral over $\mathcal{C}$ and perform the sum over $k$
\begin{align}
& \label{long_sum_eq}   \sum_{k=0}^\infty \frac{(-1)^k\Gamma(\frac{1}{2}-k-t)\Gamma(\frac{2k+r+t-s}{2})\Gamma(\frac{2k+r+s+t-1}{2})}{k!} \\
&\notag= \frac{\sin{[\pi(r-t)]}+\sin{(\pi s)}}{2 \sin(\pi r) \cos(\pi t)}\frac{ \Gamma\big(\frac{r+1-s-t}{2}\big)\Gamma\big(\frac{r+s-t}{2}\big)\Gamma\big(\frac{t+r-s}{2}\big)\Gamma\big(\frac{t+r+s-1}{2}\big)}{\Gamma(r)}\,.
\end{align}
We are then left with the expression

\begin{align} \label{fzero_contour_rep}
   \fzero{a}{b}{r}{s}{y}  = \sum_{m>0} \int_{\mathcal{C}}&\Big( \frac{\sin{[\pi(r-t)]}+\sin{(\pi s)}}{2 \sin(\pi r) \cos(\pi t)}\Big)\Big( \frac{\sigma_a(m)\sigma_{2t-1}(m)}{m^{t+r-b}}\Big) \\
&\notag   \times 
    \frac{\Gamma\big(\frac{r+1-s-t}{2}\big)\Gamma\big(\frac{r+s-t}{2}\big)\Gamma\big(\frac{t+r-s}{2}\big)\Gamma\big(\frac{t+r+s-1}{2}\big)}{\pi^{r} \Gamma(r)\xi(2-2t)}y^t \frac{ \dd t}{4\pi i}\, .
\end{align}

Since the integration contour $\mathcal{C}$ is closed, the integral is uniquely fixed by the residues at the poles in the interior of $\mathcal{C}$.
The only poles situated in the interior of the contour $\mathcal{C}$ are located at $t=-2n+s-r$ and $t=-2n+1-s-r$ for $n\in \mathbb{N}$ and come from the last two gamma functions at numerator in the above integrand.
 Furthermore, we notice that at these pole locations the ratio of trigonometric factors in \eqref{fzero_contour_rep} always evaluates to $1$.
 We conclude that this ratio of trigonometric terms can be dropped from the contour integral \eqref{fzero_contour_rep} without changing the result
\begin{align} \label{eq:aug_integral}
   \fzero{a}{b}{r}{s}{y} & = \sum_{m>0} \int_{\mathcal{C}} \Big( \frac{\sigma_a(m)\sigma_{2t-1}(m)}{m^{t+r-b}}\Big) 
    \frac{\Gamma\big(\frac{r+1-s-t}{2}\big)\Gamma\big(\frac{r+s-t}{2}\big)\Gamma\big(\frac{t+r-s}{2}\big)\Gamma\big(\frac{t+r+s-1}{2}\big)}{\pi^{r} \Gamma(r)\xi(2-2t)}y^t \frac{ \dd t}{4\pi i}\, .
\end{align}

Since the trigonometric factors have been removed, we have that the previously mentioned poles which were located on the negative $t$-axis at $t=\frac{1}{2}-\ell$, with $\ell \in \mathbb{N}$ are no longer present in \eqref{eq:aug_integral}. As depicted in figure \ref{eq:contour}, we are now free to deform the auxiliary contour of integration $\mathcal{C}$ to an infinite semi-circle. The contribution from the circle at infinity vanishes and the only non-trivial contribution to the integral comes from the line $\Re(t)=\frac{1}{2}$, hence we have managed to restore the original contour of integration. 

Finally, we turn to the sum over $m$. We notice that at large-$m$ the summand is bounded by
$$
\Big\vert \frac{\sigma_a(m)\sigma_{2t-1}(m)}{m^{t+r-b}}\Big\vert = O\big(m^{- [ \Re (r-b) -{\rm max}\{\Re(a),0\} ] }\big)\,,
$$ 
and, thanks to the conditions \eqref{eq:AbsConv} for the absolute convergence of the Poincar\'e series, we can easily see that $ \Re (r-b) -{\rm max}\{\Re(a),0\}>1$ for the range of parameters considered, hence this sum converges absolutely (note that for the convergence of this sum it is crucial we managed to reduce the contour $\mathcal{C}$ back to just the line $\Re(t)=\frac{1}{2}$). We can then use a well-known identity due to Ramanujan,
\begin{equation}\label{Ramanujan_sigma}
\sum_{m>0}\frac{\sigma_a(m)\sigma_{b}(m)}{m^{s}} =\frac{\zeta (s) \zeta (s-a) \zeta (s-b) \zeta (s-a-b)}{\zeta (2s-a-b)}\,,
\end{equation}
specialised to the case $b=2t-1\,,\,s=r+t-b$ and substitute it back in equation \eqref{fzero_contour_rep}.

Our final result is the Mellin-Barnes integral representation for the Fourier zero-mode,
\begin{equation}\label{modfn_Fourierzero_modeApp}
    \fzero{a}{b}{r}{s}{y}= \int_{\frac{1}{2}-i\infty}^{\frac{1}{2}+i\infty} U(a,b,r,s\vert t)\,y^t  \frac{\dd t}{2\pi i},
\end{equation}
where we define
\begin{multline}\label{U_definitionApp}
   U(a,b,r,s\vert t) :=    \frac{\Gamma\big(\frac{r+1-s-t}{2}\big)\Gamma\big(\frac{r+s-t}{2}\big)\Gamma\big(\frac{t+r-s}{2}\big)\Gamma\big(\frac{t+r+s-1}{2}\big)}{2\pi^r\, \Gamma(r)\xi(2-2t)}\\ \times
   \frac{\zeta(r+1-b-t)\zeta(r+1-a-b-t)
    \zeta(t+r-b)\zeta(t+r-a-b)}{\zeta(2r+1-a-2b)}\,.
\end{multline}

The Mellin-Barnes integral representation \eqref{modfn_Fourierzero_modeApp} can be analytically continued to values of parameters, $(a,b,r,s)$, for which the Poincar\'e series \eqref{f_definition} is not absolutely convergent. In general, rather than the vertical line $\Re(t) = \half$, the integration contour, $\gamma$, in \eqref{modfn_Fourierzero_modeApp} has to be chosen such that it separates two sets of poles of \eqref{U_definitionApp}.
 The contour $\gamma$ is such that the poles coming from
$$
\Gamma\big(\frac{t+r-s}{2}\big)\Gamma\big(\frac{t+r+s-1}{2}\big)
    \zeta(t+r-b)\zeta(t+r-a-b)\,,
$$
are located to the left of $\gamma$, while the remaining poles coming from
$$
 \frac{\Gamma\big(\frac{r+1-s-t}{2}\big)\Gamma\big(\frac{r+s-t}{2}\big)\zeta(r+1-b-t)\zeta(r+1-a-b-t)
}{\xi(2-2t)}\,,
$$
are located to the right of $\gamma$.

\bibliography{cites}
\bibliographystyle{utphys}

\end{document}